\documentclass{aa}  

\usepackage{graphicx}
\usepackage{booktabs}
\usepackage{txfonts}
\usepackage{hyperref}
\hypersetup{
     colorlinks = true,
     linkcolor = blue,
     anchorcolor = blue,
     citecolor = blue,
     filecolor = blue,
     urlcolor = blue
     }
\begin{document}

   \title{ALMA Chemical Evolution (ACE) survey: dust-to-gas ratios in sub-solar metallicity galaxies at cosmic noon}


    \author{N. N. Geesink
          \inst{1},
          G. Popping
          \inst{1},
          L. A. Boogaard
          \inst{2},
          I. Langan
          \inst{3},
          A. Pope
          \inst{4},
          R. Popescu
          \inst{4},
          I. Shivaei
          \inst{3},
          M. Solimano
          \inst{3},
          M. Kaasinen
          \inst{5},
          B. Mobasher
          \inst{6},
          D. Narayanan
          \inst{7,8},
          M. Parente
          \inst{7},
          N. Reddy
          \inst{6},
          R. L. Sanders
          \inst{9}}

   \institute{European Southern Observatory, Karl-Schwarzschild-Str. 2 D-85748, Garching, Germany\\
              \email{Nikki.Geesink@eso.org}
         \and
             Leiden Observatory, Leiden University, PO Box 9513, NL-2300 RA Leiden, The Netherlands 
        \and
             Centre for Astrobiology, Madrid
        \and
            Department of Astronomy, University of Massachusetts, Amherst, MA 01003, USA
         \and
             Research School of Astronomy and Astrophysics, Australian National University, Canberra, ACT 2611, Australia
        \and
             Department of Physics and Astronomy, University of California, Riverside, 900 University Avenue, Riverside, CA 92521, USA   
         \and
             Department of Astronomy, University of Florida, 211 Bryant Space Sciences Center, Gainesville, FL 32611 USA
         \and
             Cosmic Dawn Center at the Niels Bohr Institute, University of Copenhagen and DTU-Space, Technical University of Denmark
         \and
            Department of Physics and Astronomy, University of Kentucky, 505 Rose Street, Lexington, KY 40506, USA
 }

   \date{Received ; accepted}

  \abstract
{Dust is a fundamental component of the interstellar medium and provides a key tracer of the baryon cycle that regulates galaxy evolution. In particular, the dust-to-gas ratio links metals in the gas phase to those locked into dust grains, making it a sensitive diagnostic of dust production, grain growth, and destruction. We present measurements of the dust-to-molecular-gas ratio ($\rm DGR_{mol}$), for typical star-forming galaxies ($\log M_\star \approx 10$) at sub-solar metallicites, at $z\simeq2-2.5$ from the ALMA Chemical Evolution (ACE) survey. By combining ALMA CO and dust-continuum observations with robust gas-phase metallicity measurements, ACE extends direct dust and molecular-gas measurements to lower stellar masses and lower metallicities than previously available at this epoch, reaching down to $0.4\,Z_{\odot}$. This enables the first constraints on the $\rm DGR_{mol}$--metallicity relation for typical unlensed galaxies at cosmic noon. We find that $\rm DGR_{mol}$ increases with metallicity, with a log-space slope of $1.2 \pm 0.7$, indicating that metal-poor galaxies have systematically lower $\rm DGR_{mol}$  than their more metal-rich counterparts. Results on more metal rich galaxies at cosmic noon for which dust-to-gas constraints are available agree with this relation within the observed scatter. For the detected ACE galaxies, we measure a mean value of $\log_{10}(M_{\rm dust}/M_{\rm mol})=-2.37\pm0.05$ for a mean metallicity of 12+$\log (\rm O/H) = 8.45 \pm 0.02$. We find agreement with $\rm DGR_{mol}$ in the local Universe at fixed metallicities, indicating that the same dust-growth physics, likely grain growth in the ISM, dominates at metallicities of $8.3 \leq 12+\log(\rm O/H) \leq 8.7$ at cosmic noon. These measurements provide novel empirical constraints for models of dust enrichment and galaxy evolution during the peak epoch of cosmic star formation. Additionally, ACE provides a sub-solar metallicity reference for the calibration of dust continuum as tracer of molecular gas, essential for studying metal-poor, high-redshift systems.}

\titlerunning{Dust-to-gas ratios in sub-solar metallicity galaxies at cosmic noon}
\authorrunning{N.N. Geesink et al.}
\maketitle
%

\section{Introduction}

Dust is a central component of galaxy evolution, tracing the production and cycling of metals within galaxies and their surroundings, while also shaping the thermal and chemical state of the interstellar medium \citep[ISM,][]{Galliano2018}. By enabling molecular gas formation and shielding cold gas from dissociating radiation, dust links chemical enrichment to the reservoirs that fuel star formation. Cosmic noon ($z \sim 1$–3), when cosmic star-formation activity and molecular-gas density reached their peaks, therefore provides a critical window for studying the buildup of dust in rapidly evolving galaxies \citep{Madau2014, Walter2020, Zavala2021}. Over the past decade, the Atacama Large Millimeter/submillimeter Array (ALMA) has transformed our view of galaxies at cosmic noon by probing their cold gas and dust reservoirs, from intensely dusty starbursts to main-sequence galaxies \citep{Aravena2020,Hodge2020, Boogaard2020, Tacconi2020}. However, one of the fundamental quantities needed to interpret these dust and gas reservoirs remains comparatively uncertain: the gas-phase metallicity (12+log(O/H))\footnote{Hereafter, we use metallicity to refer to the gas-phase oxygen abundance for readability}. Since metallicity sets the level of chemical enrichment and is connected to dust production, grain growth, and destruction, relating dust-to-gas ratios to metallicity offers a direct way of constraining the physical processes that shape dust evolution in galaxies.

The dust-to-gas ratio (DGR) - metallicity relation is a powerful probe of dust evolution because it reflects the balance between stellar dust injection, grain growth in the dense ISM, and dust destruction by shocks \citep{Draine2007, Galliano2018}. Models suggest that, as metallicity increases, grain growth in the ISM becomes increasingly efficient, with this transition often characterized by a model-dependent critical metallicity regime, typically of order $Z_{\rm crit} \sim 0.1{-}0.5\,Z_{\odot}$  \citep{Asano2013,Feldmann2015,Popping2017,Parente2025}. In this regime the dust buildup shifts from being dominated by stellar dust production to being dominated by grain growth in the ISM. In the local Universe, the DGR of galaxies increases with metallicity but the form of this relation, especially at low metallicity, remains debated. Some studies find evidence for a critical-metallicity regime, with a pronounced steepening of the DGR–metallicity relation toward low metallicities \citep{RemyRuyer2014,Galliano2021}, whereas other analyses find that a single power law can provide an adequate description over a broad metallicity range \citep{Draine2007,Leroy2011,DeVis2019}. Resolved studies of nearby galaxies suggest that variations in the DGR are not governed by metallicity alone, but that the gas density and the molecular gas to total gas fraction also play a significant role \citep{Vilchez2019,Park2024}. Beyond its use as a probe of dust physics, the DGR is also used to to obtain gas-mass estimates from dust observations \citep[e.g.][]{Hildebrand1983}. Either through translating dust masses based on far-infrared spectral energy distribution (SED) fitting by adopting a dust to gas ratio \citep{Magdis2012, Bethermin2015} or by converting long-wavelength dust-continuum measurements to molecular gas masses \citep[Rayleigh-Jeans tail method;][]{Scoville2014, Scoville2016, Scoville2017, Groves2015,Schinnerer2016}. Cosmological zoom-in simulations also support the ability to use single-band dust luminosities as molecular gas mass estimator \citep{Privon2018, Liang2018}. While these methods typically recover gas masses with relatively modest scatter, their calibrations are largely based on massive, relatively metal-rich systems, and become increasingly uncertain at low metallicity where variations in the DGR can introduce substantially larger uncertainties.

A key question is whether the DGR–metallicity relation and its underlying physical drivers remain the same under the different ISM conditions over cosmic time. So far, our understanding of the cosmic evolution of the DGR-metallicity relation has come primarily from studies of damped Ly$\alpha$ absorbers, for which there is no evolution with redshift up to z =5 (\citealp[and references therein]{Peroux2020}; \citealp{PoppingPeroux22}). However, these systems are selected through narrow absorption sightlines toward background sources, and therefore probe individual paths through neutral gas reservoirs rather than the galaxy-wide dust, gas, and metal content of emission-selected galaxies. As a result, they cannot directly establish how the DGR relates to global galaxy properties at high redshift, nor can they be straightforwardly compared with galaxy-wide measurements in the local Universe. Addressing this gap requires spatially integrated constraints on dust, gas, and metallicity for high-redshift galaxies. In practice, this is observationally challenging, as it demands well-sampled SEDs with deep dust-continuum observations (dust), reliable cold-gas tracers together with well-calibrated conversion factors (gas), and multiple rest-frame optical emission line observations (metallicity)

Thus far, only a few studies at cosmic noon have constrained the DGR-metallicity relation using integrated galaxy observations. \citet{Saintonge2013} presented one of the earliest comparisons for seven lensed submillimeter galaxies (SMGs) at \(2\lesssim z\lesssim 3\)  over a 0.6 dex metallicity range ($12+\log(\mathrm{O/H})=8.36-8.91$). They found an increase of DGR with metallicity, but reported values below the local relation from \citet{Leroy2011} at fixed metallicity. \citet{Shapley2020} provided constraints on the DGR–metallicity relation of four galaxies at \(1.4\lesssim z\lesssim 2.5\), but over only a narrow, near-solar metallicity range, ($12+\log(\mathrm{O/H})=8.59-8.69$), and found no significant evolution in the normalization relative to $z\sim0$. \citet{Popping2023} extended the work on $z \sim 2$ unlensed galaxies to lower metallicities, probing $8.4<12+\log(\mathrm{O/H})<8.8$, but with only three individual CO and dust detections, they had to rely on upper limits and stacked measurements to constrain the trend. Their results support an increasing DGR with metallicity and are consistent with little or no evolution relative to the local relation, although the inferred slope depends on the adopted CO-to-$\rm H_2$ conversion factor. Taken together, these studies suggest that DGR increases with metallicity and show no conclusive evidence for evolution relative to $z\sim0$ relations. However, the unlensed galaxy samples remain too small, and span too limited a metallicity range, to robustly constrain the slope, scatter, or possible nonlinearity of the DGR–metallicity relation at cosmic noon.

Beyond cosmic noon ($z \geq 3$) , integrated galaxy studies remain limited but generally find DGRs below those expected from local DGR-metallicity relations \citep{Heintz2025,  Algera2026}. This offset can reflect less time for grain growth in the ISM and/or more efficient dust destruction in the early universe, but the interpretation is limited by systematic uncertainties in the gas-mass estimates. At these earlier epochs, the  brighter $[\mathrm{CII}]\,158\,\mu\mathrm{m}$ line has been more widely used as a molecular gas tracer compared to CO lines. However, $[\mathrm{CII}]$ is difficult to calibrate as a tracer of cold molecular gas since it traces multiple gas phases \citep[e.g.][]{Casavecchia2025}, leading to a wide range of values adopted in the literature and correspondingly large uncertainties in the inferred gas masses \citep{Zanella2018,Kaasinen2024,Vallini2025}. One of the best-characterized individual systems so far is HZ10 at $z=5.65$, for which \citet{Algera2025} combined a well-sampled dust SED, JWST metallicity constraints \citep{Villanueva2024, Jones2025}, and both CO- and $[\mathrm{CII}]$-based gas measurements. They found that HZ10 lies below local DGR–metallicity relations.

Despite recent progress, the DGR–metallicity relation at high redshift remains poorly constrained, especially outside the metal-rich regime. This is not only an observational gap, but also a limitation for theory, because the normalization and shape of the relation provide a direct test of how dust formation, grain growth, and destruction are implemented in models of galaxy evolution. Recent cosmological hydrodynamical and semi-analytical models now explicitly follow the buildup and processing of dust over cosmic time \citep{Popping2017,McKinnon2018,Hou2019,Li2019,Vijayan2019,Triani2020,Parente2022, Parente23} and broadly reproduce the observed trends up to $z\!\sim\!5$ \citep{PoppingPeroux22}. However some discrepancies remain: some models predict little redshift evolution in the normalization of the DGR–metallicity relation \citep{Popping2017, Li2019, Vijayan2019, Triani2020}, whereas other models \citep{McKinnon2018, Hou2019, Parente23} predict a measurable redshift dependence. These differences reflect varying prescriptions for dust processing in the ISM, which determine how rapidly dust growth can keep pace with metal enrichment at high redshift. The differences in models are most pronounced at low metallicity, where dust is dominated by stellar sources and is therefore highly sensitive to the assumed stellar yields and supernova destruction rates through shocks. Efficient grain growth in the ISM drives the increase in DGR above a critical metallicity, $Z_{\rm crit}$, but the predicted location of this turnover varies greatly, likely because of differences in grain-growth efficiency as well as in the underlying star-formation and dust-processing timescales \citep{PoppingPeroux22, Parente2022}. Constraining the low-metallicity regime is therefore key to distinguishing between competing descriptions of dust buildup and destruction in cosmological models of galaxies.

This paper aims to constrain the DGR--metallicity relation over a previously unexplored metallicity range for unlensed typical star-forming galaxies at $z\sim2-2.5$, using data from the ALMA Chemical Evolution (ACE) Survey (Shivaei et al.~(2026)). The ACE survey probes the dust and molecular gas properties of 25 galaxies down to $0.4\,Z_\odot$ and stellar masses of $10^9 \,\rm M_{\odot}$, selected from the MOSFIRE Deep Evolution Field (MOSDEF) Survey \citep{Kriek2015}. The ACE survey provides observations of ALMA Band 3 (CO(3-2)) and Band 7 (dust continuum), while the galaxies have the strong rest-frame optical emission lines from the MOSDEF survey required for robust metallicity estimates. The data is complemented by ALMA Band 6 data to further constrain the dust emission  \citep{Shivaei22}, and extensive ancillary data from the Hubble Space Telescope (HST) and the James Webb Space Telescope (JWST) \citep{Scoville2007, Shivaei22, Casey2023, Shuntov2025}. Thus, ACE provides the first sample of unlensed star-forming galaxies at cosmic noon with both dust and molecular gas constraints in the sub-solar metallicity regime, extending measurements down to $0.4\,Z_\odot$. These observations therefore begin to close the gap toward the critical-metallicity regime predicted by dust-evolution models, where the efficiency of ISM grain growth is expected to transition, providing an important new benchmark at cosmic noon. 

This paper is organized as follows. In Section~\ref{methods}, we describe the sample, the methodology used to derive dust and gas masses, and the stacking methodology. In Section~\ref{results}, we present the results. In Section~\ref{discussion}, we discuss our findings and compare them with predictions from simulations. Finally, in Section~\ref{conclusions}, we summarize our main conclusions. Throughout this paper, we adopt a flat \(\Lambda\)CDM cosmology with $H_0 = 67.7\ {\rm km\ s^{-1}\ Mpc^{-1}}$, $\Omega_{\rm m} = 0.31$, and $\Omega_{\Lambda} = 0.69$, consistent with \citet{Planck2020}. Stellar masses and star formation rates assume a \citet{Chabrier2003} initial mass function. Metallicity is defined in terms of the gas-phase oxygen abundance, $12+\log({\rm O/H})$, and throughout the paper we adopt a solar metallicity of $12+\log({\rm O/H})_\odot = 8.69$ \citep{Asplund2009}.

\section{Data \& Methodology}
\label{methods}
We study the DGR of galaxies at cosmic noon using CO(3--2) (ALMA Band~3) and dust continuum (Band~7) observations obtained as part of the ALMA Chemical Evolution (ACE) Large Program (program 2024.1.00534.L; PIs: I.~Shivaei \& G.~Popping). The ACE survey,  presented in \citet{Shivaei2026}, targets 25 main-sequence galaxies at $\sim 2$-2.5, uniquely probing a lower stellar-mass ($\log(\rm M_*/M_\odot)=9.13-10.62$)and metallicity regime (( $8.19 \leq 12+\log(\mathrm{O/H}) < 8.61$ ) ) than previously explored at this epoch. For the molecular gas measurements, ACE targeted the CO(3--2) line in 22 out of the 25 galaxies. The dataset is complemented by CO(3--2) observations from \citet{Sanders2023} for three additional galaxies.  The galaxies are drawn from the MOSDEF survey \citep{Kriek2015}, providing extensive ancillary multi-wavelength data and spectroscopic measurements that enable robust metallicity determinations. The metallicities are derived using the latest calibration for \(z>1\) based on direct-\(T_{\rm e}\) measurements \citep{Sanders2026}. For a more comprehensive description of the ACE program, we refer to Shivaei et al.~(2026), which reports updated measurements of the fundamental galaxy properties, including star formation rate, stellar mass, and metallicity. The ACE survey builds upon an earlier, shallower Band~6 ALMA program (program 2019.1.01142.S; PI: I.~Shivaei), which observed 27 galaxies and provides multi-band ALMA coverage for the full sample \citep{Shivaei22}. Since neutral atomic gas cannot be observed directly for ACE, we distinguish throughout this work between the dust-to-molecular-gas ratio, $M_{\rm dust}/M_{\rm mol}$ (DGR$_{\rm mol}$), and the dust-to-total-gas ratio, $M_{\rm dust}/M_{\rm gas}$ (DGR$_{\rm tot}$), where the latter includes both molecular and atomic gas.

\subsection{Dust-continuum observations and dust masses}
\label{subsec:dustmethod}
For the dust continuum analysis, we adopt the 873\,$\mu$m (Band~7) flux measurements from \citet{Popescu2026} and briefly summarize the main steps of the analysis here. The fluxes are measured in the image plane from maps tapered to have a 2\arcsec\ or 3\arcsec\ beamsize, based on the source extent, so that all sources are effectively unresolved and peak-flux measurements can be compared consistently. Continuum emission is detected at a signal-to-noise ratio (S/N) $>$ 3 in 17 out of 25 sources. We adopt the dust mass estimates from \citet{Solimano2026}. The masses are derived from single-band ALMA Band~7 observations assuming optically thin emission, following
\begin{align} \label{eq:Mdust}
M_{\mathrm{dust}} =
\frac{(S_\nu / f_{\mathrm{CMB}})\, D_L^2(z)}
{B_\nu(T)\, (1+z)\, \kappa_\nu(\beta)} \,,
\end{align}
where $S_\nu$ is the observed flux density at frequency $\nu$, $D_L(z)$ is the luminosity distance at redshift $z$, and $B_\nu(T)$ is the Planck function evaluated at dust temperature $T$. The dust mass absorption coefficient is parameterized as $\kappa_\nu = \kappa_0 (\nu/\nu_0)^{\beta}$, where $\beta$ is the dust emissivity spectral index and $\kappa_0$ is the normalization at reference frequency $\nu_0$. 
We adopt a dust emissivity index of $\beta = 2.08$ and a mass absorption coefficient of $\kappa_{\lambda_0} = 0.4\,\mathrm{m^2\,kg^{-1}}$ at $\lambda_0 = 250\,\mu\mathrm{m}$ \citep{Draine2003}, while adopting a dust temperature of 25$\pm$5 K. The factor $f_{\mathrm{CMB}}$ corrects for the effects of the cosmic microwave background (CMB) and follows from \cite{DaCunha2013}.  The motivation for these assumptions is discussed in detail in \citet{Solimano2026}.

\begin{figure*}[h!]
    \centering
    \includegraphics[width=\textwidth]{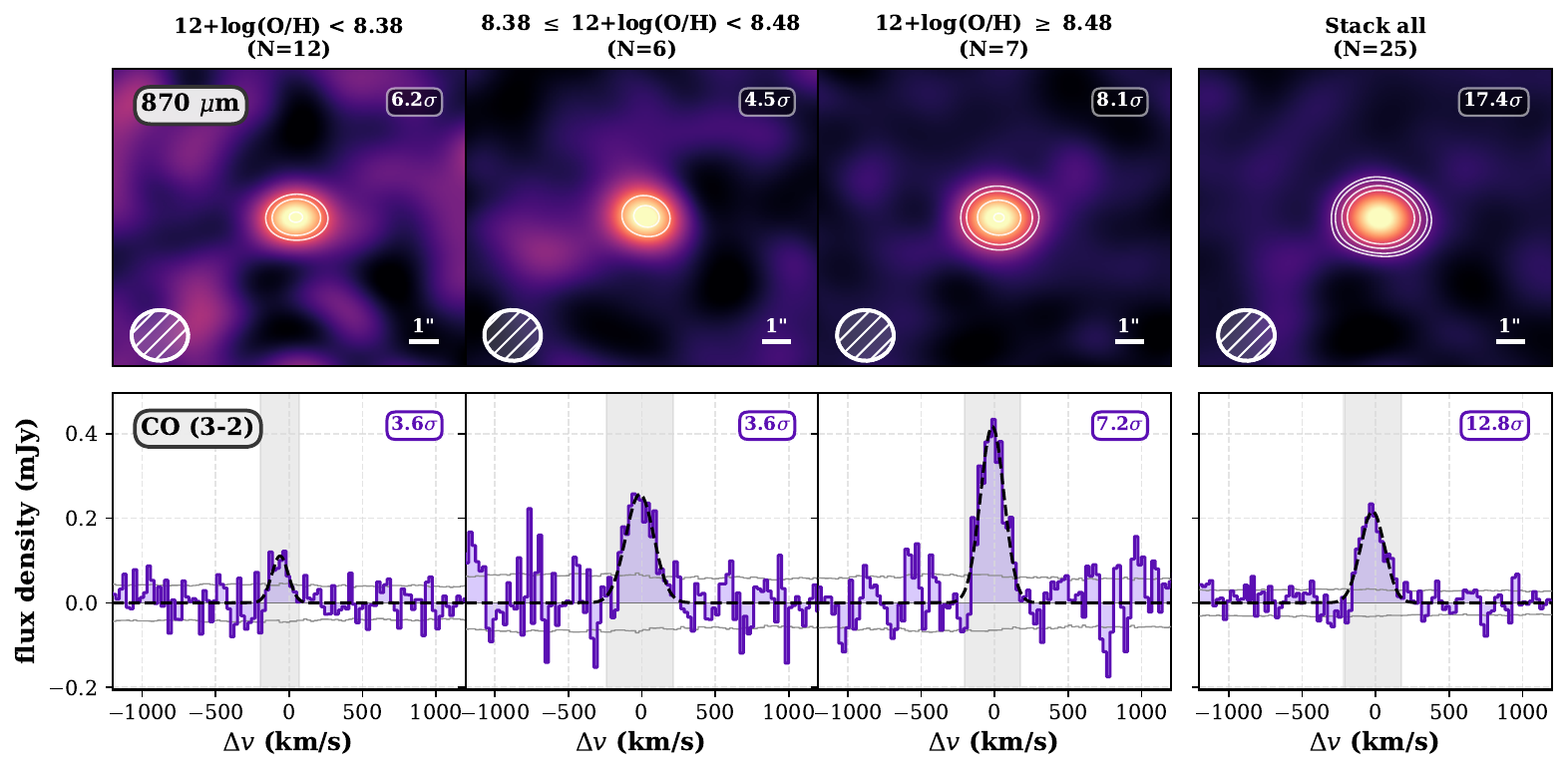}
    \caption{Weighted-mean stacked $873\,\mu{\rm m}$ continuum maps and CO(3--2) spectra in three metallicity bins: $12+\log({\rm O/H})<8.38$, $8.38 \leq 12+\log({\rm O/H})<8.48$, and $12+\log({\rm O/H})\geq 8.48$, with the full-sample stack shown in the rightmost column. In the continuum maps, white contours show the $3$, $4$, $6$, $8$, and $18\sigma$ levels. The bottom panels show the CO(3--2) spectra resulting from the cube-stacking routine. The purple histograms show the stacked spectra, the dashed black curves show the Gaussian fits used to define the optimized velocity windows, the grey shaded regions mark these windows and the fine grey lines indicate the rms per channel.}
    \label{fig:stacks per bins}
\end{figure*}

\subsection{CO line observations and molecular gas masses}
\label{subsec:CO method}
We adopt the CO flux measurements reported in \citet{Langan2026}, and briefly summarize the main steps of the analysis below. To ensure homogeneous flux measurements, and following the approach adopted for the continuum analysis, a $1\arcsec$ or $2\arcsec$ taper was applied to the CO data cubes so that all galaxies are spatially unresolved. The resulting beam sizes span $\sim1.4\arcsec$--$3.2\arcsec$ across the sample. CO fluxes are extracted iteratively by refining the peak position and velocity range from moment-0 maps and Gaussian spectral fits until the integrated flux converges; for non-detections, the spectrum is instead extracted over a single beam-sized aperture centered at the position of the source. Sources with CO(3-2) emission at $>3\sigma$ are defined as detections, yielding detections for 16 of 25 sources. To convert the observed CO(3--2) fluxes into molecular gas masses, we first compute the CO line luminosity following the standard formulation \citep{Solomon2005}:
\begin{equation} \label{eq:LCO}
\begin{split}
L'_{\rm CO} =\;& 3.25 \times 10^{7}
\left(\frac{S_{\rm CO}\Delta v}{\mathrm{Jy\,km\,s^{-1}}}\right)
\left(\frac{d_L}{\mathrm{Mpc}}\right)^{2}
\left(\frac{\nu_{\mathrm{obs}}}{\mathrm{GHz}}\right)^{-2}\\
& (1+z)^{-3} \,\mathrm{K\,km\,s^{-1}\,pc^{2}} .
\end{split}
\end{equation}
where $S_{\rm CO}\Delta v$ is the velocity-integrated CO line flux, $d_L$ is the luminosity distance, $\nu_{\mathrm{obs}}$ is the observed frequency, and $z$ is the redshift. The molecular gas mass is then derived from the CO(3--2) luminosity via
\begin{equation}\label{eq:Mmol}
M_{\mathrm{mol}} = \alpha_{\mathrm{CO}} \,
\frac{L'_{\mathrm{CO}(3-2)}}{r_{31}},
\end{equation}
where $\alpha_{\mathrm{CO}}$ is the CO--H$_2$ conversion factor and $r_{31}$ is the CO(3--2)/CO(1--0) excitation ratio. We adopt the value $r_{31} = 0.77 \pm 0.14$ from \cite{Boogaard2020}, based on a stack of typical star-forming galaxies at $\rm z=2.5$. Local dwarf galaxies show no evidence for systematically different CO excitation ratios compared to higher-metallicity systems \citep{Meier2001,Cormier2014}, supporting this assumption. However, $r_{31}$ remains unconstrained for low-mass, low-metallicity galaxies at high redshift. For the CO--H$_2$ conversion factor we follow the metallicity-dependent prescription of \cite{Accurso2017}:
\begin{equation} \label{eq:accurso}
\begin{split}
\log \alpha_{\mathrm{CO}} (\pm 0.165\,\mathrm{dex}) =\;&
14.752 - 1.632 \left[12 + \log(\mathrm{O/H})\right] \\
&+ 0.062 \log(\Delta \mathrm{MS}).
\end{split}
\end{equation}
Here, $\Delta \mathrm{MS}$ denotes the offset from the star-forming main sequence. We compute $\Delta \mathrm{MS}$ using the main-sequence parameterization of \cite{Shivaei2015}, modified for updated stellar masses (see \citet{Shivaei2026}). A dynamical mass consistency analysis (presented in \citet{Langan2026}) supports the adoption of the \cite{Accurso2017} calibration as the preferred choice for $\alpha_{\mathrm{CO}}$ in this sample.

\begin{table}
\centering
\caption{Weighted-mean stacked $873\,\mu{\rm m}$ continuum and CO(3--2) fluxes for the metallicity bins and full sample.}
\label{tab:metallicity_stack_technical}
\begin{tabular}{lrllll}
\toprule
Bin & $N$ & \shortstack{$S_{873\,\mu{\rm m}}$\\{}[mJy]} & S/N & \shortstack{$S_{\rm CO(3-2)}\Delta v$\\{}[mJy km s$^{-1}$]} & S/N \\
\midrule
Z < 8.38 & 12 & $0.09$ & $6.21$ & $15.83$ & $3.63$ \\
8.38-8.48 & 6 & $0.2$ & $4.47$ & $53.8$ & $3.63$ \\
Z > 8.48 & 7 & $0.27$ & $8.14$ & $75.67$ & $7.23$ \\
\midrule
Stack all & 25 & $0.17$ & $17.39$ & $40.97$ & $12.80$ \\
\bottomrule
\end{tabular}
\end{table}
\subsection{Stacking}
Across the full sample, ten galaxies have at least one non-detection in either CO(3–2) or Band 7 dust continuum, including seven galaxies that are undetected in both tracers. The full sample, including individual measurements and 3$\sigma$ upper limits, is listed in Table~\ref{tab:physical_sample_by_metallicity}. These non-detections occur predominantly at the low-metallicity end of the sample. To infer metallicity-dependent trends without restricting the analysis to only individually detected galaxies, we perform stacking in bins of metallicity, including both the detected and undetected sources. We divide the sample into three metallicity bins, defined as $12+\log(\mathrm{O/H}) < 8.38$ ($N=12$), $8.38 \leq 12+\log(\mathrm{O/H}) < 8.48$ ($N=6$), and $12+\log(\mathrm{O/H}) \geq 8.48$ ($N=7$). In the lowest-metallicity bin, six galaxies are undetected in both CO(3--2) and Band 7 dust continuum, while one additional source is detected only in CO(3--2) and another only in Band 7 dust continuum. The intermediate metallicity bin contains one galaxy undetected in both tracers and one further source with a CO(3--2) non-detection. Stacking allows us to include sources without individual detections in one or both tracers, which would bias the inferred trends toward the more readily detected, typically higher metallicity systems. 
\begin{table*}
\centering
\caption{Physical quantities inferred from the weighted-mean continuum and CO(3-2) cube stacks.}
\label{tab:metallicity_stack_physical}
\renewcommand{\arraystretch}{1.3}
\begin{tabular}{llllllll}
\toprule
Bin & \shortstack{$12+\log({\rm O/H})$} & \shortstack{$\log M_\star$\\{}[$M_\odot$]} & \shortstack{$\log {\rm SFR}$\\{}[$M_\odot$ yr$^{-1}$]} & \shortstack{$\log L_\nu(873\,\mu{\rm m})$\\{}[erg s$^{-1}$ Hz$^{-1}$]} & \shortstack{$\log M_{\rm dust}$\\{}[$M_\odot$]} & \shortstack{$\log L'_{\rm CO(1-0)}$\\{}[K km s$^{-1}$ pc$^2$]} & \shortstack{$\log M_{\rm mol}$\\{}[$\,M_\odot$]} \\
\midrule
Z < 8.38 & $8.29_{-0.02}^{+0.01}$ & $9.86_{-0.09}^{+0.07}$ & $1.78_{-0.06}^{+0.05}$ & $31.06_{-0.09}^{+0.07}$ & $7.69_{-0.09}^{+0.07}$ & $8.79_{-0.14}^{+0.12}$ & $10.12_{-0.20}^{+0.2}$ \\
8.38-8.48 & $8.44_{-0.02}^{+0.01}$ & $9.99_{-0.10}^{+0.15}$ & $1.80_{-0.15}^{+0.11}$ & $31.43_{-0.10}^{+0.09}$ & $8.03_{-0.1}^{+0.09}$ & $9.33_{-0.15}^{+0.11}$ & $10.40_{-0.23}^{+0.19}$ \\
Z > 8.48 & $8.53_{-0.02}^{+0.02}$ & $10.18_{-0.08}^{+0.07}$ & $1.85_{-0.10}^{+0.11}$ & $31.56_{-0.06}^{+0.06}$ & $8.16_{-0.05}^{+0.05}$ & $9.48_{-0.07}^{+0.05}$ & $10.40_{-0.2}^{+0.17}$ \\
\midrule
Stack all & $8.40_{-0.02}^{+0.02}$ & $9.98_{-0.04}^{+0.05}$ & $1.81_{-0.05}^{+0.04}$ & $31.35_{-0.03}^{+0.03}$ & $7.95_{-0.03}^{+0.02}$ & $9.20_{-0.04}^{+0.03}$ & $10.36_{-0.18}^{+0.18}$ \\
\bottomrule
\end{tabular}
\end{table*}
To derive conservative error estimates, we account for three contributions: intrinsic measurement uncertainty, uncertainty in metallicity bin assignment, and sensitivity to the galaxies included in each stack. We implement a combined Monte Carlo and $N-1$ bootstrap procedure. Specifically, bin membership is determined by randomly sampling each galaxy's metallicity from its associated uncertainty distribution and assigning it to one of the three predefined metallicity bins. Simultaneously, we perform an $N-1$ bootstrap resampling of the galaxies within each bin, randomly removing one galaxy per bin in each realization and stacking the remaining sources. This procedure is repeated 500 times, yielding distributions of stacked flux measurements and corresponding measurement uncertainties. The final reported flux is the median of this distribution. The 16th and 84th percentiles capture the uncertainty from bin assignment and $N-1$sampling, while the median measurement uncertainty is added in quadrature to these percentile-based uncertainties.

We adopt weighted mean stacking to account for variations in the root-mean-square (rms) noise across the ALMA observations. This weighting ensures that measurements with lower noise contribute more strongly to the final stacked signal, while preserving the statistical properties of the sample. We also perform median stacking and unweighted mean stacking. The resulting fluxes are consistent with those obtained from the weighted mean stacking, with differences within $\sim 1\sigma$, indicating that our results are not sensitive to the specific stacking methodology.

A key requirement for stacking is that the data correspond to a common physical scale, which requires homogeneous beam sizes across the sample. \citet{Popescu2026} present Band~7 dust continuum images tapered to $2''$ and $3''$, with the choice based on the spatial extent of each galaxy from a curve of growth analysis. To ensure a consistent physical scale, we perform image-plane stacking using both tapered datasets. We find that the curve of growth of the stacked images reaches a plateau at $2''$, indicating that this taper recovers the total emission. We therefore adopt the $2''$ tapered images for the stacking analysis. Sigma clipping is applied when estimating the rms noise to reduce contamination from serendipitous background sources. Fluxes are measured following \citet{Popescu2026}, adopting the peak flux density from the unresolved $2''$ tapered stacks. 

For the CO(3--2) observations, we perform cube stacking at a common angular resolution. We adopt a common beam size of approximately $3''$, corresponding to the lowest-resolution observation in the sample. To this aim, we re-image each object following the prescriptions of \citet{Langan2026}, using a Briggs robust parameter of 2 and applying a $uv$-taper in the range 1.5--2 to ensure the requested beam size.

Prior to stacking, each cube is shifted to its systemic velocity based on the spectroscopic redshift, and only the overlapping spectral channels are retained. The cubes are then stacked channel by channel. To account for differences in spectral resolution, all cubes are interpolated to the coarsest channel width in the sample, corresponding to $\sim 23$\,km\,s$^{-1}$.

CO(3--2) fluxes are measured from the stacked cubes using an optimized velocity window. An initial moment-0 map is generated over $\pm250$\,km\,s$^{-1}$ around the systemic velocity, and the spectrum at the peak pixel is fitted with a Gaussian. The final integration window is centered on the fitted centroid and spans $\pm\mathrm{FWHM}$, with $\mathrm{FWHM}=2.355\,\sigma$. We then generate a moment-0 map over this optimized window and measure the CO flux from its peak pixel. The uncertainty is obtained by propagating the channel rms over the same velocity range. The optimized windows are visualized in the spectra shown in Fig.~\ref{fig:stacks per bins}, and the resulting CO(3--2) fluxes and corresponding S/N ratios are reported together with the continuum stacking results in Table~\ref{tab:metallicity_stack_technical}. We detect both the CO(3--2) emission and the Band 7 continuum flux with a S/N > 3 in all three metallicity bins. In Fig.~\ref{fig:stacks per bins} and Table~\ref{tab:metallicity_stack_technical} we also report the stack of the complete sample. For the stack all, including all 25 galaxies, there is no bin-assignment uncertainty present. We therefore derive the flux and uncertainty from 500 $N-1$ realizations, combining the resulting percentile spread with the median measurement uncertainty in quadrature. The full sample stacks yield a significant detection in both the CO(3--2) emission and the Band 7 continuum. 

In Table~\ref{tab:metallicity_stack_physical}, we report the median metallicity, stellar mass, and star formation rate (SFR) for each metallicity-stacked bin and for the total stack. The SFR follows from the dust corrected H$\alpha$ luminosities. The computation of the SFR, as well as the metallicity and stellar mass, is discussed in more detail in \citet{Shivaei2026}. These quantities are derived using the same Monte Carlo and $N-1$ resampling framework used for the stacked fluxes, but with measurement uncertainties sampled directly in each realization rather than added in quadrature afterward. We do this because, unlike for the stacked fluxes, the measurement uncertainties are not remeasured independently for each realization. We then report the median of these realization medians as the representative value, with uncertainties given by the 16th and 84th percentiles.
\begin{figure*}[h!]
    \centering
    \includegraphics[width=0.9\textwidth]{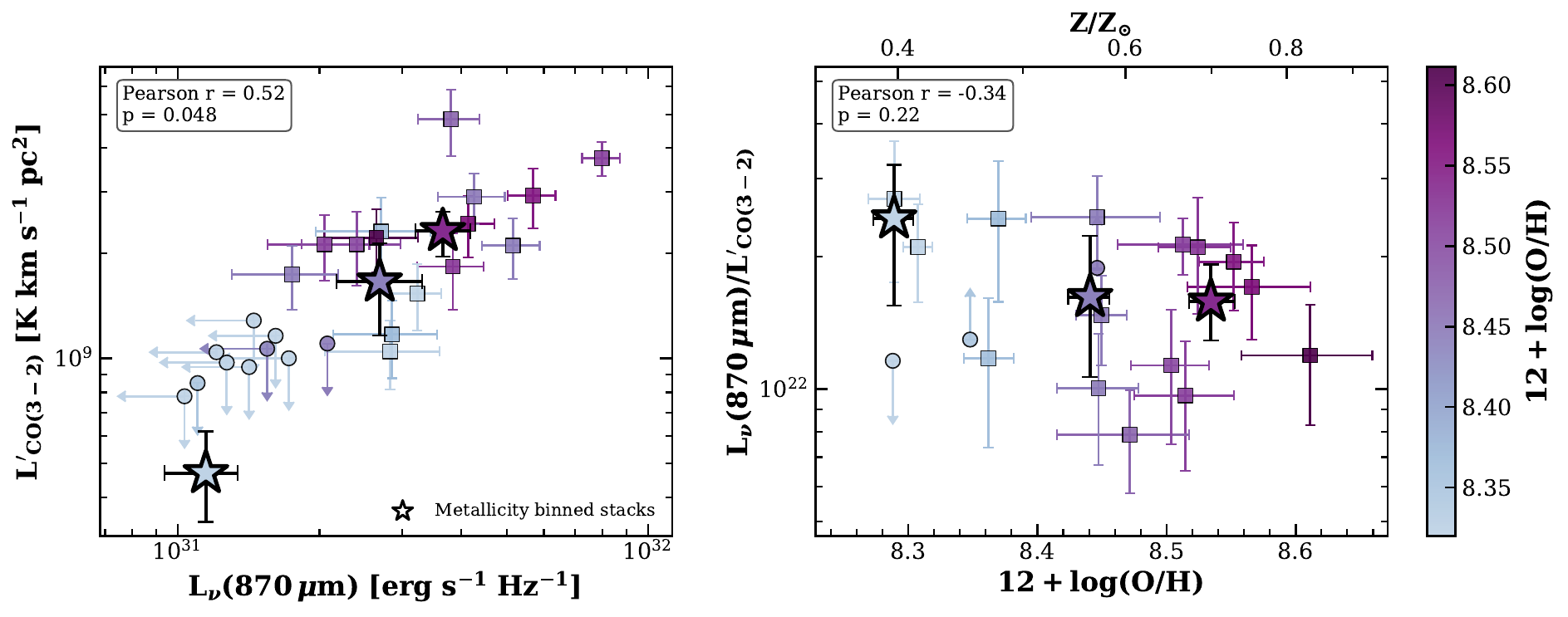}
    \caption{Observed dust-continuum and CO(3--2) luminosities for the ACE sample. \textbf{Left: }CO(3--2) luminosity as a function of rest-frame \(873\,\mu{\rm m}\) luminosity, with symbols colour-coded by metallicity, \(12+\log({\rm O/H})\). \textbf{Right:} observed luminosity ratio \(L_\nu(873\,\mu{\rm m})/L'_{\rm CO(3-2)}\) as a function of metallicity. Squares denote galaxies detected in both tracers, circles show upper and lower limits and large stars show metallicity-binned stacks. The annotated Pearson coefficients and \(p\)-values are computed for the individual detections only.}
    \label{fig:2panelobservables}
\end{figure*}
\section{Results}
\label{results}
\subsection{Observable ratios}
\label{sec:observableratios}
Before converting the measurements into physical quantities, we first examine the directly observed CO and dust luminosities. The left panel of Figure~\ref{fig:2panelobservables} shows the observed CO(3--2) line luminosity as a function of the $873\,\mu\mathrm{m}$ dust luminosity for the ACE sample. We convert the measured line fluxes to $L'_{\rm CO(3-2)}$ using Eq.~\ref{eq:LCO}. The observed Band-7 continuum flux densities are converted to rest-frame monochromatic luminosities using the luminosity distance and the appropriate redshift-dependent frequency correction. The individual luminosity measurements and corresponding upper limits are reported in Table~\ref{tab:physical_sample_by_metallicity}. The metallicity-binned stacked fluxes are converted to CO and dust luminosities in the same way, and the resulting stacked luminosities are listed in Table~\ref{tab:metallicity_stack_physical}.

The individual detections show a moderate positive correlation between CO and dust luminosity, with Pearson $r=0.52$ and $p=0.048$. This indicates that galaxies with higher $873\,\mu\mathrm{m}$ dust luminosities tend to have higher CO(3--2) luminosities, although the relation exhibits substantial scatter, and the correlation is significant at the 2$\sigma$ level. This trend is consistent with the use of dust continuum emission as a tracer of molecular gas, which we further explore in section ~\ref{sec:scoville}. A handful of galaxies, primarily at the low-metallicity end of the sample, remain undetected in one or both tracers and therefore appear as limits in Figure~\ref{fig:2panelobservables}. To retain information from these fainter systems and probe the full dynamic range of the relation, we also include metallicity-binned stacks. With a $p$-- value of 0.048 the correlation among individual detections is only marginally significant, and it may be influenced by the detection limit. However, the metallicity-binned stacks reveal a clear trend, the stacked CO luminosity increases by $\sim0.7$ dex (a factor of $\sim5$) from the lowest- to highest-metallicity bin.

Next we examine the ratio of the observed dust and CO luminosities. The right panel of Figure~\ref{fig:2panelobservables} shows the ratio of $873\,\mu\mathrm{m}$ dust luminosity to observed CO(3--2) luminosity as a function of metallicity. Among the individual detections, this ratio shows substantial scatter and no statistically significant dependence on metallicity: the detection-only Pearson test gives $r=-0.34$ and $p=0.22$. The metallicity-binned stacks recover information from the fainter part of the sample and suggest that the lowest-metallicity bin has a higher $L_\nu(873\,\mu\mathrm{m})/L'_{\rm CO(3-2)}$ ratio than the higher-metallicity bins. However, the stacked measurements are consistent with a constant value within their uncertainties. We therefore cannot conclude any metallicity dependence for the observable dust-to-CO luminosity ratio. 

\begin{figure*}[h!]
    \centering
    \includegraphics[width=\textwidth]{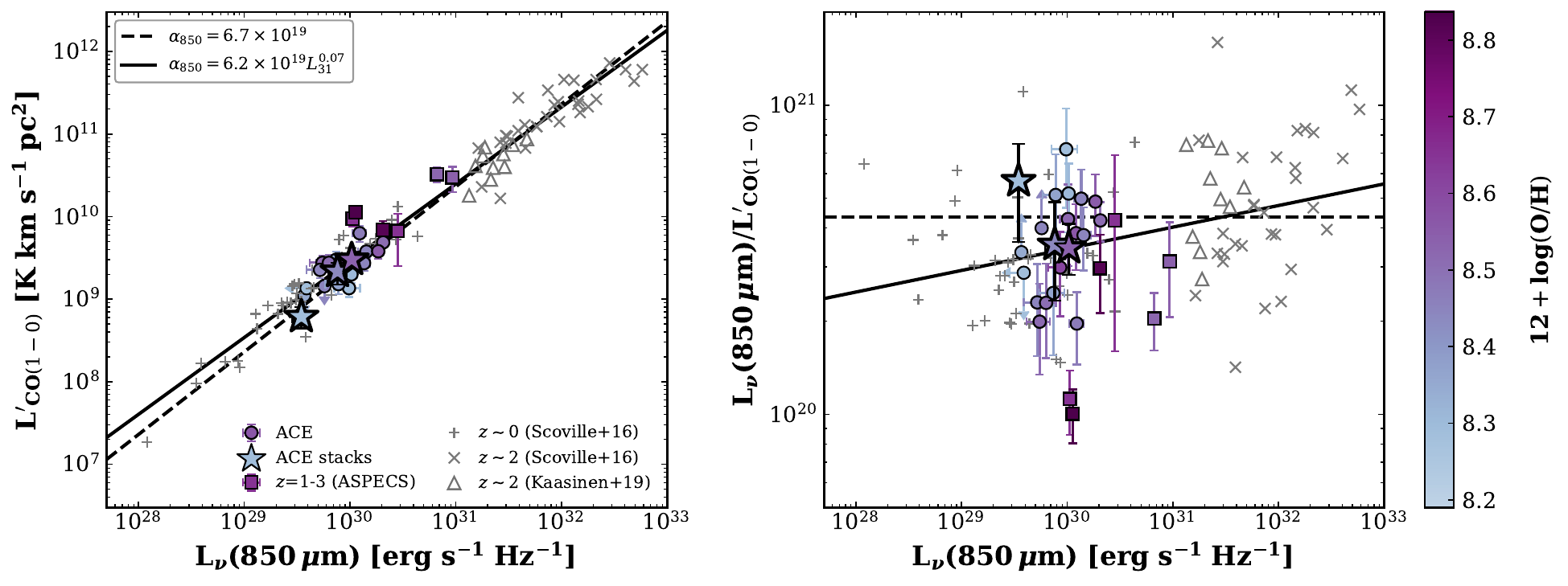}
    \caption{Comparison between rest-frame $850\,\mu{\rm m}$ dust continuum luminosity and CO(1--0) luminosity. \textbf{Left}: $L'_{\rm CO(1-0)}$ as a function of $L_\nu(850\,\mu{\rm m})$ for the ACE galaxies, compared to the empirical calibration of \citet{Scoville2016}, the local universe (plusses) and $z\sim 2$ measurements (crosses) of \citet{Scoville2016} and to individual measurements from \citet{Kaasinen2019} (triangles) and ASPECS \citep{Boogaard2020} (squares). The dashed line shows the constant $\alpha_{850}=6.7\times10^{19}\,{\rm erg\,s^{-1}\,Hz^{-1}\,M_\odot^{-1}}$ calibration from \citet{Scoville2016}, while the solid line shows their luminosity-dependent form, $\alpha_{850}=6.2\times10^{19}L_{\nu,31}(850\,\mu{\rm m})^{0.07}$. \textbf{Right}: ratio $L_\nu(850\,\mu{\rm m})/L'_{\rm CO(1-0)}$ as a function of $L_\nu(850\,\mu{\rm m})$, showing the same calibrations. ACE and ASPECS galaxies are colour-coded by metallicity. The ACE sources lie broadly within the scatter of the literature relation, and the ratio plot shows no clear systematic trend with $L_\nu(850\,\mu{\rm m})$ or metallicity over the range probed here.}
    \label{fig:Scoville/Boogaaard}
\end{figure*}

\subsection{CO and dust continuum emission relation}
\label{sec:scoville}  
\citet{Scoville2016} presented an empirical calibration of Rayleigh--Jeans dust continuum emission as a molecular gas tracer using 70 galaxies with global CO(1--0) and long-wavelength continuum measurements, spanning local star-forming galaxies, low-redshift ULIRGs, and $z\sim2$ SMGs. They found an approximately linear relation between rest-frame $850\,\mu{\rm m}$ luminosity ($L_\nu(850\,\mu{\rm m})$) and CO(1--0) line luminosity, ($L'_{\rm CO(1-0)}$), with a scatter of $\sim0.1$ dex. This calibration, however, was constructed for relatively massive, metal-rich systems, and was not designed to provide a metallicity-independent conversion applicable across the full galaxy population. Likewise, \citet{Kaasinen2019} tested the dust--CO relation in 12 massive $z\sim2$ star-forming galaxies, again deliberately targeting systems where the metallicity-dependent dust-to-gas ratio was not expected to affect the conversion between dust continuum luminosity and molecular gas mass.

By comparing the ACE sample to the \citet{Scoville2016} empirical $L_\nu(850\,\mu{\rm m})$--$L'_{\rm CO(1-0)}$ calibrations, based on massive galaxies. We can test if it remains valid for lower mass galaxies at $z\sim2 - 2.5$ in the sub-solar metallicity regime. Because the ACE observations target CO(3--2), we convert the measured line luminosities to equivalent CO(1--0) luminosities using $r_{31}\equiv L'_{\rm CO(3-2)}/L'_{\rm CO(1-0)}=0.77\pm0.14$ \citep{Boogaard2020}. Similarly to obtain a fair comparison, we convert the continuum measurements to rest-frame $850\,\mu{\rm m}$ luminosities using the modified-blackbody assumptions adopted by \citet{Scoville2016}; a dust emissivity index $\beta=1.8$ and a mass-weighted dust temperature $T_{\rm dust}=25\,{\rm K}$.

Figure~\ref{fig:Scoville/Boogaaard} compares the ACE galaxies to the constant $\alpha_{850}=6.7\times10^{19}\,{\rm erg\,s^{-1}\,Hz^{-1}\,M_\odot^{-1}}$ calibration from \citet{Scoville2016} (dashed line), and the luminosity-dependent form, $\alpha_{850}=6.2\times10^{19}L_{\nu,31}(850\,\mu{\rm m})^{0.07}$ (solid line). In the left-hand panel, we show individual high-redshift detections from \citet{Boogaard2020} at $z\sim1$--3, for which gas-phase metallicity measurements are available, together with individual detections at $z\sim2$ from \citet{Kaasinen2019} and \citet{Scoville2016}, for which metallicity information is not available. We additionally include the local comparison sample from \citet{Scoville2016}, which contains nearby star-forming galaxies and (U)LIRGs without metallicity measurements. The ACE galaxies lie broadly along the established $L'_{\rm CO(1-0)}$--$L_\nu(850,\mu{\rm m})$ relation and overlap with the range of dust and CO luminosities spanned by the local \citet{Scoville2016} sample. At high redshift, ACE extends the existing individual detections from \citet{Kaasinen2019}, \citet{Scoville2016}, and \citet{Boogaard2020} toward substantially lower dust and CO luminosities, allowing the dust--CO relation to be explored in lower-luminosity galaxies than previously accessible at these redshifts. The right-hand panel shows the corresponding luminosity ratio, \(L_\nu(850\,\mu{\rm m})/L'_{\rm CO(1-0)}\), which removes the dominant near-linear scaling and makes deviations from a constant dust-to-CO luminosity ratio more apparent.
The ACE galaxies span a range of metallicities but do not show a clear monotonic trend in \(L_\nu(850\,\mu{\rm m})/L'_{\rm CO(1-0)}\) with \(12+\log({\rm O/H})\), nor a strong systematic offset from the Scoville relation. This suggests that the empirical coupling between Rayleigh--Jeans dust continuum luminosity and CO(1-0) luminosity persists into the lower-metallicity regime sampled by ACE ( $8.3 \leq 12+\log(\mathrm{O/H}) < 8.7$ ) .

\begin{figure*}[h!]
    \centering
    \includegraphics[width=0.7\textwidth]{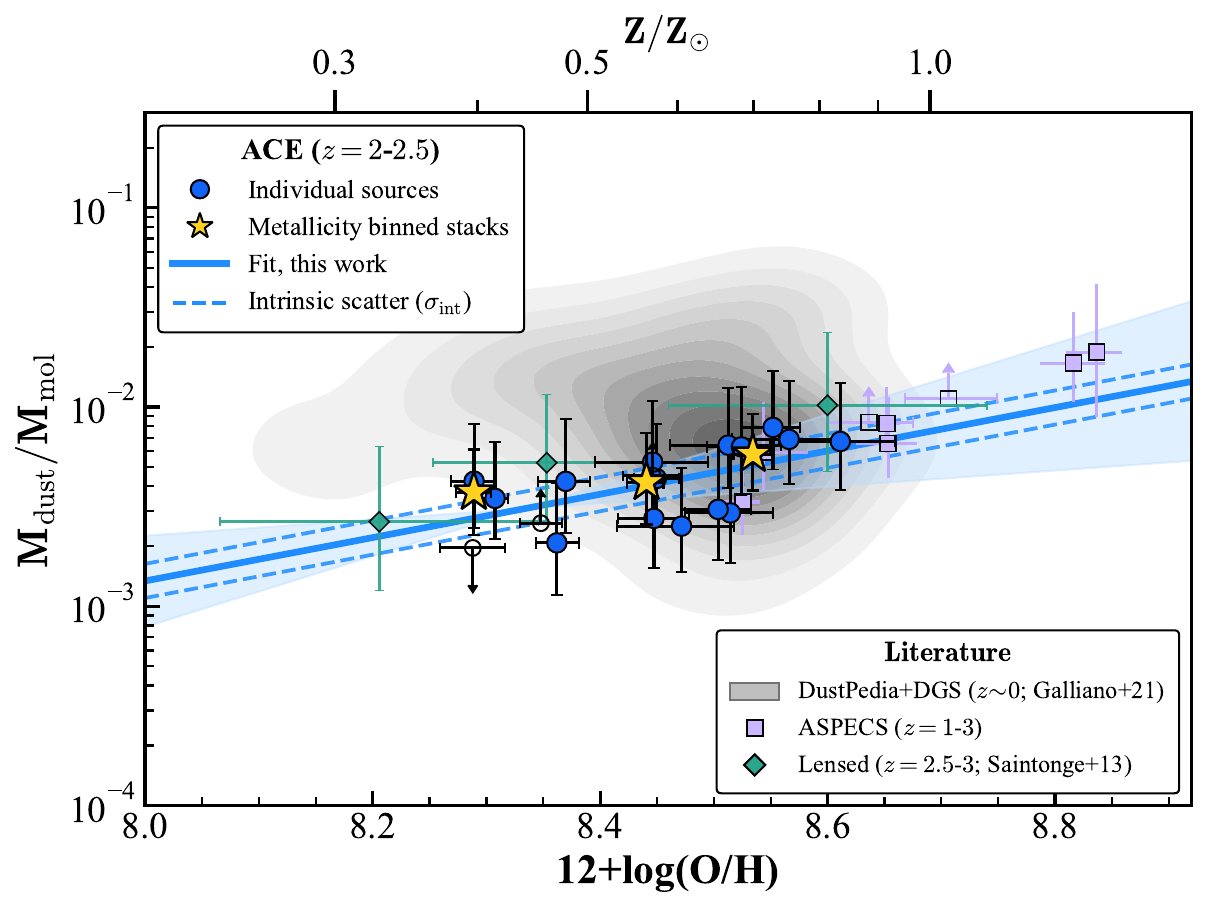}
    \caption{Dust-to-gas ratio, expressed as \(M_{\rm dust}/M_{\rm mol}\), as a function of  metallicity for the ACE sample at $z=2-2.5$. Blue circles show individual ACE galaxies, with open symbols indicating sources with upper limits. Yellow stars mark the metallicity-binned stacks. The blue line shows the best-fitting relation derived in this work, with the shaded region indicating the error on the fitting parameters and the dashed lines the intrinsic scatter. We compare to literature measurements from DustPedia+DGS at $z\sim0$ (grey contours), ASPECS at $z=1-3$ (purple squares) and lensed galaxies from \citet{Saintonge2013} at $z=2.5-3$ (green diamonds).}
    \label{fig:MdustMmol}
\end{figure*}

\subsection{Physical DGR$_{\rm mol}$-metallicity relation}
\label{sec:DRGmolphysical}
While the previous section examined the metallicity dependence of the observed luminosity ratios alone, here we restrict the analysis to inferred physical properties obtained by converting the dust and CO luminosities into \(M_{\rm dust}\) and $M_{\rm mol}$, respectively, using the assumptions outlined in Sections~\ref{subsec:dustmethod} and ~\ref{subsec:CO method}. Figure~\ref{fig:MdustMmol} shows the resulting $\rm DGR_{mol}$, as a function of metallicity for the ACE sample. In contrast to the observed luminosity ratios shown in Figure~\ref{fig:2panelobservables}, the inferred dust-to-molecular gas ratios display a positive correlation with metallicity. For the detected sources alone, we find a Pearson correlation coefficient of \(r=0.53\), with \(p=0.0433\), indicating a moderate positive correlation. The metallicity-binned stacks are consistent with the same trend and help extend the analysis to fainter sources. We fit the observed DGR$_{\rm mol}$--metallicity relation with a Bayesian power-law model using the \texttt{PyMC} framework \citep{Patil2010}. Specifically, we model the data as
\begin{equation}
\log\left(\frac{M_{\rm dust}}{M_{\rm mol}}\right)
=
a\,\log\left(\frac{Z}{Z_0}\right) + b,
\end{equation}
where \(Z\) denotes the observed metallicity and \(Z_0 = 8.45\) is the mean metallicity of the sources included in the fit. The fit includes intrinsic scatter ($\rm \sigma_{int}$), measurement uncertainties in both \(\log(Z)\) and \(\log(M_{\rm dust}/M_{\rm mol})\), and upper and lower limits through cumulative Gaussian likelihoods. We obtain
\begin{align*}
a &= 1.20^{+0.69}_{-0.68}, \\
b &= -2.38^{+0.07}_{-0.07}, \\
\sigma_{\rm int} &= 0.10^{+0.06}_{-0.04}.
\end{align*}
The blue line shows the best-fitting relation derived using this method, with the intrinsic scatter indicated by the dashed lines.

In Figure~\ref{fig:MdustMmol}, we compare our measurements to both local-Universe and high-redshift literature samples. To ensure a fair comparison, we only include studies with molecular gas masses based on CO observations and place the comparison data on our adopted assumptions as described in Section~\ref{methods} where possible. The \cite{Galliano2021} sample combines DustPedia \citep{Davies17} with the Dwarf Galaxy Survey \citep[DGS;][]{Madden13}, which comprises $\sim$ 800 nearby galaxies spanning a broad range of galaxy properties and morphologies. To be able to compare their $\rm DGR_{mol}$ ratios we select their subsample for which CO observations are available, reducing the sample size to $\sim$ 130 galaxies. We convert the CO-based molecular gas masses in that sample to our adopted assumptions, for the dust masses we only apply a correction for the adopted $\kappa_{\lambda_0}$ since their masses follow from fitting the THEMIS model \citep{Jones2017}. We do not convert their metallicities to a different abundance scale, since they were derived using the \cite{Pilyugin2016} S-calibration, an empirical H II-region calibration appropriate for local-universe ISM conditions. For the high-redshift comparison, we show 3 lensed galaxies from \citet{Saintonge2013} and 8 galaxies from the ALMA Spectroscopic Survey in the Hubble Ultra Deep Field (ASPECS; \citet{Walter2016}). We expand the earlier subset of ASPECS targets studied for the DGR-metallicity relation in \citet{Shapley2020}, by compiling low J CO and Band 6 continuum measurements from \citep{GonzalezLopez2020, Aravena2020, Boogaard2020}. We recalculate all dust and molecular gas masses under the same assumptions adopted in this work. For metallicity, we recalibrate to the strong-line calibration used in this work \citet{Sanders2026}. For ASPECS we adopt the strong-line measurements from \citet{Kiyota2026}. For the \citet{Saintonge2013} lenses we only selected targets with 1.2 mm dust measurements and strong-line measurements available. This results in the subsample of \textit{8.00arc} and \textit{Eye} with strong-lines from \cite{Richard2011} and \textit{cB58} with strong lines from \cite{Teplitz2000}. 

Notably, the ACE measurements do not show a strong offset relative to the \(z\sim0\) DustPedia+DGS observations at fixed metallicity, suggesting little evidence for strong redshift evolution in the $\rm DGR_{mol}$--metallicity relation over the probed metallicity range. The ASPECS galaxies, which probe a broadly similar redshift regime, also lie close to the ACE best-fit relation, however at higher metallicities. The detected ACE galaxies have a mean ratio of $\log(M_{\rm dust}/M_{\rm mol})=-2.37\pm0.05$ and a mean metallicity of $12+\log(\rm O/H)=8.45\pm0.02$, where the uncertainties are the standard error on the mean. Because this average excludes upper and lower limits, we report it only for reference; the quantitative relation is derived from the censored Bayesian fit. We explore the offsets from the CO-based literature samples in further detail in Section~\ref{sec:DTMolcomp}. 

\begin{figure*}[h!]
    \centering
    \includegraphics[width=\textwidth]{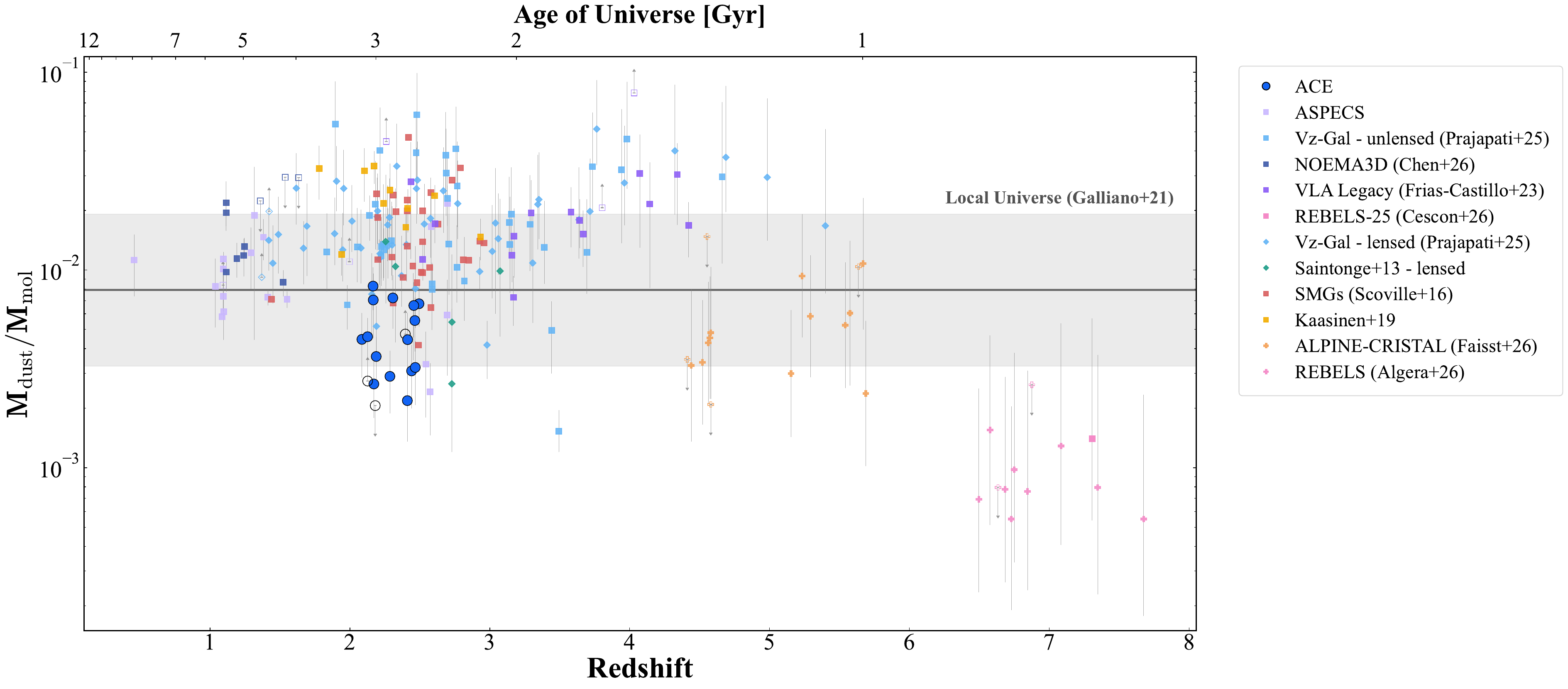}
    \caption{Dust-to-molecular-gas mass ratio (DGR$_{\rm mol}$), \(M_{\rm dust}/M_{\rm mol}\), as a function of redshift. ACE galaxies are shown in blue; filled circles denote measured ratios, while downward arrows indicate upper limits on $\rm DGR_{mol}$. Literature samples with CO-based molecular gas measurements are shown for comparison, including ASPECS, \citet{Kaasinen2019}, \citet{Scoville2016}, \citet{Saintonge2013}, NOEMA3D \citep{Chen2026}, the VLA Legacy survey \citep{FriasCastillo2023}, and Vz-GAL \citep{Prajapati2026}. At higher redshift, we include [C\,\textsc{ii}]-based estimates from ALPINE-CRISTAL and REBELS-IFU \citep{Faist2026,Algera2026}, with REBELS-25 highlighted separately using the available CO-based molecular gas constraint \citep{Cescon2026}. The grey band marks the local-universe reference range from \citet{Galliano2021}. ACE extends the observed distribution at \(z\sim1\)--3 toward lower dust-to-molecular-gas ratios than most existing CO-selected comparison samples, while the \(z\gtrsim4\)[C\,\textsc{ii}]-based galaxies occupy similarly low values or even lower values. }
    \label{fig:MdustMmolredshift}
\end{figure*}

\subsection{Cosmic evolution of the DGR$_{\rm mol}$}
\label{redshiftevo}
To look at the broader cosmic evolution of dust we place the results of ACE in the context of other surveys with both CO and dust continuum measurements available in Fig.~\ref{fig:MdustMmolredshift}, which also includes surveys without metallicity information. We show again the work from ASPECS and \citet{Kaasinen2019}, both samples of massive star-forming galaxies at cosmic noon, the $z\sim 2$ sample from \citet{Scoville2016}, which consists of sub-millimeter galaxies and the lensed galaxies from \citet{Saintonge2013}. Additional surveys shown are the NOEMA3D survey, which targets main-sequence galaxies in the redshift range of $z=1-1.6$ \citep{Chen2026}. As well as the VLA Legacy survey of molecular gas \citep{FriasCastillo2023}, and the Vz-gal survey, both probing CO(1-O) in dusty star-forming galaxies \citep{Prajapati2026}, respectively for redshifts $z=2-5$ and $z=1-6$.  

We extend our comparison to higher redshifts by including surveys that infer the molecular gas masses from [C\,\textsc{ii}] rather than CO, namely ALPINE-CRISTAL ($z=4-6$) \citep{Faist2026} and REBELS-IFU ($z=6-8$) \citep{Algera2026}. For ALPINE we adopt the molecular gas masses originally presented in \citet{Dessauges2020} and for REBELS originally presented in \citet{Aravena2024}. In both works, the [C\,\textsc{ii}] luminosities are converted to molecular gas masses using the $\alpha_{\rm [CII]}$ conversion factor calibrated by \citet{Zanella2018}. For ALPINE-CRISTAL, we recompute the dust masses from the single-band ALMA observations using the same assumptions adopted in this work, whereas for REBELS--IFU we use the dust masses reported by \citet{Algera2026}.

The ACE galaxies push the observed $\rm DGR_{mol}$ distribution at $z\sim1-3$ down to $\log(M_{\rm dust}/M_{\rm mol})\sim-3$, populating a low-ratio regime that is largely absent from previous samples. By contrast, most galaxies in the CO literature samples lie above $\log(M_{\rm dust}/M_{\rm mol})\sim-2$. ACE is therefore particularly important for revealing the full dynamic range of dust-to-molecular-gas ratios present in star-forming galaxies at these redshifts. Both the works from \cite{Scoville2016} and \cite{Kaasinen2019} target massive galaxies with stellar masses $>2 \times10^{10} M_{\odot}$. No stellar mass measurements are available for the Vz-gal sample, however the VLA Legacy survey targets galaxies similar to the Vz-gal sample and report median stellar masses of $25 \times 10^{10} M_{\odot}$. Based on the mass--metallicity relation \citep{Curti2020, Sanders2021}, these galaxies are expected to have higher metallicities than the ACE galaxies. Their preferential location at higher $\rm DGR_{mol}$ than ACE therefore provides additional, support for a positive dependence of the dust-to-molecular-gas ratio on metallicity. Taken together, the current measurements suggest no clear redshift evolution out to \(z\sim3\), with the ACE and the literature samples remaining broadly consistent with the local reference range. The higher-redshift [C\,{\sc ii}]-based samples, however, tend to occupy lower $\rm DGR_{mol}$ values. This apparent decline at $z\gtrsim5$ may indicate evolution in the dust-to-molecular-gas ratio, although it is also sensitive to the calibration of [C\,{\sc ii}] as a molecular gas tracer, to differences in sample selection and to their lower metallicities. For one of the brightest REBELS galaxies, REBELS-25 at $z=7.3$, there is a CO-based molecular gas mass available which we adopt from \citet{Cescon2026} ($\alpha_{\rm CO}=3$). REBELS-25 is indicated with a different symbol than the [C\,{\sc ii}] REBELS sample, but it still falls in the same regime as the rest of the REBELS galaxies in the dust-to-molecular gas plane. We discuss the comparison and uncertainties associated with the higher-redshift [C\,{\sc ii}]-based samples in more detail in Appendix~\ref{appendixCII}.

\subsection{Dust-to-metal ratio}
\label{sec:DMR}
A closely related quantity often considered alongside the dust-to-gas ratio is the dust-to-metal ratio (DMR), which quantifies the fraction of metals associated with the molecular gas reservoir that is locked into dust, defined as
\begin{align*}
\mathrm{DMR} = \frac{M_{\rm dust}}{M_{\rm dust} + M_{\rm metal}}
= \frac{M_{\rm dust}}{M_{\rm dust} + (M_{\rm mol} \times Z_{\rm gas})},
\end{align*}
where
\begin{align*}
Z_{\rm gas} = Z_{\odot} \times 10^{(12+\log (\rm O/H))-8.69}.
\end{align*}
For the solar metal mass fraction, we adopt $Z_{\odot}=0.0134$ \citep{Asplund2009}. Here, $12+\log(\mathrm{O/H})$ is measured from nebular emission lines tracing the ionized gas, while $M_{\rm mol}$ traces the molecular gas reservoir. We therefore assume that the oxygen abundance measured in the ionized gas is representative of the metallicity of the molecular gas, allowing us to apply $Z_{\rm gas}$ to $M_{\rm mol}$ when estimating its associated metal mass. The implications of this assumption are discussed further in Section~\ref{sec:discussioncaveats}.

\begin{figure}[h!]
    \centering
    \includegraphics[width=\linewidth]{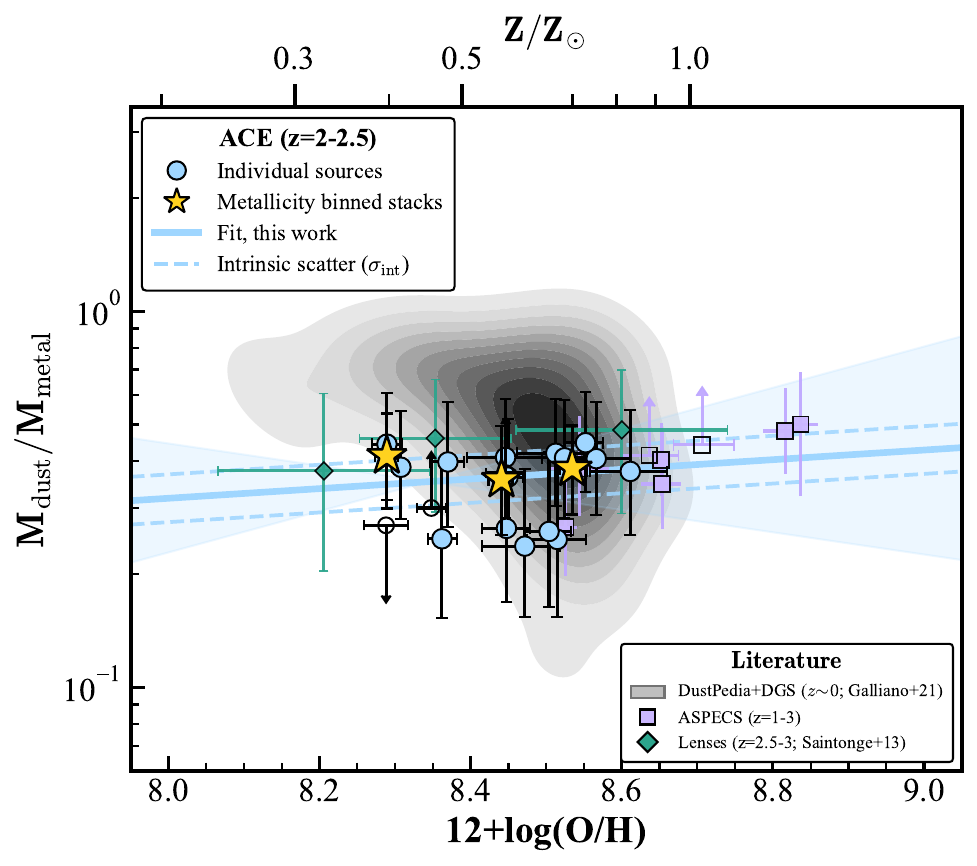}
    \caption{Dust-to-metal ratio (DMR), expressed as $M_{\rm dust}/M_{\rm metal}$, as a function of metallicity for the ACE sample at $z=2-2.5$. Light blue circles show individual ACE galaxies, with smaller symbols indicating sources with upper limits. Yellow stars mark the metallicity-binned stacks. The light blue line shows the best-fitting relation derived in this work, with the shaded region indicating the error on the fitting parameters and the dashed lines the intrinsic scatter. We compare to literature measurements from DustPedia+DGS at $z\sim0$ (grey contours), ASPECS at $z=1-3$ (purple squares)  and lensed galaxies from \citet{Saintonge2013} at $z=2.5-3$ (green diamonds). We find no metallicity dependence in the DMR of the ACE galaxies.}
    \label{fig:MdustMmetal}
\end{figure}

The resulting DMR results are shown as a function of metallicity in Figure~\ref{fig:MdustMmetal}. Neither the ACE galaxies nor the literature samples show an obvious dependence of DMR on metallicity. For the ACE detections, we measure a Pearson correlation coefficient of $r=0.001$ and a $p$-value of $p=0.1$, indicating no statistically significant correlation over the metallicity range probed. To quantify this result, we fit the ACE galaxies in the DMR--metallicity plane using the single-power-law model described in Section~\ref{sec:DRGmolphysical}, resulting in a slope of $0.13 \pm 0.4$. The inferred slope is consistent with zero, supporting the absence of a measurable metallicity dependence in the DMR of the ACE galaxies. The detected ACE galaxies have a mean $\log(\rm DMR)=-0.46\pm0.03$ and a mean metallicity of $12+\log({\rm O/H})=8.45\pm0.02$, where the quoted uncertainties are the standard errors on the mean. 

\section{Discussion}
\label{discussion}
We present constraints on the $\rm DGR_{mol}$--metallicity relation using 18 unlensed sub-solar metallicity galaxies at $z\sim2-2.5$ going down to 0.4 $Z_{\odot}$. This provides a direct comparison between local and high-redshift dust-to-gas measurements to typical galaxies at cosmic noon, and offers new constraints on models of galaxy evolution in a previously unexplored regime.
\subsection{Dust-to-molecular-gas versus metallicity}
\label{sec:DTMolcomp}
The ACE sample probes the $\rm DGR_{mol}$ relation in typical galaxies at $z\sim2-2.5$ across metallicities ranging between $8.3 \leq 12+\log(\mathrm{O/H}) < 8.7$. For this metallicity range, we find a metallicity dependence consistent with a single power law. We do not find evidence for a clear turnover or a critical metallicity regime in the $\rm DGR_{mol}$ relation, which has been observed in the local universe for $\rm DGR_{tot}$ \citep{RemyRuyer2014} and shown in theoretical models \citep{Galliano2021, Parente2025}. We further explore the lack of a turnover in the ACE sample against theoretical predictions in Section~\ref{sec:simulations}. The $\rm DGR_{mol}$ of the ACE sample can be best described by a slope of $\approx 1$, corresponding to a constant dust-to-metal ratio, as shown in Figure~\ref{fig:MdustMmetal}. This implies that variations in $\rm DGR_{mol}$ over the regime probed by ACE, primarily reflect changes in metal content rather than by variations in the fraction of metals locked into dust grains. The absence of a turnover in Figure~\ref{fig:MdustMmol} may indicate that it occurs at lower metallicities than those probed by our sample ($\rm 12+\log({\rm O/H}) \leq 8.2$). However, unlike for the total cold gas, it is also possible that no turnover is present at all. The tight coupling between dust and molecular gas may prevent the low-metallicity turnover expected for the total cold gas reservoir. It is also possible that the absence of a clear turnover reflects a selection effect, since ACE targets actively star-forming galaxies that may preferentially trace systems with already enriched, molecule-rich ISM conditions. Distinguishing between these possibilities will require extending such measurements to lower metallicities and to more diverse galaxy populations. 

To place the ACE results in context, we compare to literature samples using homogeneous dust- and molecular-gas calibrations, including only systems with CO-based molecular gas measurements. Under this consistent comparison, the more massive ASPECS galaxies and the lensed \citet{Saintonge2013} galaxies at similar redshift follow the same relation as ACE extending the metallicity range of galaxies at cosmic noon to $8.2 \leq 12+\log(\mathrm{O/H}) < 9$. Combining ASPECS and the sample of \citet{Saintonge2013}, we fit a single power-law, following the method as described in Section~\ref{sec:DRGmolphysical}. This results in a slope of $1.37^{+0.63}_{-0.57}$, which agrees within the uncertainties for the fit to the ACE data, supporting the trend we derive for the ACE galaxies.

The local universe sample of \cite{Galliano2021} shows a striking overlap with the ACE sample at fixed metallicity. The residuals of the ACE $\rm DGR_{mol}$ measurements relative to the results from \cite{Galliano2021} are consistent with zero within the scatter, indicating no significant difference in normalization. In contrast, the \cite{Galliano2021} sample does not exhibit the clear increase of  $\rm DGR_{mol}$ with metallicity seen in ACE, despite adopting the same metallicity-dependent $\alpha_{\rm CO}$. This may be a sample selection effect, since restricting the comparison to galaxies with available CO measurements causes the \cite{Galliano2021} sample to cluster within $8.4 \lesssim 12+\log({\rm O/H}) \lesssim 8.6$, which is a relatively narrow range to constrain a metallicity trend. Alternatively this could also suggest the stronger metallicity dependence of $\rm DGR_{tot}$ in the local universe is primarily driven by the atomic-gas component that we exclude here, we further explore the ACE sample in the context of the $\rm DGR_{tot}$ relations in Appendix~\ref{totalgasappendix}.

The agreement in normalization between the $z=0$ sample and ACE suggests that the same dust-growth physics, in this regime likely grain growth in the ISM, dominates across the metallicities of $8.3 \leq 12+\log(\rm O/H) \leq 8.7$ at cosmic noon. Extending this analysis to lower metallicity regimes than explored in this work remains challenging with CO, since it becomes increasingly difficult to detect in metal-poor systems \citep{Madden2020}. In addition, uncertainties in the CO-to-H$_2$ conversion factor, $\alpha_{\rm CO}$, particularly in low-metallicity environments, complicate the interpretation of CO-based molecular gas measurements \citep{Bolatto2013, Schinnerer2024}. These limitations motivate future work with complementary tracers, in particular atomic carbon ([C I]), which may remain an effective probe of molecular gas in low-metallicity environments \citep{Glover2016}.

\subsection{Caveats}
\label{sec:discussioncaveats}
In Section~\ref{sec:scoville}, we compare the ACE dust and CO luminosities with the empirical relations of \citet{Scoville2016}, which are calibrated primarily on massive local star-forming galaxies and ULIRGs, together with SMGs at $z\sim2$. Despite the metallicity dependence of $\mathrm{DGR}_{\rm mol}$ in the ACE sample, no corresponding trend is evident in the observed dust and CO luminosities: the ACE galaxies remain within the scatter of the \citet{Scoville2016} relations in both panels of Figure~\ref{fig:2panelobservables}. This indicates that the dust--CO luminosity relation extends to the lower-mass, sub-solar-metallicity galaxies probed by ACE. However, a roughly constant $L_{\nu}(850,\mu{\rm m})/L'_{\rm CO(1-0)}$ across these metallicities does not imply metallicity should not be taken into account when converting dust masses to molecular gas masses. At lower metallicity, a decreasing dust-to-gas ratio reduces dust emission per unit gas mass, while an increasing CO-dark gas fraction reduces CO luminosity per unit molecular gas mass (i.e., increases $\alpha_{\rm CO}$). These effects can partially compensate in the dust-to-CO luminosity ratio. Thus, while dust and CO luminosities remain closely correlated in ACE, metallicity-dependent corrections to both $\mathrm{DGR}_{\rm mol}$ and $\alpha_{\rm CO}$ may still be required to infer $M_{\rm mol}$.

The lack of a clear metallicity dependence in the observable dust-to-CO luminosity ratio emphasizes the importance of the conversion factors used to infer dust and molecular gas masses. Focusing first on the dust mass, in this work we have adopted a constant dust temperature of 25~K, see also \citet{Popescu2026} for more details on the dust temperatures in the ACE sample. However, observations in the local Universe suggest that low-metallicity galaxies tend to have warmer dust \citep{RemyRuyer2013,RemyRuyer2015}, raising the question of whether a fixed $T_{\rm dust}$ affects the metallicity dependence inferred for $\mathrm{DGR}_{\rm mol}$. If the lower-metallicity ACE galaxies are similarly warmer, assuming 25~K would overestimate their dust masses. Accounting for this variation would therefore steepen the inferred $\mathrm{DGR}_{\rm mol}$--metallicity relation. However, because our observations probe the Rayleigh--Jeans tail and are therefore sensitive primarily to the cold dust component, the inferred dust mass depends is only inversely proportional to temperature, $M_{\rm dust}\propto T_{\rm dust}^{-1}$. In comparison, the dust masses used by both local $\mathrm{DGR}_{\rm tot}$ studies of  \citet{RemyRuyer2014} and \citet{Galliano2021} were derived from SED fitting, that accounts for dust spanning a range of temperatures, instead of assuming a single dust temperature.

For the conversion from CO to molecular gas mass, we adopt a metallicity-dependent $\alpha_{\rm CO}$ for ACE. If instead we assume a constant Milky Way value \citep{Bolatto2013}, the positive $\mathrm{DGR}_{\rm mol}$--metallicity correlation disappears, with no statistically significant correlation remaining, consistent with the nearly constant observed dust-to-CO luminosity ratio. Thus, the choice of $\alpha_{\rm CO}$ strongly affects the recovered slope of the $\mathrm{DGR}_{\rm mol}$ relation. In the local Universe, \citet{RemyRuyer2014} found that $\mathrm{DGR}_{\rm tot}$, including both atomic and molecular gas, increases with metallicity when adopting either a constant Galactic or a metallicity-dependent CO-to-H$_2$ conversion factor. This suggests that the local $\mathrm{DGR}_{\rm tot}$--metallicity relation is not driven solely by the adopted $\alpha_{\rm CO}$. However, because H\,{\sc i} can dominate the total gas reservoir, particularly at low metallicity, the sensitivity to $\alpha_{\rm CO}$ is expected to be weaker than for the molecular $\mathrm{DGR}_{\rm mol}$ considered for ACE. At cosmic noon, where direct measurements of atomic gas in emission remain largely inaccessible, we are limited to the molecular gas based measurements only, further underscoring the importance of constraining $\alpha_{\rm CO}$ at $z>1$.

As emphasized above, our analysis focuses on the molecular gas phase and assumes that the measured gas-phase metallicity is representative of the material traced by the dust and CO emission. However, these quantities do not necessarily probe the same ISM phase: CO and Rayleigh-Jeans tail dust emission trace the predominantly neutral ISM, whereas $12+\log(\mathrm{O/H})$ is inferred from emission lines arising from ionized gas in H\,{\sc ii} regions. We therefore implicitly assume that the metallicity measured in the ionized gas is representative of the molecular gas considered here. This assumption is particularly important for $\mathrm{DTM}$, where the ionized-gas metallicity is combined with the CO-based molecular gas mass to infer the metal mass. Spatial offsets may further complicate this comparison, as the optical, CO, and dust emission may not originate from the same regions. The ALMA and JWST maps show tentative differences in their spatial distributions, as discussed further in \citet{Shivaei2026}. Higher-resolution observations will be needed to better understand how these different ISM tracers are spatially related.
\begin{figure}[h!]
    \centering
    \includegraphics[width=\linewidth]{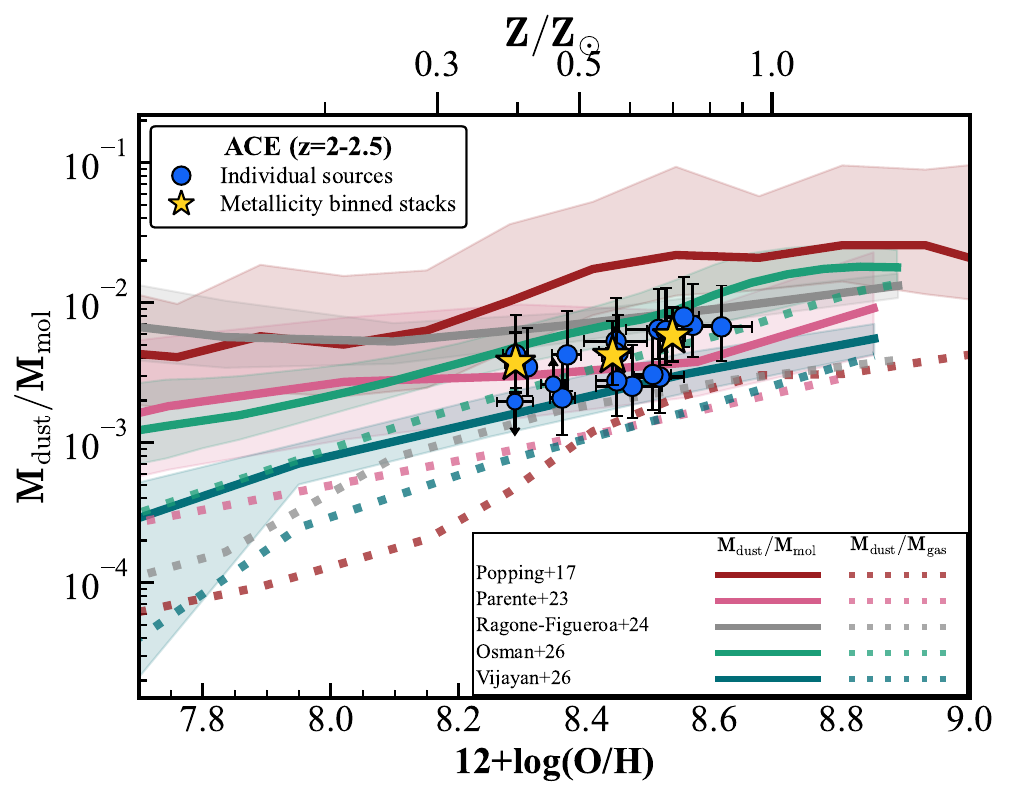}
    \caption{$M_{\rm dust}/M_{\rm mol}$ as a function of metallicity, ACE compared to simulations predictions. The ACE measurements are the same as in Figure~\ref{fig:MdustMmol}. Solid curves show model predictions for $M_{\rm dust}/M_{\rm mol}$, while dotted curves show the corresponding $M_{\rm dust}/M_{\rm gas}$ predictions from \cite{Popping2017}, \cite{Parente23},\cite{Ragone2024}, \cite{Osman2026} and \cite{Vijayan2026}.}
    \label{fig:ratiosvssims}
\end{figure}
\subsection{Simulation perspective}
\label{sec:simulations}
The ACE sample allows us to test galaxy-evolution models with explicit dust physics in the sub-solar metallicity regime at cosmic noon. In Figure \ref{fig:ratiosvssims}, we compare the observed $\rm DGR_{mol}$ with predictions for both $\rm DGR_{mol}$ and $\rm DGR_{tot}$ from semi-analytic models and hydrodynamical simulations at redshifts $z=2-3$. We show the semi-analytical models of \citet{Popping2017}, \citet{Parente23} and \citet{Osman2026}, which all implement detailed dust evolution physics, but differ in their underlying galaxy formation models and prescriptions for processes such as grain growth, dust destruction and the H\,{\sc i} and H$_2$ partitioning. Focusing on $\rm DGR_{mol}$, we find the best agreement with \citet{Parente23}, while both \citet{Osman2026} and \citet{Popping2017} predict systematically higher ratios. \citet{Popping2017} shows the biggest discrepancy, which may indicate that it produces dust too efficiently or does not destroy it rapidly enough in this regime. The ordering changes when we consider the $\rm DGR_{tot}$, including the atomic gas component: \citet{Popping2017} predicts the lowest ratios, while \citet{Parente23} and \citet{Osman2026} remain relatively close, with the latter predicting somewhat higher values. This contrasting behaviour of \citet{Popping2017} indicates that the high $\rm DGR_{mol}$ values might not be driven by the dust abundance but differences in the molecular gas fraction. In terms of shape all three semi-analytical models show similar trends in $\rm DGR_{mol}$, with no clear turnover over the metallicity range probed by ACE. For $\rm DGR_{tot}$, however, the models show more pronounced differences: \citet{Popping2017} shows an apparent turnover around $12+\log({\rm O/H})\sim 8.4$, \citet{Parente23} instead predicts a turnover only at lower metallicities, below $12+\log({\rm O/H})\sim7.8$. \citet{Osman2026} does not predict a distinct turnover, but rather a steepening of the $\rm DGR_{tot}$--metallicity relation toward higher metallicities. Given that we cannot quantify  $\rm DGR_{tot}$ for ACE we can only place upper limits in this context, which does not constrain the shape in comparison to the models. 

Additionally, we compare the ACE results to predictions of two hydrodynamical simulations: the COLIBRE cosmological simulations \citep{Schaye2026, Vijayan2026} and the MUPPI-based hydrodynamical models presented by \citet{Ragone2024}. Both frameworks self-consistently model the production, growth, and destruction of dust, including the evolution of different grain sizes and chemical compositions. However, they differ in their treatment of the multiphase ISM: COLIBRE uniquely tracks the cold ISM, allowing gas to cool to $\sim10$ K. Figure \ref{fig:ratiosvssims}, shows that for the $\rm DGR_{mol}$ \citet{Ragone2024} predicts a flat relationship with values above the ACE observations whereas \citet{Vijayan2026} predictions agree with ACE in the lower end of the $\rm DGR_{mol}$ detections, reaching a turnover regime at lower metallicities, below $12+\log({\rm O/H})\sim8$, not reached by ACE. If instead we focus on their $\rm DGR_{tot}$ predictions, both hydrodyimcal models predict values below the ACE data, which is expected from the inclusion of atomic gas that lowers the total dust-to-gas ratio. Adittionally, the shapes of the $\rm DGR_{tot}$ models are more similar, showing apparent turnovers around metallicities below $12+\log({\rm O/H})\sim8.1$, a metallicity regime not probed by ACE. 

Overall, the models indicate a relatively tight coupling between dust mass and molecular gas mass, with variations increasing at lower metallicities, below $12+\log({\rm O/H})\sim8$. Although there is some scatter in the predicted $\rm DGR_{mol}$ relation, the models generally do not show the strong turnover, or pronounced S-shaped behaviour, seen when atomic gas is included in $\rm DGR_{tot}$. This suggests that the main difference between the two ratios is not only their normalization, but the shape of the relation itself. In turn, this implies that the change in shape is driven primarily by the inclusion of atomic gas, whereas dust and molecular gas appear much more closely linked. For $\rm DGR_{mol}$ the slopes of the predictions appear to remain constant, wheras the normalization varies between the models, with the models from \citet{Parente23} and \citet{Vijayan2026} closest to the ACE data in the metallicity regime probed here, whereas \citet{Osman2026}, \citet{Ragone2024} and \citet{Popping2017} predicts higher values than ACE for $\rm DGR_{mol}$ at fixed metallicity.

\section{Conclusions}
\label{conclusions}
We present measurements of the dust-to-molecular-gas ratio, $\rm DGR_{mol}$, for galaxies from the ALMA Large Program ACE. By targeting typical star-forming galaxies at $z\simeq2-2.5$, ACE uniquely extends direct dust and CO measurements at cosmic noon down to stellar masses of $10^9\, \rm M_{\odot}$ and metallicities of 0.4 $\rm Z_{\odot}$. This allows us to constrain the relation between $\rm DGR_{mol}$ and metallicity in the sub-solar metallicity regime at $z\sim2-2.5$. Our main conclusions are as follows.

\begin{enumerate}

    \item We find that $\rm DGR_{mol}$ increases with metallicity in the ACE sample. Fitting the relation in log space, we measure a slope of $1.2 \pm 0.7$. This trend is supported by the metallicity binned stacks and extends to more-massive galaxies taken from the ASPECS survey and a sample of lensed galaxies from \cite{Saintonge2013}, suggesting that metal-poor galaxies have systematically lower dust-to-molecular-gas ratios than their more metal-rich counterparts. In agreement with the slope for the $\rm DGR_{mol}$ we find a constant dust-to-metal ratio with metallicity. 

    \item The observed dust and CO luminosities follow the \cite{Scoville2016} relation down to the lower masses and metallicities probed by ACE, supporting dust continuum as a tracer of CO and molecular gas in this regime.

    \item  The inferred slope of the $\rm DGR_{mol}$ -- metallicity relation is driven primarily by the adopted metallicity-dependent $\alpha_{\rm CO}$; highlighting the need to constrain $\alpha_{\rm CO}$ at $z>1$. 

    \item The ACE galaxies at $z\sim2-2.5$ are consistent with the normalization of the $\rm DGR_{mol}$ of $z\sim0$ observations over the metallicity range probed here. The agreement in normalization between the local universe galaxies and ACE suggests that the same dust-growth physics, in this regime likely grain growth in the ISM, dominates across the metallicities of $8.3 \leq 12+\log(\rm O/H) \leq 8.7$ at cosmic noon.

    \item The ACE measurements provide new empirical constraints for models of galaxy evolution and dust enrichment at cosmic noon. By extending to sub-solar metallicities (0.4 $\rm Z_{\odot}$), ACE directly tests the regime where ISM grain growth is expected to dominate dust production. Current semi-analytic models and hydrodynamical simulations reproduce some qualitative features of the data, but often predict $\rm DGR_{mol}$ relations that do not show the same slope or normalization as ACE, suggesting differences in how dust growth, destruction, and the molecular gas reservoir are treated.
\end{enumerate}
Further progress at high redshift requires extending measurements to lower metallicities and resolving the dust and gas distributions within galaxies. Reaching faint, compact, metal-poor systems will demand greater sensitivity, longer integrations and potentially next-generation facilities such as major upgraded ALMA with significantly higher spectral line sensitivity. At the same time alternative molecular gas tracers such as [CI] should also be considered to constrain the cold gas reservoir. To measure the total dust-to-gas ratio constraints on the atomic gas at $z>0.1$ from future planned facilities like SKA will be essential. Higher resolution observations can then connect the global ratios to dust and gas surface densities, providing more direct constraints on dust growth and destruction. 

\begin{acknowledgements}
We want to thank Benedetta Casavecchia for valuable discussion, and Aswin Vijayan, Omima Osman and Cinthia Ragone-Figueroa for making their models available. This work has been funded by the European Research Council (ERC) under the European Union's Horizon 2020 research and innovation programme (DistantDust, Grant agreement No. 101117541). L.A.B. acknowledges support from the Dutch Research Council (NWO) under grant VI.Veni.242.055 (\url{https://doi.org/10.61686/LAJVP77714}). IL acknowledges the grant Atracción de Talento Grant No. 2022-T1/TIC-20472, funded by the Comunidad de Madrid, Spain. MK acknowledges support from the Australian Research Council via the Discovery Early Career Researcher Award DE250100709. DN is grateful for support from NASA via grants ATP-21-0013 and ATP-23-0002. MP is funded by NASA grant ATP-23-0002. This paper makes use of the following ALMA data: ADS/JAO.ALMA\#2018.1.01128.S, 2019.1.01142.S, 2024.1.00534.L. ALMA is a partnership of ESO (representing its member states), NSF (USA) and NINS (Japan), together with NRC (Canada), MOST and ASIAA (Taiwan), and KASI (Republic of Korea), in cooperation with the Republic of Chile. We acknowledge assistance and computational support provided by Allegro, the European ALMA Regional Center node in the Netherlands.
\end{acknowledgements}

\bibliographystyle{aa}
\bibliography{bib}

@ARTICLE{Parente2025,
       author = {{Parente}, Massimiliano},
        title = "{Modeling Dust in Galaxy Evolution Simulations}",
      journal = {arXiv e-prints},
         year = 2025,
        month = apr,
          eid = {arXiv:2504.10585},
        pages = {arXiv:2504.10585},
          doi = {10.48550/arXiv.2504.10585},
archivePrefix = {arXiv},
       eprint = {2504.10585},
 primaryClass = {astro-ph.GA},
       adsurl = {https://ui.adsabs.harvard.edu/abs/2025arXiv250410585P}
}

@ARTICLE{Schaye2026,
       author = {{Schaye}, Joop and {Chaikin}, Evgenii and {Schaller}, Matthieu and {Ploeckinger}, Sylvia and {Hu{\v{s}}ko}, Filip and {McGibbon}, Robert J. and {Trayford}, James W. and {Ben{\'\i}tez-Llambay}, Alejandro and {Correa}, Camila and {Frenk}, Carlos S. and et al.},
        title = "{The COLIBRE project: cosmological hydrodynamical simulations of galaxy formation and evolution}",
      journal = {\mnras},
         year = 2026,
        month = may,
       volume = {548},
       number = {1},
          eid = {stag375},
        pages = {stag375},
          doi = {10.1093/mnras/stag375},
archivePrefix = {arXiv},
       eprint = {2508.21126},
 primaryClass = {astro-ph.GA},
       adsurl = {https://ui.adsabs.harvard.edu/abs/2026MNRAS.548ag375S}
}

@ARTICLE{Vijayan2026,
       author = {{Vijayan}, Aswin P. and {Trayford}, James W. and {Schaye}, Joop and {Ploeckinger}, Sylvia and {Gebek}, Andrea and {Andreadis}, Nick and {Baes}, Maarten and {Ben{\'\i}tez-Llambay}, Alejandro and {Chaikin}, Evgenii and {Frenk}, Carlos S. and et al.},
        title = "{The evolution of galaxy dust scaling relations in the COLIBRE simulations}",
      journal = {arXiv e-prints},
         year = 2026,
        month = jul,
          eid = {arXiv:2607.26058},
        pages = {arXiv:2607.26058},
          doi = {10.48550/arXiv.2607.26058},
archivePrefix = {arXiv},
       eprint = {2607.26058},
 primaryClass = {astro-ph.GA},
       adsurl = {https://ui.adsabs.harvard.edu/abs/2026arXiv260726058V}
}

@ARTICLE{Osman2026,
       author = {{Osman}, Omima and {De Lucia}, Gabriella and {Fontanot}, Fabio and {Xie}, Lizhi and {Hirschmann}, Michaela},
        title = "{Dust evolution across cosmic times as seen through DUSTY-GAEA}",
      journal = {\aap},
         year = 2026,
        month = jun,
       volume = {710},
          eid = {A306},
        pages = {A306},
          doi = {10.1051/0004-6361/202558572},
       adsurl = {https://ui.adsabs.harvard.edu/abs/2026A&A...710A.306O}
}

@ARTICLE{MessiasSKA2024,
       author = {{Messias}, Hugo and {Guerrero}, Andrea and {Nagar}, Neil and {Regueiro}, Jack and {Impellizzeri}, Violette and {Orellana}, Gustavo and {Vioque}, Miguel},
        title = "{H I content at cosmic noon - a millimetre-wavelength perspective}",
      journal = {\mnras},
         year = 2024,
        month = oct,
       volume = {533},
       number = {4},
        pages = {3937-3956},
          doi = {10.1093/mnras/stae1807},
archivePrefix = {arXiv},
       eprint = {2312.02782},
 primaryClass = {astro-ph.GA},
       adsurl = {https://ui.adsabs.harvard.edu/abs/2024MNRAS.533.3937M}
}

@ARTICLE{Saintonge2022,
       author = {{Saintonge}, Am{\'e}lie and {Catinella}, Barbara},
        title = "{The Cold Interstellar Medium of Galaxies in the Local Universe}",
      journal = {\araa},
         year = 2022,
        month = aug,
       volume = {60},
        pages = {319-361},
          doi = {10.1146/annurev-astro-021022-043545},
archivePrefix = {arXiv},
       eprint = {2202.00690},
 primaryClass = {astro-ph.GA},
       adsurl = {https://ui.adsabs.harvard.edu/abs/2022ARA&A..60..319S}
}

@unpublished{Shivaei2026,
  author = {{Shivaei}, Irene and {Popping}, Gergo and {Boogaard}, Leindert A. and {Pope}, Alexandra and {Langan}, Ivanna and {Popescu}, Roxana and {Geesink}, Nikki N. and {Solimano}, Manuel and {Casey}, Caitlin M. and {Franco}, Maximilien and {Kaasinen}, Melanie and {Narayanan}, Desika and {Perez-Gonzalez}, Pablo G. and {Reddy}, Naveen and {Sanders}, Ryan},
  journal = {\aap},
  year   = {2026},
  note   = {submitted (ALMA Chemical Evolution (ACE) survey)}
}

@unpublished{Langan2026,
  author = {{Langan}, Ivanna and {Shivaei}, Irene and {Popping}, Gergo and {Boogaard}, Leindert A. and {Geesink}, Nikki N. and {Kaasinen}, Melanie and {Pope}, Alexandra and {Popescu}, Roxana and {Solimano}, Manuel and {Arriscado}, Leonor and {Narayanan}, Desika and {Parente}, Massimiliano and {Reddy}, Naveen and {Sanders}, Ryan },
  journal = {\aap},
  year   = {2026},
  note   = {submitted (ALMA Chemical Evolution (ACE) survey)}
}

@unpublished{Solimano2026,
  author = {{Solimano}, Manuel and {Shivaei}, Irene and {Popping}, Gergo and {Boogaard}, Leindert A. and {Langan}, Ivanna and {Popescu}, Roxana and {Geesink}, Nikki N. and  {Pope}, Alexandra and {Arriscado}, Leonor and {Mobasher}, Bahram and {Narayanan}, Desika and {Parente}, Massimiliano and {Reddy}, Naveen and {Sanders}, Ryan},
  journal = {\aap},
  year   = {2026},
  note   = {submitted (ALMA Chemical Evolution (ACE) survey)}
}

@unpublished{Popescu2026,
  author = {{Popescu}, Roxana and {Pope}, Alexandra and {Shivaei}, Irene and {Popping}, Gergo and {Boogaard}, Leindert A. and  {Geesink}, Nikki N. and  {Langan}, Ivanna and  {Solimano}, Manuel},
  journal = {\aap},
  year   = {2026},
  note   = {submitted (ALMA Chemical Evolution (ACE) survey)}
}

@ARTICLE{Groves2015,
       author = {{Groves}, Brent A. and {Schinnerer}, Eva and {Leroy}, Adam and {Galametz}, Maud and {Walter}, Fabian and {Bolatto}, Alberto and {Hunt}, Leslie and {Dale}, Daniel and {Calzetti}, Daniela and {Croxall}, Kevin and et al.},
        title = "{Dust Continuum Emission as a Tracer of Gas Mass in Galaxies}",
      journal = {\apj},
         year = 2015,
        month = jan,
       volume = {799},
       number = {1},
          eid = {96},
        pages = {96},
          doi = {10.1088/0004-637X/799/1/96},
archivePrefix = {arXiv},
       eprint = {1411.2975},
 primaryClass = {astro-ph.GA},
       adsurl = {https://ui.adsabs.harvard.edu/abs/2015ApJ...799...96G}
}

@ARTICLE{Schinnerer2016,
       author = {{Schinnerer}, E. and {Groves}, B. and {Sargent}, M.~T. and {Karim}, A. and {Oesch}, P.~A. and {Magnelli}, B. and {LeFevre}, O. and {Tasca}, L. and {Civano}, F. and {Cassata}, P. and et al.},
        title = "{Gas Fraction and Depletion Time of Massive Star-forming Galaxies at z \raisebox{-0.5ex}\textasciitilde 3.2: No Change in Global Star Formation Process out to z > 3}",
      journal = {\apj},
         year = 2016,
        month = dec,
       volume = {833},
       number = {1},
          eid = {112},
        pages = {112},
          doi = {10.3847/1538-4357/833/1/112},
archivePrefix = {arXiv},
       eprint = {1610.03656},
 primaryClass = {astro-ph.GA},
       adsurl = {https://ui.adsabs.harvard.edu/abs/2016ApJ...833..112S}
}

@ARTICLE{Bethermin2015,
       author = {{B{\'e}thermin}, Matthieu and {Daddi}, Emanuele and {Magdis}, Georgios and {Lagos}, Claudia and {Sargent}, Mark and {Albrecht}, Marcus and {Aussel}, Herv{\'e} and {Bertoldi}, Frank and {Buat}, V{\'e}ronique and {Galametz}, Maud and et al.},
        title = "{Evolution of the dust emission of massive galaxies up to z = 4 and constraints on their dominant mode of star formation}",
      journal = {\aap},
         year = 2015,
        month = jan,
       volume = {573},
          eid = {A113},
        pages = {A113},
          doi = {10.1051/0004-6361/201425031},
archivePrefix = {arXiv},
       eprint = {1409.5796},
 primaryClass = {astro-ph.GA},
       adsurl = {https://ui.adsabs.harvard.edu/abs/2015A&A...573A.113B}
}

@ARTICLE{Magdis2012,
       author = {{Magdis}, Georgios E. and {Daddi}, E. and {B{\'e}thermin}, M. and {Sargent}, M. and {Elbaz}, D. and {Pannella}, M. and {Dickinson}, M. and {Dannerbauer}, H. and {da Cunha}, E. and {Walter}, F. and et al.},
        title = "{The Evolving Interstellar Medium of Star-forming Galaxies since z = 2 as Probed by Their Infrared Spectral Energy Distributions}",
      journal = {\apj},
         year = 2012,
        month = nov,
       volume = {760},
       number = {1},
          eid = {6},
        pages = {6},
          doi = {10.1088/0004-637X/760/1/6},
archivePrefix = {arXiv},
       eprint = {1210.1035},
 primaryClass = {astro-ph.CO},
       adsurl = {https://ui.adsabs.harvard.edu/abs/2012ApJ...760....6M}
}

@ARTICLE{RemyRuyer2013,
       author = {{R{\'e}my-Ruyer}, A. and {Madden}, S.~C. and {Galliano}, F. and {Hony}, S. and {Sauvage}, M. and {Bendo}, G.~J. and {Roussel}, H. and {Pohlen}, M. and {Smith}, M.~W.~L. and {Galametz}, M. and et al.},
        title = "{Revealing the cold dust in low-metallicity environments. I. Photometry analysis of the Dwarf Galaxy Survey with Herschel}",
      journal = {\aap},
         year = 2013,
        month = sep,
       volume = {557},
          eid = {A95},
        pages = {A95},
          doi = {10.1051/0004-6361/201321602},
archivePrefix = {arXiv},
       eprint = {1309.1371},
 primaryClass = {astro-ph.CO},
       adsurl = {https://ui.adsabs.harvard.edu/abs/2013A&A...557A..95R}
}

@ARTICLE{RemyRuyer2015,
       author = {{R{\'e}my-Ruyer}, A. and {Madden}, S.~C. and {Galliano}, F. and {Lebouteiller}, V. and {Baes}, M. and {Bendo}, G.~J. and {Boselli}, A. and {Ciesla}, L. and {Cormier}, D. and {Cooray}, A. and et al.},
        title = "{Linking dust emission to fundamental properties in galaxies: the low-metallicity picture}",
      journal = {\aap},
         year = 2015,
        month = oct,
       volume = {582},
          eid = {A121},
        pages = {A121},
          doi = {10.1051/0004-6361/201526067},
archivePrefix = {arXiv},
       eprint = {1507.05432},
 primaryClass = {astro-ph.GA},
       adsurl = {https://ui.adsabs.harvard.edu/abs/2015A&A...582A.121R}
}

@ARTICLE{Sanders2021,
       author = {{Sanders}, Ryan L. and {Shapley}, Alice E. and {Jones}, Tucker and {Reddy}, Naveen A. and {Kriek}, Mariska and {Siana}, Brian and {Coil}, Alison L. and {Mobasher}, Bahram and {Shivaei}, Irene and {Dav{\'e}}, Romeel and et al.},
        title = "{The MOSDEF Survey: The Evolution of the Mass-Metallicity Relation from z = 0 to z 3.3}",
      journal = {\apj},
         year = 2021,
        month = jun,
       volume = {914},
       number = {1},
          eid = {19},
        pages = {19},
          doi = {10.3847/1538-4357/abf4c1},
archivePrefix = {arXiv},
       eprint = {2009.07292},
 primaryClass = {astro-ph.GA},
       adsurl = {https://ui.adsabs.harvard.edu/abs/2021ApJ...914...19S}
}

@ARTICLE{Curti2020,
       author = {{Curti}, Mirko and {Mannucci}, Filippo and {Cresci}, Giovanni and {Maiolino}, Roberto},
        title = "{The mass-metallicity and the fundamental metallicity relation revisited on a fully T$_{e}$-based abundance scale for galaxies}",
      journal = {\mnras},
         year = 2020,
        month = jan,
       volume = {491},
       number = {1},
        pages = {944-964},
          doi = {10.1093/mnras/stz2910},
archivePrefix = {arXiv},
       eprint = {1910.00597},
 primaryClass = {astro-ph.GA},
       adsurl = {https://ui.adsabs.harvard.edu/abs/2020MNRAS.491..944C}
}

@ARTICLE{Shuntov2025,
       author = {{Shuntov}, Marko and {Akins}, Hollis B. and {Paquereau}, Louise and {Casey}, Caitlin M. and {Ilbert}, Olivier and {Arango-Toro}, Rafael C. and {McCracken}, Henry Joy and {Franco}, Maximilien and {Harish}, Santosh and {Kartaltepe}, Jeyhan S. and et al.},
        title = "{COSMOS2025: The COSMOS-Web galaxy catalog of photometry, morphology, redshifts, and physical parameters from JWST, HST, and ground-based imaging}",
      journal = {\aap},
         year = 2025,
        month = dec,
       volume = {704},
          eid = {A339},
        pages = {A339},
          doi = {10.1051/0004-6361/202555799},
archivePrefix = {arXiv},
       eprint = {2506.03243},
 primaryClass = {astro-ph.GA},
       adsurl = {https://ui.adsabs.harvard.edu/abs/2025A&A...704A.339S}
}

@ARTICLE{Bolatto2013,
       author = {{Bolatto}, Alberto D. and {Wolfire}, Mark and {Leroy}, Adam K.},
        title = "{The CO-to-H$_{2}$ Conversion Factor}",
      journal = {\araa},
         year = 2013,
        month = aug,
       volume = {51},
       number = {1},
        pages = {207-268},
          doi = {10.1146/annurev-astro-082812-140944},
archivePrefix = {arXiv},
       eprint = {1301.3498},
 primaryClass = {astro-ph.GA},
       adsurl = {https://ui.adsabs.harvard.edu/abs/2013ARA&A..51..207B}
}

@ARTICLE{Hildebrand1983,
       author = {{Hildebrand}, R.~H.},
        title = "{The determination of cloud masses and dust characteristics from submillimetre thermal emission.}",
      journal = {\qjras},
         year = 1983,
        month = sep,
       volume = {24},
        pages = {267-282},
       adsurl = {https://ui.adsabs.harvard.edu/abs/1983QJRAS..24..267H}
}

@ARTICLE{Meier2001,
       author = {{Meier}, David S. and {Turner}, Jean L. and {Crosthwaite}, Lucian P. and {Beck}, Sara C.},
        title = "{Warm Molecular Gas in Dwarf Starburst Galaxies: CO(3-2) Observations}",
      journal = {\aj},
         year = 2001,
        month = feb,
       volume = {121},
       number = {2},
        pages = {740-752},
          doi = {10.1086/318782},
archivePrefix = {arXiv},
       eprint = {astro-ph/0011582},
 primaryClass = {astro-ph},
       adsurl = {https://ui.adsabs.harvard.edu/abs/2001AJ....121..740M}
}

@ARTICLE{Cormier2014,
       author = {{Cormier}, D. and {Madden}, S.~C. and {Lebouteiller}, V. and {Hony}, S. and {Aalto}, S. and {Costagliola}, F. and {Hughes}, A. and {R{\'e}my-Ruyer}, A. and {Abel}, N. and {Bayet}, E. and et al.},
        title = "{The molecular gas reservoir of 6 low-metallicity galaxies from the Herschel Dwarf Galaxy Survey. A ground-based follow-up survey of CO(1-0), CO(2-1), and CO(3-2)}",
      journal = {\aap},
         year = 2014,
        month = apr,
       volume = {564},
          eid = {A121},
        pages = {A121},
          doi = {10.1051/0004-6361/201322096},
archivePrefix = {arXiv},
       eprint = {1401.0563},
 primaryClass = {astro-ph.GA},
       adsurl = {https://ui.adsabs.harvard.edu/abs/2014A&A...564A.121C}
}

@ARTICLE{Sommovigo2022,
       author = {{Sommovigo}, L. and {Ferrara}, A. and {Pallottini}, A. and {Dayal}, P. and {Bouwens}, R.~J. and {Smit}, R. and {da Cunha}, E. and {De Looze}, I. and {Bowler}, R.~A.~A. and {Hodge}, J. and et al.},
        title = "{The ALMA REBELS Survey: cosmic dust temperature evolution out to z   7}",
      journal = {\mnras},
         year = 2022,
        month = jul,
       volume = {513},
       number = {3},
        pages = {3122-3135},
          doi = {10.1093/mnras/stac302},
archivePrefix = {arXiv},
       eprint = {2202.01227},
 primaryClass = {astro-ph.GA},
       adsurl = {https://ui.adsabs.harvard.edu/abs/2022MNRAS.513.3122S}
}

@ARTICLE{Rowland2026,
       author = {{Rowland}, Lucie E. and {Stefanon}, Mauro and {Bouwens}, Rychard and {Hodge}, Jacqueline and {Algera}, Hiddo and {Fisher}, Rebecca and {Dayal}, Pratika and {Pallottini}, Andrea and {Stark}, Daniel P. and {Heintz}, Kasper E. and et al.},
        title = "{REBELS-IFU: evidence for metal-rich massive galaxies at z {\ensuremath{\sim}} 6{\ensuremath{-}}8}",
      journal = {\mnras},
         year = 2026,
        month = feb,
       volume = {546},
       number = {2},
          eid = {staf2023},
        pages = {staf2023},
          doi = {10.1093/mnras/staf2023},
archivePrefix = {arXiv},
       eprint = {2501.10559},
 primaryClass = {astro-ph.GA},
       adsurl = {https://ui.adsabs.harvard.edu/abs/2026MNRAS.546f2023R}
}

@ARTICLE{Aravena2024,
       author = {{Aravena}, M. and {Heintz}, K. and {Dessauges-Zavadsky}, M. and {Oesch}, P. and {Algera}, H. and {Bouwens}, R. and {da Cunha}, E. and {Dayal}, P. and {De Looze}, I. and {Ferrara}, A. and et al.},
        title = "{The ALMA Reionization Era Bright Emission Line Survey: The molecular gas content of galaxies at z 7}",
      journal = {\aap},
         year = 2024,
        month = feb,
       volume = {682},
          eid = {A24},
        pages = {A24},
          doi = {10.1051/0004-6361/202347281},
archivePrefix = {arXiv},
       eprint = {2309.15948},
 primaryClass = {astro-ph.GA},
       adsurl = {https://ui.adsabs.harvard.edu/abs/2024A&A...682A..24A}
}

@ARTICLE{Dessauges2020,
       author = {{Dessauges-Zavadsky}, M. and {Ginolfi}, M. and {Pozzi}, F. and {B{\'e}thermin}, M. and {Le F{\`e}vre}, O. and {Fujimoto}, S. and {Silverman}, J.~D. and {Jones}, G.~C. and {Vallini}, L. and {Schaerer}, D. and et al.},
        title = "{The ALPINE-ALMA [C II] survey. Molecular gas budget in the early Universe as traced by [C II]}",
      journal = {\aap},
         year = 2020,
        month = nov,
       volume = {643},
          eid = {A5},
        pages = {A5},
          doi = {10.1051/0004-6361/202038231},
archivePrefix = {arXiv},
       eprint = {2004.10771},
 primaryClass = {astro-ph.GA},
       adsurl = {https://ui.adsabs.harvard.edu/abs/2020A&A...643A...5D}
}

@ARTICLE{Schinnerer2024,
       author = {{Schinnerer}, E. and {Leroy}, A.~K.},
        title = "{Molecular Gas and the Star-Formation Process on Cloud Scales in Nearby Galaxies}",
      journal = {\araa},
         year = 2024,
        month = sep,
       volume = {62},
       number = {1},
        pages = {369-436},
          doi = {10.1146/annurev-astro-071221-052651},
archivePrefix = {arXiv},
       eprint = {2403.19843},
 primaryClass = {astro-ph.GA},
       adsurl = {https://ui.adsabs.harvard.edu/abs/2024ARA&A..62..369S}
}

@ARTICLE{Glover2016,
       author = {{Glover}, Simon C.~O. and {Clark}, Paul C.},
        title = "{Is atomic carbon a good tracer of molecular gas in metal-poor galaxies?}",
      journal = {\mnras},
         year = 2016,
        month = mar,
       volume = {456},
       number = {4},
        pages = {3596-3609},
          doi = {10.1093/mnras/stv2863},
archivePrefix = {arXiv},
       eprint = {1509.01939},
 primaryClass = {astro-ph.GA},
       adsurl = {https://ui.adsabs.harvard.edu/abs/2016MNRAS.456.3596G}
}

@ARTICLE{Madden2020,
       author = {{Madden}, S.~C. and {Cormier}, D. and {Hony}, S. and {Lebouteiller}, V. and {Abel}, N. and {Galametz}, M. and {De Looze}, I. and {Chevance}, M. and {Polles}, F.~L. and {Lee}, M.-Y. and et al.},
        title = "{Tracing the total molecular gas in galaxies: [CII] and the CO-dark gas}",
      journal = {\aap},
         year = 2020,
        month = nov,
       volume = {643},
          eid = {A141},
        pages = {A141},
          doi = {10.1051/0004-6361/202038860},
archivePrefix = {arXiv},
       eprint = {2009.00649},
 primaryClass = {astro-ph.GA},
       adsurl = {https://ui.adsabs.harvard.edu/abs/2020A&A...643A.141M}
}

@ARTICLE{FriasCastillo2023,
       author = {{Frias Castillo}, Marta and {Hodge}, Jacqueline and {Rybak}, Matus and {van der Werf}, Paul and {Smail}, Ian and {Birkin}, Jack E. and {Chen}, Chian-Chou and {Chapman}, Scott C. and {Hill}, Ryley and {Lagos}, Claudia del P. and et al.},
        title = "{VLA Legacy Survey of Molecular Gas in Massive Star-forming Galaxies at High Redshift}",
      journal = {\apj},
         year = 2023,
        month = mar,
       volume = {945},
       number = {2},
          eid = {128},
        pages = {128},
          doi = {10.3847/1538-4357/acb931},
archivePrefix = {arXiv},
       eprint = {2302.03713},
 primaryClass = {astro-ph.GA},
       adsurl = {https://ui.adsabs.harvard.edu/abs/2023ApJ...945..128F}
}

@ARTICLE{Chen2026,
       author = {{Chen}, Jianhang and {Tacconi}, Linda J. and {Genzel}, Reinhard and {Neri}, Roberto and {Schuster}, Karl and {F{\"o}rster Schreiber}, Natascha M. and {Jolly}, Jean-Baptiste and {Pastras}, Stavros and {Scaloni}, Letizia and {Tozzi}, Giulia and et al.},
        title = "{NOEMA3D: Spatially resolved dust, CO, and [C I] in massive star-forming main sequence galaxies at cosmic noon}",
      journal = {arXiv e-prints},
         year = 2026,
        month = apr,
          eid = {arXiv:2604.18504},
        pages = {arXiv:2604.18504},
          doi = {10.48550/arXiv.2604.18504},
archivePrefix = {arXiv},
       eprint = {2604.18504},
 primaryClass = {astro-ph.GA},
       adsurl = {https://ui.adsabs.harvard.edu/abs/2026arXiv260418504C}
}

@ARTICLE{Cescon2026,
       author = {{Cescon}, Karin and {Hodge}, Jacqueline A. and {Boogaard}, Leindert A. and {Algera}, Hiddo S.~B. and {Rowland}, Lucie E. and {Riechers}, Dominik A. and {Smit}, Renske and {De Looze}, Ilse and {Bouwens}, Rychard and {van der Werf}, Paul and et al.},
        title = "{Direct detection of cool molecular gas in a star-forming galaxy at z=7.31}",
      journal = {\mnras},
         year = 2026,
        month = jul,
       volume = {549},
       number = {3},
          eid = {stag924},
        pages = {stag924},
          doi = {10.1093/mnras/stag924},
archivePrefix = {arXiv},
       eprint = {2606.13393},
 primaryClass = {astro-ph.GA},
       adsurl = {https://ui.adsabs.harvard.edu/abs/2026MNRAS.549ag924C}
}

@ARTICLE{Privon2018,
       author = {{Privon}, G.~C. and {Narayanan}, D. and {Dav{\'e}}, R.},
        title = "{On the Interpretation of Far-infrared Spectral Energy Distributions. I. The 850 {\ensuremath{\mu}}m Molecular Mass Estimator}",
      journal = {\apj},
         year = 2018,
        month = nov,
       volume = {867},
       number = {2},
          eid = {102},
        pages = {102},
          doi = {10.3847/1538-4357/aae485},
archivePrefix = {arXiv},
       eprint = {1805.03649},
 primaryClass = {astro-ph.GA},
       adsurl = {https://ui.adsabs.harvard.edu/abs/2018ApJ...867..102P}
}

@ARTICLE{Liang2018,
       author = {{Liang}, Lichen and {Feldmann}, Robert and {Faucher-Gigu{\`e}re}, Claude-Andr{\'e} and {Kere{\v{s}}}, Du{\v{s}}an and {Hopkins}, Philip F. and {Hayward}, Christopher C. and {Quataert}, Eliot and {Scoville}, Nick Z.},
        title = "{Submillimetre flux as a probe of molecular ISM mass in high-z galaxies}",
      journal = {\mnras},
         year = 2018,
        month = jul,
       volume = {478},
       number = {1},
        pages = {L83-L88},
          doi = {10.1093/mnrasl/sly071},
archivePrefix = {arXiv},
       eprint = {1804.02403},
 primaryClass = {astro-ph.GA},
       adsurl = {https://ui.adsabs.harvard.edu/abs/2018MNRAS.478L..83L}
}

@ARTICLE{Teplitz2000,
       author = {{Teplitz}, Harry I. and {McLean}, Ian S. and {Becklin}, E.~E. and {Figer}, Donald F. and {Gilbert}, Andrea M. and {Graham}, James R. and {Larkin}, James E. and {Levenson}, N.~A. and {Wilcox}, Mavourneen K.},
        title = "{The Rest-Frame Optical Spectrum of MS 1512-CB58}",
      journal = {\apjl},
         year = 2000,
        month = apr,
       volume = {533},
       number = {1},
        pages = {L65-L68},
          doi = {10.1086/312595},
archivePrefix = {arXiv},
       eprint = {astro-ph/0002508},
 primaryClass = {astro-ph},
       adsurl = {https://ui.adsabs.harvard.edu/abs/2000ApJ...533L..65T}
}

@ARTICLE{Richard2011,
       author = {{Richard}, Johan and {Jones}, Tucker and {Ellis}, Richard and {Stark}, Daniel P. and {Livermore}, Rachael and {Swinbank}, Mark},
        title = "{The emission line properties of gravitationally lensed 1.5 < z < 5 galaxies}",
      journal = {\mnras},
         year = 2011,
        month = may,
       volume = {413},
       number = {1},
        pages = {643-658},
          doi = {10.1111/j.1365-2966.2010.18161.x},
archivePrefix = {arXiv},
       eprint = {1011.6413},
 primaryClass = {astro-ph.CO},
       adsurl = {https://ui.adsabs.harvard.edu/abs/2011MNRAS.413..643R}
}

@ARTICLE{Khare2007,
       author = {{Khare}, P. and {Kulkarni}, V.~P. and {P{\'e}roux}, C. and {York}, D.~G. and {Lauroesch}, J.~T. and {Meiring}, J.~D.},
        title = "{The nature of damped Lyman {\ensuremath{\alpha}} and sub-damped Lyman {\ensuremath{\alpha}} absorbers}",
      journal = {\aap},
         year = 2007,
        month = mar,
       volume = {464},
       number = {2},
        pages = {487-493},
          doi = {10.1051/0004-6361:20066186},
archivePrefix = {arXiv},
       eprint = {astro-ph/0608127},
 primaryClass = {astro-ph},
       adsurl = {https://ui.adsabs.harvard.edu/abs/2007A&A...464..487K}
}

@ARTICLE{Aravena2020,
       author = {{Aravena}, Manuel and {Boogaard}, Leindert and {G{\'o}nzalez-L{\'o}pez}, Jorge and {Decarli}, Roberto and {Walter}, Fabian and {Carilli}, Chris L. and {Smail}, Ian and {Weiss}, Axel and {Assef}, Roberto J. and {Bauer}, Franz Erik and et al.},
        title = "{The ALMA Spectroscopic Survey in the Hubble Ultra Deep Field: The Nature of the Faintest Dusty Star-forming Galaxies}",
      journal = {\apj},
         year = 2020,
        month = sep,
       volume = {901},
       number = {1},
          eid = {79},
        pages = {79},
          doi = {10.3847/1538-4357/ab99a2},
archivePrefix = {arXiv},
       eprint = {2006.04284},
 primaryClass = {astro-ph.GA},
       adsurl = {https://ui.adsabs.harvard.edu/abs/2020ApJ...901...79A}
}

@ARTICLE{GonzalezLopez2020,
       author = {{Gonz{\'a}lez-L{\'o}pez}, Jorge and {Novak}, Mladen and {Decarli}, Roberto and {Walter}, Fabian and {Aravena}, Manuel and {Carilli}, Chris and {Boogaard}, Leindert and {Popping}, Gerg{\"o} and {Weiss}, Axel and {Assef}, Roberto J. and et al.},
        title = "{The ALMA Spectroscopic Survey in the HUDF: Deep 1.2 mm Continuum Number Counts}",
      journal = {\apj},
         year = 2020,
        month = jul,
       volume = {897},
       number = {1},
          eid = {91},
        pages = {91},
          doi = {10.3847/1538-4357/ab765b},
archivePrefix = {arXiv},
       eprint = {2002.07199},
 primaryClass = {astro-ph.GA},
       adsurl = {https://ui.adsabs.harvard.edu/abs/2020ApJ...897...91G}
}

@ARTICLE{Walter2016,
       author = {{Walter}, Fabian and {Decarli}, Roberto and {Aravena}, Manuel and {Carilli}, Chris and {Bouwens}, Rychard and {da Cunha}, Elisabete and {Daddi}, Emanuele and {Ivison}, R.~J. and {Riechers}, Dominik and {Smail}, Ian and et al.},
        title = "{ALMA Spectroscopic Survey in the Hubble Ultra Deep Field: Survey Description}",
      journal = {\apj},
         year = 2016,
        month = dec,
       volume = {833},
       number = {1},
          eid = {67},
        pages = {67},
          doi = {10.3847/1538-4357/833/1/67},
archivePrefix = {arXiv},
       eprint = {1607.06768},
 primaryClass = {astro-ph.GA},
       adsurl = {https://ui.adsabs.harvard.edu/abs/2016ApJ...833...67W}
}

@ARTICLE{Jones2017,
       author = {{Jones}, A.~P. and {K{\"o}hler}, M. and {Ysard}, N. and {Bocchio}, M. and {Verstraete}, L.},
        title = "{The global dust modelling framework THEMIS}",
      journal = {\aap},
         year = 2017,
        month = jun,
       volume = {602},
          eid = {A46},
        pages = {A46},
          doi = {10.1051/0004-6361/201630225},
archivePrefix = {arXiv},
       eprint = {1703.00775},
 primaryClass = {astro-ph.GA},
       adsurl = {https://ui.adsabs.harvard.edu/abs/2017A&A...602A..46J}
}

@ARTICLE{Draine2003,
       author = {{Draine}, B.~T.},
        title = "{Interstellar Dust Grains}",
      journal = {\araa},
         year = 2003,
        month = jan,
       volume = {41},
        pages = {241-289},
          doi = {10.1146/annurev.astro.41.011802.094840},
archivePrefix = {arXiv},
       eprint = {astro-ph/0304489},
 primaryClass = {astro-ph},
       adsurl = {https://ui.adsabs.harvard.edu/abs/2003ARA&A..41..241D}
}

@ARTICLE{Scoville2007,
       author = {{Scoville}, N. and {Aussel}, H. and {Brusa}, M. and {Capak}, P. and {Carollo}, C.~M. and {Elvis}, M. and {Giavalisco}, M. and {Guzzo}, L. and {Hasinger}, G. and {Impey}, C. and et al.},
        title = "{The Cosmic Evolution Survey (COSMOS): Overview}",
      journal = {\apjs},
         year = 2007,
        month = sep,
       volume = {172},
       number = {1},
        pages = {1-8},
          doi = {10.1086/516585},
archivePrefix = {arXiv},
       eprint = {astro-ph/0612305},
 primaryClass = {astro-ph},
       adsurl = {https://ui.adsabs.harvard.edu/abs/2007ApJS..172....1S}
}

@ARTICLE{Casey2023,
       author = {{Casey}, Caitlin M. and {Kartaltepe}, Jeyhan S. and {Drakos}, Nicole E. and {Franco}, Maximilien and {Harish}, Santosh and {Paquereau}, Louise and {Ilbert}, Olivier and {Rose}, Caitlin and {Cox}, Isabella G. and {Nightingale}, James W. and et al.},
        title = "{COSMOS-Web: An Overview of the JWST Cosmic Origins Survey}",
      journal = {\apj},
         year = 2023,
        month = sep,
       volume = {954},
       number = {1},
          eid = {31},
        pages = {31},
          doi = {10.3847/1538-4357/acc2bc},
archivePrefix = {arXiv},
       eprint = {2211.07865},
 primaryClass = {astro-ph.GA},
       adsurl = {https://ui.adsabs.harvard.edu/abs/2023ApJ...954...31C}
}

@ARTICLE{Draine2007,
       author = {{Draine}, B.~T. and {Dale}, D.~A. and {Bendo}, G. and {Gordon}, K.~D. and {Smith}, J.~D.~T. and {Armus}, L. and {Engelbracht}, C.~W. and {Helou}, G. and {Kennicutt}, Jr., R.~C. and {Li}, A. and et al.},
        title = "{Dust Masses, PAH Abundances, and Starlight Intensities in the SINGS Galaxy Sample}",
      journal = {\apj},
         year = 2007,
        month = jul,
       volume = {663},
       number = {2},
        pages = {866-894},
          doi = {10.1086/518306},
archivePrefix = {arXiv},
       eprint = {astro-ph/0703213},
 primaryClass = {astro-ph},
       adsurl = {https://ui.adsabs.harvard.edu/abs/2007ApJ...663..866D}
}

@ARTICLE{Jones2025,
       author = {{Jones}, Gareth C. and {Bunker}, Andrew J. and {Telikova}, Kseniia and {Arribas}, Santiago and {Carniani}, Stefano and {Charlot}, Stephane and {D'Eugenio}, Francesco and {Maiolino}, Roberto and {Perna}, Michele and {Rodr{\'\i}guez Del Pino}, Bruno and et al.},
        title = "{GA-NIFS: witnessing the complex assembly of a star-forming system at z = 5.7}",
      journal = {\mnras},
         year = 2025,
        month = jul,
       volume = {540},
       number = {4},
        pages = {3311-3329},
          doi = {10.1093/mnras/staf899},
archivePrefix = {arXiv},
       eprint = {2405.12955},
 primaryClass = {astro-ph.GA},
       adsurl = {https://ui.adsabs.harvard.edu/abs/2025MNRAS.540.3311J}
}

@ARTICLE{Kaasinen2024,
       author = {{Kaasinen}, Melanie and {Venemans}, Bram and {Harrington}, Kevin C. and {Boogaard}, Leindert A. and {Meyer}, Romain A. and {Ba{\~n}ados}, Eduardo and {Decarli}, Roberto and {Walter}, Fabian and {Neeleman}, Marcel and {Calistro Rivera}, Gabriela and et al.},
        title = "{The cold molecular gas in z {\ensuremath{\gtrsim}} 6 quasar host galaxies}",
      journal = {\aap},
         year = 2024,
        month = apr,
       volume = {684},
          eid = {A33},
        pages = {A33},
          doi = {10.1051/0004-6361/202348463},
archivePrefix = {arXiv},
       eprint = {2402.05165},
 primaryClass = {astro-ph.GA},
       adsurl = {https://ui.adsabs.harvard.edu/abs/2024A&A...684A..33K}
}

@ARTICLE{Villanueva2024,
       author = {{Villanueva}, V. and {Herrera-Camus}, R. and {Gonz{\'a}lez-L{\'o}pez}, J. and {Aravena}, M. and {Assef}, R.~J. and {Baeza-Garay}, M. and {Barcos-Mu{\~n}oz}, L. and {Bovino}, S. and {Bowler}, R.~A.~A. and {da Cunha}, E. and et al.},
        title = "{The ALMA-CRISTAL survey: Dust temperature and physical conditions of the interstellar medium in a typical galaxy at z = 5.66}",
      journal = {\aap},
         year = 2024,
        month = nov,
       volume = {691},
          eid = {A133},
        pages = {A133},
          doi = {10.1051/0004-6361/202451490},
archivePrefix = {arXiv},
       eprint = {2407.09681},
 primaryClass = {astro-ph.GA},
       adsurl = {https://ui.adsabs.harvard.edu/abs/2024A&A...691A.133V}
}

@ARTICLE{Leroy2011,
       author = {{Leroy}, Adam K. and {Bolatto}, Alberto and {Gordon}, Karl and {Sandstrom}, Karin and {Gratier}, Pierre and {Rosolowsky}, Erik and {Engelbracht}, Charles W. and {Mizuno}, Norikazu and {Corbelli}, Edvige and {Fukui}, Yasuo and et al.},
        title = "{The CO-to-H$_{2}$ Conversion Factor from Infrared Dust Emission across the Local Group}",
      journal = {\apj},
         year = 2011,
        month = aug,
       volume = {737},
       number = {1},
          eid = {12},
        pages = {12},
          doi = {10.1088/0004-637X/737/1/12},
archivePrefix = {arXiv},
       eprint = {1102.4618},
 primaryClass = {astro-ph.CO},
       adsurl = {https://ui.adsabs.harvard.edu/abs/2011ApJ...737...12L}
}

@ARTICLE{Saintonge2013,
       author = {{Saintonge}, Am{\'e}lie and {Lutz}, Dieter and {Genzel}, Reinhard and {Magnelli}, Benjamin and {Nordon}, Raanan and {Tacconi}, Linda J. and {Baker}, Andrew J. and {Bandara}, Kaushala and {Berta}, Stefano and {F{\"o}rster Schreiber}, Natascha M. and et al.},
        title = "{Validation of the Equilibrium Model for Galaxy Evolution to z \raisebox{-0.5ex}\textasciitilde 3 through Molecular Gas and Dust Observations of Lensed Star-forming Galaxies}",
      journal = {\apj},
         year = 2013,
        month = nov,
       volume = {778},
       number = {1},
          eid = {2},
        pages = {2},
          doi = {10.1088/0004-637X/778/1/2},
archivePrefix = {arXiv},
       eprint = {1309.3281},
 primaryClass = {astro-ph.CO},
       adsurl = {https://ui.adsabs.harvard.edu/abs/2013ApJ...778....2S}
}

@ARTICLE{Kaasinen2019,
       author = {{Kaasinen}, M. and {Scoville}, N. and {Walter}, F. and {Da Cunha}, E. and {Popping}, G. and {Pavesi}, R. and {Darvish}, B. and {Casey}, C.~M. and {Riechers}, D.~A. and {Glover}, S.},
        title = "{The Molecular Gas Reservoirs of z {\ensuremath{\sim}} 2 Galaxies: A Comparison of CO(1-0) and Dust-based Molecular Gas Masses}",
      journal = {\apj},
         year = 2019,
        month = jul,
       volume = {880},
       number = {1},
          eid = {15},
        pages = {15},
          doi = {10.3847/1538-4357/ab253b},
archivePrefix = {arXiv},
       eprint = {1905.11417},
 primaryClass = {astro-ph.GA},
       adsurl = {https://ui.adsabs.harvard.edu/abs/2019ApJ...880...15K}
}

@ARTICLE{Scoville2014,
       author = {{Scoville}, N. and {Aussel}, H. and {Sheth}, K. and {Scott}, K.~S. and {Sanders}, D. and {Ivison}, R. and {Pope}, A. and {Capak}, P. and {Vanden Bout}, P. and {Manohar}, S. and et al.},
        title = "{The Evolution of Interstellar Medium Mass Probed by Dust Emission: ALMA Observations at z = 0.3-2}",
      journal = {\apj},
         year = 2014,
        month = mar,
       volume = {783},
       number = {2},
          eid = {84},
        pages = {84},
          doi = {10.1088/0004-637X/783/2/84},
archivePrefix = {arXiv},
       eprint = {1401.2987},
 primaryClass = {astro-ph.GA},
       adsurl = {https://ui.adsabs.harvard.edu/abs/2014ApJ...783...84S}
}

@ARTICLE{Scoville2017,
       author = {{Scoville}, N. and {Lee}, N. and {Vanden Bout}, P. and {Diaz-Santos}, T. and {Sanders}, D. and {Darvish}, B. and {Bongiorno}, A. and {Casey}, C.~M. and {Murchikova}, L. and {Koda}, J. and et al.},
        title = "{Evolution of Interstellar Medium, Star Formation, and Accretion at High Redshift}",
      journal = {\apj},
         year = 2017,
        month = mar,
       volume = {837},
       number = {2},
          eid = {150},
        pages = {150},
          doi = {10.3847/1538-4357/aa61a0},
archivePrefix = {arXiv},
       eprint = {1702.04729},
 primaryClass = {astro-ph.GA},
       adsurl = {https://ui.adsabs.harvard.edu/abs/2017ApJ...837..150S}
}

@ARTICLE{Scoville2016,
       author = {{Scoville}, N. and {Sheth}, K. and {Aussel}, H. and {Vanden Bout}, P. and {Capak}, P. and {Bongiorno}, A. and {Casey}, C.~M. and {Murchikova}, L. and {Koda}, J. and {{\'A}lvarez-M{\'a}rquez}, J. and et al.},
        title = "{ISM Masses and the Star formation Law at Z = 1 to 6: ALMA Observations of Dust Continuum in 145 Galaxies in the COSMOS Survey Field}",
      journal = {\apj},
         year = 2016,
        month = apr,
       volume = {820},
       number = {2},
          eid = {83},
        pages = {83},
          doi = {10.3847/0004-637X/820/2/83},
archivePrefix = {arXiv},
       eprint = {1511.05149},
 primaryClass = {astro-ph.GA},
       adsurl = {https://ui.adsabs.harvard.edu/abs/2016ApJ...820...83S}
}

@ARTICLE{Solomon2005,
       author = {{Solomon}, P.~M. and {Vanden Bout}, P.~A.},
        title = "{Molecular Gas at High Redshift}",
      journal = {\araa},
         year = 2005,
        month = sep,
       volume = {43},
       number = {1},
        pages = {677-725},
          doi = {10.1146/annurev.astro.43.051804.102221},
archivePrefix = {arXiv},
       eprint = {astro-ph/0508481},
 primaryClass = {astro-ph},
       adsurl = {https://ui.adsabs.harvard.edu/abs/2005ARA&A..43..677S}
}

@ARTICLE{Asplund2009,
       author = {{Asplund}, Martin and {Grevesse}, Nicolas and {Sauval}, A. Jacques and {Scott}, Pat},
        title = "{The Chemical Composition of the Sun}",
      journal = {\araa},
         year = 2009,
        month = sep,
       volume = {47},
       number = {1},
        pages = {481-522},
          doi = {10.1146/annurev.astro.46.060407.145222},
archivePrefix = {arXiv},
       eprint = {0909.0948},
 primaryClass = {astro-ph.SR},
       adsurl = {https://ui.adsabs.harvard.edu/abs/2009ARA&A..47..481A}
}

@ARTICLE{Chabrier2003,
       author = {{Chabrier}, Gilles},
        title = "{Galactic Stellar and Substellar Initial Mass Function}",
      journal = {\pasp},
         year = 2003,
        month = jul,
       volume = {115},
       number = {809},
        pages = {763-795},
          doi = {10.1086/376392},
archivePrefix = {arXiv},
       eprint = {astro-ph/0304382},
 primaryClass = {astro-ph},
       adsurl = {https://ui.adsabs.harvard.edu/abs/2003PASP..115..763C}
}

@ARTICLE{Popping2015model,
       author = {{Popping}, Gerg{\"o} and {Behroozi}, Peter S. and {Peeples}, Molly S.},
        title = "{Evolution of the atomic and molecular gas content of galaxies in dark matter haloes}",
      journal = {\mnras},
         year = 2015,
        month = may,
       volume = {449},
       number = {1},
        pages = {477-493},
          doi = {10.1093/mnras/stv318},
archivePrefix = {arXiv},
       eprint = {1409.1574},
 primaryClass = {astro-ph.GA},
       adsurl = {https://ui.adsabs.harvard.edu/abs/2015MNRAS.449..477P}
}

@ARTICLE{Popping2014,
       author = {{Popping}, Gerg{\"o} and {Somerville}, Rachel S. and {Trager}, Scott C.},
        title = "{Evolution of the atomic and molecular gas content of galaxies}",
      journal = {\mnras},
         year = 2014,
        month = aug,
       volume = {442},
       number = {3},
        pages = {2398-2418},
          doi = {10.1093/mnras/stu991},
archivePrefix = {arXiv},
       eprint = {1308.6764},
 primaryClass = {astro-ph.CO},
       adsurl = {https://ui.adsabs.harvard.edu/abs/2014MNRAS.442.2398P}
}

@ARTICLE{Morselli2021,
       author = {{Morselli}, Laura and {Renzini}, A. and {Enia}, A. and {Rodighiero}, G.},
        title = "{Redshift evolution of the H$_{2}$/H I mass ratio in galaxies}",
      journal = {\mnras},
         year = 2021,
        month = mar,
       volume = {502},
       number = {1},
        pages = {L85-L89},
          doi = {10.1093/mnrasl/slab007},
archivePrefix = {arXiv},
       eprint = {2101.10372},
 primaryClass = {astro-ph.GA},
       adsurl = {https://ui.adsabs.harvard.edu/abs/2021MNRAS.502L..85M}
}

@ARTICLE{Lagos2011,
       author = {{Lagos}, Claudia Del P. and {Baugh}, Carlton M. and {Lacey}, Cedric G. and {Benson}, Andrew J. and {Kim}, Han-Seek and {Power}, Chris},
        title = "{Cosmic evolution of the atomic and molecular gas contents of galaxies}",
      journal = {\mnras},
         year = 2011,
        month = dec,
       volume = {418},
       number = {3},
        pages = {1649-1667},
          doi = {10.1111/j.1365-2966.2011.19583.x},
archivePrefix = {arXiv},
       eprint = {1105.2294},
 primaryClass = {astro-ph.CO},
       adsurl = {https://ui.adsabs.harvard.edu/abs/2011MNRAS.418.1649L}
}

@INPROCEEDINGS{StaveleySmith2015,
       author = {{Staveley-Smith}, L. and {Oosterloo}, T.},
        title = "{HI Science with the Square Kilometre Array}",
    booktitle = {Advancing Astrophysics with the Square Kilometre Array (AASKA14)},
         year = 2015,
        month = apr,
          eid = {167},
        pages = {167},
          doi = {10.22323/1.215.0167},
archivePrefix = {arXiv},
       eprint = {1506.04473},
 primaryClass = {astro-ph.GA},
       adsurl = {https://ui.adsabs.harvard.edu/abs/2015aska.confE.167S}
}

@ARTICLE{Galliano2011,
       author = {{Galliano}, F. and {Hony}, S. and {Bernard}, J.-P. and {Bot}, C. and {Madden}, S.~C. and {Roman-Duval}, J. and {Galametz}, M. and {Li}, A. and {Meixner}, M. and {Engelbracht}, C.~W. and et al.},
        title = "{Non-standard grain properties, dark gas reservoir, and extended submillimeter excess, probed by Herschel in the Large Magellanic Cloud}",
      journal = {\aap},
         year = 2011,
        month = dec,
       volume = {536},
          eid = {A88},
        pages = {A88},
          doi = {10.1051/0004-6361/201117952},
archivePrefix = {arXiv},
       eprint = {1110.1260},
 primaryClass = {astro-ph.CO},
       adsurl = {https://ui.adsabs.harvard.edu/abs/2011A&A...536A..88G}
}

@ARTICLE{Ragone2024,
       author = {{Ragone-Figueroa}, Cinthia and {Granato}, Gian Luigi and {Parente}, Massimiliano and {Murante}, Giuseppe and {Valentini}, Milena and {Borgani}, Stefano and {Maio}, Umberto},
        title = "{Intertwined formation of H$_{2}$, dust, and stars in cosmological simulations}",
      journal = {\aap},
         year = 2024,
        month = nov,
       volume = {691},
          eid = {A200},
        pages = {A200},
          doi = {10.1051/0004-6361/202451344},
archivePrefix = {arXiv},
       eprint = {2407.06269},
 primaryClass = {astro-ph.GA},
       adsurl = {https://ui.adsabs.harvard.edu/abs/2024A&A...691A.200R}
}

@ARTICLE{Faist2026,
       author = {{Faisst}, A.~L. and {Fujimoto}, S. and {Tsujita}, A. and {Wang}, W. and {Nezhad}, N. and {Loiacono}, F. and {{\"U}bler}, H. and {B{\'e}thermin}, M. and {Cassata}, P. and {Dessauges-Zavadsky}, M. and et al.},
        title = "{The ALPINE-CRISTAL-JWST Survey: JWST/IFU Optical Observations for 18 Main-sequence Galaxies at z = 4─6}",
      journal = {\apjs},
         year = 2026,
        month = jan,
       volume = {282},
       number = {1},
          eid = {19},
        pages = {19},
          doi = {10.3847/1538-4365/ae0928},
archivePrefix = {arXiv},
       eprint = {2510.16111},
 primaryClass = {astro-ph.GA},
       adsurl = {https://ui.adsabs.harvard.edu/abs/2026ApJS..282...19F}
}

@ARTICLE{Prajapati2026,
       author = {{Prajapati}, Prachi and {Riechers}, Dominik and {Cox}, Pierre and {Weiss}, Axel and {Saintonge}, Am{\'e}lie and {Jones}, Bethany and {Bakx}, Tom J.~L.~C. and {Berta}, Stefano and {van der Werf}, Paul and {Neri}, Roberto and et al.},
        title = "{Vz-GAL: Probing Cold Molecular Gas in Dusty Star-forming Galaxies at z = 1─6}",
      journal = {\apjs},
         year = 2026,
        month = feb,
       volume = {282},
       number = {2},
          eid = {40},
        pages = {40},
          doi = {10.3847/1538-4365/ae27d4},
archivePrefix = {arXiv},
       eprint = {2509.25167},
 primaryClass = {astro-ph.GA},
       adsurl = {https://ui.adsabs.harvard.edu/abs/2026ApJS..282...40P}
}

@ARTICLE{Patil2010,
       author = {{Patil}, Anand and {Huard}, David and {Fonnesbeck}, Christopher J.},
        title = "{PyMC: Bayesian Stochastic Modelling in Python}",
      journal = {Journal of Statistical Software},
         year = 2010,
        month = jul,
       volume = {35},
        pages = {1},
          doi = {10.18637/jss.v035.i04},
       adsurl = {https://ui.adsabs.harvard.edu/abs/2010JSS....35....1P}
}

@ARTICLE{Kiyota2026,
       author = {{Kiyota}, Tomokazu and {Ouchi}, Masami and {Iono}, Daisuke and {Fujimoto}, Seiji and {Kohno}, Kotaro and {Ueda}, Yoshihiro and {Nakajima}, Kimihiko and {Nishigaki}, Moka and {Yajima}, Hidenobu},
        title = "{JWST Spectroscopic Census of ALMA Faint Submillimeter Galaxies in the Hubble Ultra Deep Field}",
      journal = {arXiv e-prints},
         year = 2026,
        month = jan,
          eid = {arXiv:2601.18149},
        pages = {arXiv:2601.18149},
          doi = {10.48550/arXiv.2601.18149},
archivePrefix = {arXiv},
       eprint = {2601.18149},
 primaryClass = {astro-ph.GA},
       adsurl = {https://ui.adsabs.harvard.edu/abs/2026arXiv260118149K}
}

@ARTICLE{Pilyugin2016,
       author = {{Pilyugin}, L.~S. and {Grebel}, E.~K.},
        title = "{New calibrations for abundance determinations in H II regions}",
      journal = {\mnras},
         year = 2016,
        month = apr,
       volume = {457},
       number = {4},
        pages = {3678-3692},
          doi = {10.1093/mnras/stw238},
archivePrefix = {arXiv},
       eprint = {1601.08217},
 primaryClass = {astro-ph.GA},
       adsurl = {https://ui.adsabs.harvard.edu/abs/2016MNRAS.457.3678P}
}

@ARTICLE{Galametz2011,
       author = {{Galametz}, M. and {Madden}, S.~C. and {Galliano}, F. and {Hony}, S. and {Bendo}, G.~J. and {Sauvage}, M.},
        title = "{Probing the dust properties of galaxies up to submillimetre wavelengths. II. Dust-to-gas mass ratio trends with metallicity and the submm excess in dwarf galaxies}",
      journal = {\aap},
         year = 2011,
        month = aug,
       volume = {532},
          eid = {A56},
        pages = {A56},
          doi = {10.1051/0004-6361/201014904},
archivePrefix = {arXiv},
       eprint = {1104.0827},
 primaryClass = {astro-ph.CO},
       adsurl = {https://ui.adsabs.harvard.edu/abs/2011A&A...532A..56G}
}

@ARTICLE{Kennicutt2011,
       author = {{Kennicutt}, R.~C. and {Calzetti}, D. and {Aniano}, G. and {Appleton}, P. and {Armus}, L. and {Beir{\~a}o}, P. and {Bolatto}, A.~D. and {Brandl}, B. and {Crocker}, A. and {Croxall}, K. and et al.},
        title = "{KINGFISH{\textemdash}Key Insights on Nearby Galaxies: A Far-Infrared Survey with Herschel: Survey Description and Image Atlas}",
      journal = {\pasp},
         year = 2011,
        month = dec,
       volume = {123},
       number = {910},
        pages = {1347},
          doi = {10.1086/663818},
archivePrefix = {arXiv},
       eprint = {1111.4438},
 primaryClass = {astro-ph.CO},
       adsurl = {https://ui.adsabs.harvard.edu/abs/2011PASP..123.1347K}
}

@ARTICLE{Madden2013,
       author = {{Madden}, S.~C. and {R{\'e}my-Ruyer}, A. and {Galametz}, M. and {Cormier}, D. and {Lebouteiller}, V. and {Galliano}, F. and {Hony}, S. and {Bendo}, G.~J. and {Smith}, M.~W.~L. and {Pohlen}, M. and et al.},
        title = "{An Overview of the Dwarf Galaxy Survey}",
      journal = {\pasp},
         year = 2013,
        month = jun,
       volume = {125},
       number = {928},
        pages = {600},
          doi = {10.1086/671138},
archivePrefix = {arXiv},
       eprint = {1305.2628},
 primaryClass = {astro-ph.GA},
       adsurl = {https://ui.adsabs.harvard.edu/abs/2013PASP..125..600M}
}

@ARTICLE{Schruba2012,
       author = {{Schruba}, Andreas and {Leroy}, Adam K. and {Walter}, Fabian and {Bigiel}, Frank and {Brinks}, Elias and {de Blok}, W.~J.~G. and {Kramer}, Carsten and {Rosolowsky}, Erik and {Sandstrom}, Karin and {Schuster}, Karl and et al.},
        title = "{Low CO Luminosities in Dwarf Galaxies}",
      journal = {\aj},
         year = 2012,
        month = jun,
       volume = {143},
       number = {6},
          eid = {138},
        pages = {138},
          doi = {10.1088/0004-6256/143/6/138},
archivePrefix = {arXiv},
       eprint = {1203.4231},
 primaryClass = {astro-ph.CO},
       adsurl = {https://ui.adsabs.harvard.edu/abs/2012AJ....143..138S}
}

@ARTICLE{Planck2020,
       author = {{Planck Collaboration} and {Aghanim}, N. and {Akrami}, Y. and {Ashdown}, M. and {Aumont}, J. and {Baccigalupi}, C. and {Ballardini}, M. and {Banday}, A.~J. and {Barreiro}, R.~B. and {Bartolo}, N. and {Basak}, S. and {Battye}, R. and {Benabed}, K. and {Bernard}, J.-P. and {Bersanelli}, M. and {Bielewicz}, P. and {Bock}, J.~J. and {Bond}, J.~R. and {Borrill}, J. and {Bouchet}, F.~R. and {Boulanger}, F. and {Bucher}, M. and {Burigana}, C. and {Butler}, R.~C. and {Calabrese}, E. and {Cardoso}, J.-F. and {Carron}, J. and {Challinor}, A. and {Chiang}, H.~C. and {Chluba}, J. and {Colombo}, L.~P.~L. and {Combet}, C. and {Contreras}, D. and {Crill}, B.~P. and {Cuttaia}, F. and {de Bernardis}, P. and {de Zotti}, G. and {Delabrouille}, J. and {Delouis}, J.-M. and {Di Valentino}, E. and {Diego}, J.~M. and {Dor{\'e}}, O. and {Douspis}, M. and {Ducout}, A. and {Dupac}, X. and {Dusini}, S. and {Efstathiou}, G. and {Elsner}, F. and {En{\ss}lin}, T.~A. and {Eriksen}, H.~K. and {Fantaye}, Y. and {Farhang}, M. and {Fergusson}, J. and {Fernandez-Cobos}, R. and {Finelli}, F. and {Forastieri}, F. and {Frailis}, M. and {Fraisse}, A.~A. and {Franceschi}, E. and {Frolov}, A. and {Galeotta}, S. and {Galli}, S. and {Ganga}, K. and {G{\'e}nova-Santos}, R.~T. and {Gerbino}, M. and {Ghosh}, T. and {Gonz{\'a}lez-Nuevo}, J. and {G{\'o}rski}, K.~M. and {Gratton}, S. and {Gruppuso}, A. and {Gudmundsson}, J.~E. and {Hamann}, J. and {Handley}, W. and {Hansen}, F.~K. and {Herranz}, D. and {Hildebrandt}, S.~R. and {Hivon}, E. and {Huang}, Z. and {Jaffe}, A.~H. and {Jones}, W.~C. and {Karakci}, A. and {Keih{\"a}nen}, E. and {Keskitalo}, R. and {Kiiveri}, K. and {Kim}, J. and {Kisner}, T.~S. and {Knox}, L. and {Krachmalnicoff}, N. and {Kunz}, M. and {Kurki-Suonio}, H. and {Lagache}, G. and {Lamarre}, J.-M. and {Lasenby}, A. and {Lattanzi}, M. and {Lawrence}, C.~R. and {Le Jeune}, M. and {Lemos}, P. and {Lesgourgues}, J. and {Levrier}, F. and {Lewis}, A. and {Liguori}, M. and {Lilje}, P.~B. and {Lilley}, M. and {Lindholm}, V. and {L{\'o}pez-Caniego}, M. and {Lubin}, P.~M. and {Ma}, Y.-Z. and {Mac{\'\i}as-P{\'e}rez}, J.~F. and {Maggio}, G. and {Maino}, D. and {Mandolesi}, N. and {Mangilli}, A. and {Marcos-Caballero}, A. and {Maris}, M. and {Martin}, P.~G. and {Martinelli}, M. and {Mart{\'\i}nez-Gonz{\'a}lez}, E. and {Matarrese}, S. and {Mauri}, N. and {McEwen}, J.~D. and {Meinhold}, P.~R. and {Melchiorri}, A. and {Mennella}, A. and {Migliaccio}, M. and {Millea}, M. and {Mitra}, S. and {Miville-Desch{\^e}nes}, M.-A. and {Molinari}, D. and {Montier}, L. and {Morgante}, G. and {Moss}, A. and {Natoli}, P. and {N{\o}rgaard-Nielsen}, H.~U. and {Pagano}, L. and {Paoletti}, D. and {Partridge}, B. and {Patanchon}, G. and {Peiris}, H.~V. and {Perrotta}, F. and {Pettorino}, V. and {Piacentini}, F. and {Polastri}, L. and {Polenta}, G. and {Puget}, J.-L. and {Rachen}, J.~P. and {Reinecke}, M. and {Remazeilles}, M. and {Renzi}, A. and {Rocha}, G. and {Rosset}, C. and {Roudier}, G. and {Rubi{\~n}o-Mart{\'\i}n}, J.~A. and {Ruiz-Granados}, B. and {Salvati}, L. and {Sandri}, M. and {Savelainen}, M. and {Scott}, D. and {Shellard}, E.~P.~S. and {Sirignano}, C. and {Sirri}, G. and {Spencer}, L.~D. and {Sunyaev}, R. and {Suur-Uski}, A.-S. and {Tauber}, J.~A. and {Tavagnacco}, D. and {Tenti}, M. and {Toffolatti}, L. and {Tomasi}, M. and {Trombetti}, T. and {Valenziano}, L. and {Valiviita}, J. and {Van Tent}, B. and {Vibert}, L. and {Vielva}, P. and {Villa}, F. and {Vittorio}, N. and {Wandelt}, B.~D. and {Wehus}, I.~K. and {White}, M. and {White}, S.~D.~M. and {Zacchei}, A. and {Zonca}, A.},
        title = "{Planck 2018 results. VI. Cosmological parameters}",
      journal = {\aap},
         year = 2020,
        month = sep,
       volume = {641},
          eid = {A6},
        pages = {A6},
          doi = {10.1051/0004-6361/201833910},
archivePrefix = {arXiv},
       eprint = {1807.06209},
 primaryClass = {astro-ph.CO},
       adsurl = {https://ui.adsabs.harvard.edu/abs/2020A&A...641A...6P}
}

@ARTICLE{Sanders2026,
       author = {{Sanders}, Ryan L. and {Shapley}, Alice E. and {Topping}, Michael W. and {Reddy}, Naveen A. and {Berg}, Danielle A. and {Khostovan}, Ali Ahmad and {Bouwens}, Rychard J. and {Brammer}, Gabriel and {Carnall}, Adam C. and {Cullen}, Fergus and et al.},
        title = "{The AURORA Survey: High-redshift Empirical Metallicity Calibrations from Electron Temperature Measurements at z = 2─10}",
      journal = {\apj},
         year = 2026,
        month = jun,
       volume = {1003},
       number = {2},
          eid = {228},
        pages = {228},
          doi = {10.3847/1538-4357/ae66e2},
archivePrefix = {arXiv},
       eprint = {2508.10099},
 primaryClass = {astro-ph.GA},
       adsurl = {https://ui.adsabs.harvard.edu/abs/2026ApJ..1003..228S}
}

@ARTICLE{Parente2022,
       author = {{Parente}, Massimiliano and {Ragone-Figueroa}, Cinthia and {Granato}, Gian Luigi and {Borgani}, Stefano and {Murante}, Giuseppe and {Valentini}, Milena and {Bressan}, Alessandro and {Lapi}, Andrea},
        title = "{Dust evolution with MUPPI in cosmological volumes}",
      journal = {\mnras},
         year = 2022,
        month = sep,
       volume = {515},
       number = {2},
        pages = {2053-2071},
          doi = {10.1093/mnras/stac1913},
archivePrefix = {arXiv},
       eprint = {2204.11884},
 primaryClass = {astro-ph.GA},
       adsurl = {https://ui.adsabs.harvard.edu/abs/2022MNRAS.515.2053P}
}

@ARTICLE{Triani2020,
       author = {{Triani}, Dian P. and {Sinha}, Manodeep and {Croton}, Darren J. and {Pacifici}, Camilla and {Dwek}, Eli},
        title = "{The origin of dust in galaxies across cosmic time}",
      journal = {\mnras},
         year = 2020,
        month = apr,
       volume = {493},
       number = {2},
        pages = {2490-2505},
          doi = {10.1093/mnras/staa446},
archivePrefix = {arXiv},
       eprint = {2002.05343},
 primaryClass = {astro-ph.GA},
       adsurl = {https://ui.adsabs.harvard.edu/abs/2020MNRAS.493.2490T}
}

@ARTICLE{Vijayan2019,
       author = {{Vijayan}, Aswin P. and {Clay}, Scott J. and {Thomas}, Peter A. and {Yates}, Robert M. and {Wilkins}, Stephen M. and {Henriques}, Bruno M.},
        title = "{Detailed dust modelling in the L-GALAXIES semi-analytic model of galaxy formation}",
      journal = {\mnras},
         year = 2019,
        month = nov,
       volume = {489},
       number = {3},
        pages = {4072-4089},
          doi = {10.1093/mnras/stz1948},
archivePrefix = {arXiv},
       eprint = {1904.02196},
 primaryClass = {astro-ph.GA},
       adsurl = {https://ui.adsabs.harvard.edu/abs/2019MNRAS.489.4072V}
}

@ARTICLE{Li2019,
       author = {{Li}, Qi and {Narayanan}, Desika and {Dav{\'e}}, Romeel},
        title = "{The dust-to-gas and dust-to-metal ratio in galaxies from z = 0 to 6}",
      journal = {\mnras},
         year = 2019,
        month = nov,
       volume = {490},
       number = {1},
        pages = {1425-1436},
          doi = {10.1093/mnras/stz2684},
archivePrefix = {arXiv},
       eprint = {1906.09277},
 primaryClass = {astro-ph.GA},
       adsurl = {https://ui.adsabs.harvard.edu/abs/2019MNRAS.490.1425L}
}

@ARTICLE{Hou2019,
       author = {{Hou}, Kuan-Chou and {Aoyama}, Shohei and {Hirashita}, Hiroyuki and {Nagamine}, Kentaro and {Shimizu}, Ikkoh},
        title = "{Dust scaling relations in a cosmological simulation}",
      journal = {\mnras},
         year = 2019,
        month = may,
       volume = {485},
       number = {2},
        pages = {1727-1744},
          doi = {10.1093/mnras/stz121},
archivePrefix = {arXiv},
       eprint = {1901.02886},
 primaryClass = {astro-ph.GA},
       adsurl = {https://ui.adsabs.harvard.edu/abs/2019MNRAS.485.1727H}
}

@ARTICLE{McKinnon2018,
       author = {{McKinnon}, Ryan and {Vogelsberger}, Mark and {Torrey}, Paul and {Marinacci}, Federico and {Kannan}, Rahul},
        title = "{Simulating galactic dust grain evolution on a moving mesh}",
      journal = {\mnras},
         year = 2018,
        month = aug,
       volume = {478},
       number = {3},
        pages = {2851-2886},
          doi = {10.1093/mnras/sty1248},
archivePrefix = {arXiv},
       eprint = {1805.04521},
 primaryClass = {astro-ph.GA},
       adsurl = {https://ui.adsabs.harvard.edu/abs/2018MNRAS.478.2851M}
}

@ARTICLE{Algera2025,
       author = {{Algera}, H.~S.~B. and {Herrera-Camus}, R. and {Aravena}, M. and {Assef}, R. and {Bakx}, T.~L.~J.~C. and {Bolatto}, A. and {Cescon}, K. and {Chen}, C.-C. and {da Cunha}, E. and {Dayal}, P. and et al.},
        title = "{How much gas and dust is in the $z=5.7$ Lyman Break Galaxy HZ10? An ALMA Band 10 to 4 and JWST/NIRSpec study of its interstellar medium}",
      journal = {arXiv e-prints},
         year = 2025,
        month = dec,
          eid = {arXiv:2512.02320},
        pages = {arXiv:2512.02320},
          doi = {10.48550/arXiv.2512.02320},
archivePrefix = {arXiv},
       eprint = {2512.02320},
 primaryClass = {astro-ph.GA},
       adsurl = {https://ui.adsabs.harvard.edu/abs/2025arXiv251202320A}
}

@ARTICLE{Vallini2025,
       author = {{Vallini}, L. and {Pallottini}, A. and {Kohandel}, M. and {Sommovigo}, L. and {Ferrara}, A. and {Bethermin}, M. and {Herrera-Camus}, R. and {Carniani}, S. and {Faisst}, A. and {Zanella}, A. and et al.},
        title = "{Spatially resolved [CII]{\textendash}gas conversion factor in early galaxies}",
      journal = {\aap},
         year = 2025,
        month = aug,
       volume = {700},
          eid = {A117},
        pages = {A117},
          doi = {10.1051/0004-6361/202555179},
archivePrefix = {arXiv},
       eprint = {2504.14001},
 primaryClass = {astro-ph.GA},
       adsurl = {https://ui.adsabs.harvard.edu/abs/2025A&A...700A.117V}
}

@ARTICLE{Zanella2018,
       author = {{Zanella}, A. and {Daddi}, E. and {Magdis}, G. and {Diaz Santos}, T. and {Cormier}, D. and {Liu}, D. and {Cibinel}, A. and {Gobat}, R. and {Dickinson}, M. and {Sargent}, M. and et al.},
        title = "{The [C II] emission as a molecular gas mass tracer in galaxies at low and high redshifts}",
      journal = {\mnras},
         year = 2018,
        month = dec,
       volume = {481},
       number = {2},
        pages = {1976-1999},
          doi = {10.1093/mnras/sty2394},
archivePrefix = {arXiv},
       eprint = {1808.10331},
 primaryClass = {astro-ph.GA},
       adsurl = {https://ui.adsabs.harvard.edu/abs/2018MNRAS.481.1976Z}
}

@ARTICLE{Feldmann2015,
       author = {{Feldmann}, Robert},
        title = "{The equilibrium view on dust and metals in galaxies: Galactic outflows drive low dust-to-metal ratios in dwarf galaxies}",
      journal = {\mnras},
         year = 2015,
        month = may,
       volume = {449},
       number = {3},
        pages = {3274-3292},
          doi = {10.1093/mnras/stv552},
archivePrefix = {arXiv},
       eprint = {1412.2755},
 primaryClass = {astro-ph.GA},
       adsurl = {https://ui.adsabs.harvard.edu/abs/2015MNRAS.449.3274F}
}

@ARTICLE{Vilchez2019,
       author = {{V{\'\i}lchez}, J.~M. and {Rela{\~n}o}, M. and {Kennicutt}, R. and {De Looze}, I. and {Moll{\'a}}, M. and {Galametz}, M.},
        title = "{Metals and dust content across the galaxies M 101 and NGC 628}",
      journal = {\mnras},
         year = 2019,
        month = mar,
       volume = {483},
       number = {4},
        pages = {4968-4983},
          doi = {10.1093/mnras/sty3455},
archivePrefix = {arXiv},
       eprint = {1811.10262},
 primaryClass = {astro-ph.GA},
       adsurl = {https://ui.adsabs.harvard.edu/abs/2019MNRAS.483.4968V}
}

@ARTICLE{Tacconi2020,
       author = {{Tacconi}, Linda J. and {Genzel}, Reinhard and {Sternberg}, Amiel},
        title = "{The Evolution of the Star-Forming Interstellar Medium Across Cosmic Time}",
      journal = {\araa},
         year = 2020,
        month = aug,
       volume = {58},
        pages = {157-203},
          doi = {10.1146/annurev-astro-082812-141034},
archivePrefix = {arXiv},
       eprint = {2003.06245},
 primaryClass = {astro-ph.GA},
       adsurl = {https://ui.adsabs.harvard.edu/abs/2020ARA&A..58..157T}
}

@ARTICLE{Galliano2018,
       author = {{Galliano}, Fr{\'e}d{\'e}ric and {Galametz}, Maud and {Jones}, Anthony P.},
        title = "{The Interstellar Dust Properties of Nearby Galaxies}",
      journal = {\araa},
         year = 2018,
        month = sep,
       volume = {56},
        pages = {673-713},
          doi = {10.1146/annurev-astro-081817-051900},
archivePrefix = {arXiv},
       eprint = {1711.07434},
 primaryClass = {astro-ph.GA},
       adsurl = {https://ui.adsabs.harvard.edu/abs/2018ARA&A..56..673G}
}

@ARTICLE{Parente23,
       author = {{Parente}, Massimiliano and {Ragone-Figueroa}, Cinthia and {Granato}, Gian Luigi and {Lapi}, Andrea},
        title = "{The z {\ensuremath{\lesssim}} 1 drop of cosmic dust abundance in a semi-analytic framework}",
      journal = {\mnras},
         year = 2023,
        month = jun,
       volume = {521},
       number = {4},
        pages = {6105-6123},
          doi = {10.1093/mnras/stad907},
archivePrefix = {arXiv},
       eprint = {2302.03058},
 primaryClass = {astro-ph.GA},
       adsurl = {https://ui.adsabs.harvard.edu/abs/2023MNRAS.521.6105P}
}

@ARTICLE{Popping2017,
       author = {{Popping}, Gerg{\"o} and {Somerville}, Rachel S. and {Galametz}, Maud},
        title = "{The dust content of galaxies from z = 0 to z = 9}",
      journal = {\mnras},
         year = 2017,
        month = nov,
       volume = {471},
       number = {3},
        pages = {3152-3185},
          doi = {10.1093/mnras/stx1545},
archivePrefix = {arXiv},
       eprint = {1609.08622},
 primaryClass = {astro-ph.GA},
       adsurl = {https://ui.adsabs.harvard.edu/abs/2017MNRAS.471.3152P}
}

@ARTICLE{Casavecchia2025,
       author = {{Casavecchia}, Benedetta and {Maio}, Umberto and {P{\'e}roux}, C{\'e}line and {Ciardi}, Benedetta},
        title = "{Atomic and molecular gas as traced by [C II] emission}",
      journal = {\aap},
         year = 2025,
        month = jan,
       volume = {693},
          eid = {A119},
        pages = {A119},
          doi = {10.1051/0004-6361/202452282},
archivePrefix = {arXiv},
       eprint = {2410.14284},
 primaryClass = {astro-ph.GA},
       adsurl = {https://ui.adsabs.harvard.edu/abs/2025A&A...693A.119C}
}

@ARTICLE{Heintz2025,
       author = {{Heintz}, Kasper E. and {Watson}, Darach and {Valentino}, Francesco and {Gottumukkala}, Rashmi and {Narayanan}, Desika and {Yates}, Robert M. and {Terp}, Chamilla and {Nezhad}, Negin and {Weaver}, John R. and {Witstok}, Joris and et al.},
        title = "{Inefficient dust production in a massive, metal-rich galaxy at $z=7.13$ uncovered by JWST and ALMA}",
      journal = {arXiv e-prints},
         year = 2025,
        month = oct,
          eid = {arXiv:2510.07936},
        pages = {arXiv:2510.07936},
          doi = {10.48550/arXiv.2510.07936},
archivePrefix = {arXiv},
       eprint = {2510.07936},
 primaryClass = {astro-ph.GA},
       adsurl = {https://ui.adsabs.harvard.edu/abs/2025arXiv251007936H}
}

@ARTICLE{Algera2026,
       author = {{Algera}, Hiddo S.~B. and {Rowland}, Lucie and {Stefanon}, Mauro and {Palla}, Marco and {Sommovigo}, Laura and {Inami}, Hanae and {Bouwens}, Rychard and {Aravena}, Manuel and {Bowler}, Rebecca A.~A. and {Dayal}, Pratika and et al.},
        title = "{REBELS-IFU: dust build-up in massive galaxies at redshift 7}",
      journal = {\mnras},
         year = 2026,
        month = jan,
       volume = {545},
       number = {2},
          eid = {staf1897},
        pages = {staf1897},
          doi = {10.1093/mnras/staf1897},
archivePrefix = {arXiv},
       eprint = {2501.10508},
 primaryClass = {astro-ph.GA},
       adsurl = {https://ui.adsabs.harvard.edu/abs/2026MNRAS.545f1897A}
}

@ARTICLE{Shapley2020,
       author = {{Shapley}, Alice E. and {Cullen}, Fergus and {Dunlop}, James S. and {McLure}, Ross J. and {Kriek}, Mariska and {Reddy}, Naveen A. and {Sanders}, Ryan L.},
        title = "{The First Robust Constraints on the Relationship between Dust-to-gas Ratio and Metallicity in Luminous Star-forming Galaxies at High Redshift}",
      journal = {\apjl},
         year = 2020,
        month = nov,
       volume = {903},
       number = {1},
          eid = {L16},
        pages = {L16},
          doi = {10.3847/2041-8213/abc006},
archivePrefix = {arXiv},
       eprint = {2009.10091},
 primaryClass = {astro-ph.GA},
       adsurl = {https://ui.adsabs.harvard.edu/abs/2020ApJ...903L..16S}
}

@ARTICLE{Popping2023,
       author = {{Popping}, Gerg{\"o} and {Shivaei}, Irene and {Sanders}, Ryan L. and {Jones}, Tucker and {Pope}, Alexandra and {Reddy}, Naveen A. and {Shapley}, Alice E. and {Coil}, Alison L. and {Kriek}, Mariska},
        title = "{The dust-to-gas mass ratio of luminous galaxies as a function of their metallicity at cosmic noon}",
      journal = {\aap},
         year = 2023,
        month = feb,
       volume = {670},
          eid = {A138},
        pages = {A138},
          doi = {10.1051/0004-6361/202243817},
archivePrefix = {arXiv},
       eprint = {2204.08483},
 primaryClass = {astro-ph.GA},
       adsurl = {https://ui.adsabs.harvard.edu/abs/2023A&A...670A.138P}
}

@ARTICLE{PoppingPeroux22,
       author = {{Popping}, Gerg{\"o} and {P{\'e}roux}, C{\'e}line},
        title = "{Observed cosmic evolution of galaxy dust properties with metallicity and tensions with models}",
      journal = {\mnras},
         year = 2022,
        month = jun,
       volume = {513},
       number = {1},
        pages = {1531-1543},
          doi = {10.1093/mnras/stac695},
archivePrefix = {arXiv},
       eprint = {2203.03686},
 primaryClass = {astro-ph.GA},
       adsurl = {https://ui.adsabs.harvard.edu/abs/2022MNRAS.513.1531P}
}

@ARTICLE{Asano2013,
       author = {{Asano}, Ryosuke S. and {Takeuchi}, Tsutomu T. and {Hirashita}, Hiroyuki and {Inoue}, Akio K.},
        title = "{Dust formation history of galaxies: A critical role of metallicity* for the dust mass growth by accreting materials in the interstellar medium}",
      journal = {Earth, Planets and Space},
         year = 2013,
        month = mar,
       volume = {65},
       number = {3},
        pages = {213-222},
          doi = {10.5047/eps.2012.04.014},
archivePrefix = {arXiv},
       eprint = {1206.0817},
 primaryClass = {astro-ph.GA},
       adsurl = {https://ui.adsabs.harvard.edu/abs/2013EP&S...65..213A}
}

@ARTICLE{Park2024,
       author = {{Park}, Hye-Jin and {Battisti}, Andrew J. and {Wisnioski}, Emily and {Cortese}, Luca and {Seibert}, Mark and {Grasha}, Kathryn and {Madore}, Barry F. and {Groves}, Brent and {Rich}, Jeff A. and {Beaton}, Rachael L. and et al.},
        title = "{The spatially resolved relation between dust, gas, and metal abundance with the TYPHOON survey}",
      journal = {\mnras},
         year = 2024,
        month = nov,
       volume = {535},
       number = {1},
        pages = {729-752},
          doi = {10.1093/mnras/stae2298},
archivePrefix = {arXiv},
       eprint = {2410.02222},
 primaryClass = {astro-ph.GA},
       adsurl = {https://ui.adsabs.harvard.edu/abs/2024MNRAS.535..729P}
}

@ARTICLE{Peroux2020,
       author = {{P{\'e}roux}, C{\'e}line and {Howk}, J. Christopher},
        title = "{The Cosmic Baryon and Metal Cycles}",
      journal = {\araa},
         year = 2020,
        month = aug,
       volume = {58},
        pages = {363-406},
          doi = {10.1146/annurev-astro-021820-120014},
archivePrefix = {arXiv},
       eprint = {2011.01935},
 primaryClass = {astro-ph.GA},
       adsurl = {https://ui.adsabs.harvard.edu/abs/2020ARA&A..58..363P}
}

@ARTICLE{DeVis2019,
       author = {{De Vis}, P. and {Jones}, A. and {Viaene}, S. and {Casasola}, V. and {Clark}, C.~J.~R. and {Baes}, M. and {Bianchi}, S. and {Cassara}, L.~P. and {Davies}, J.~I. and {De Looze}, I. and et al.},
        title = "{A systematic metallicity study of DustPedia galaxies reveals evolution in the dust-to-metal ratios}",
      journal = {\aap},
         year = 2019,
        month = mar,
       volume = {623},
          eid = {A5},
        pages = {A5},
          doi = {10.1051/0004-6361/201834444},
archivePrefix = {arXiv},
       eprint = {1901.09040},
 primaryClass = {astro-ph.GA},
       adsurl = {https://ui.adsabs.harvard.edu/abs/2019A&A...623A...5D}
}

@ARTICLE{Galliano2021,
       author = {{Galliano}, Fr{\'e}d{\'e}ric and {Nersesian}, Angelos and {Bianchi}, Simone and {De Looze}, Ilse and {Roychowdhury}, Sambit and {Baes}, Maarten and {Casasola}, Viviana and {Cassar{\'a}}, Letizia P. and {Dobbels}, Wouter and {Fritz}, Jacopo and et al.},
        title = "{A nearby galaxy perspective on dust evolution. Scaling relations and constraints on the dust build-up in galaxies with the DustPedia and DGS samples}",
      journal = {\aap},
         year = 2021,
        month = may,
       volume = {649},
          eid = {A18},
        pages = {A18},
          doi = {10.1051/0004-6361/202039701},
archivePrefix = {arXiv},
       eprint = {2101.00456},
 primaryClass = {astro-ph.GA},
       adsurl = {https://ui.adsabs.harvard.edu/abs/2021A&A...649A..18G}
}

@ARTICLE{RemyRuyer2014,
       author = {{R{\'e}my-Ruyer}, A. and {Madden}, S.~C. and {Galliano}, F. and {Galametz}, M. and {Takeuchi}, T.~T. and {Asano}, R.~S. and {Zhukovska}, S. and {Lebouteiller}, V. and {Cormier}, D. and {Jones}, A. and et al.},
        title = "{Gas-to-dust mass ratios in local galaxies over a 2 dex metallicity range}",
      journal = {\aap},
         year = 2014,
        month = mar,
       volume = {563},
          eid = {A31},
        pages = {A31},
          doi = {10.1051/0004-6361/201322803},
archivePrefix = {arXiv},
       eprint = {1312.3442},
 primaryClass = {astro-ph.GA},
       adsurl = {https://ui.adsabs.harvard.edu/abs/2014A&A...563A..31R}
}

@ARTICLE{Hodge2020,
       author = {{Hodge}, J.~A. and {da Cunha}, E.},
        title = "{High-redshift star formation in the Atacama large millimetre/submillimetre array era}",
      journal = {Royal Society Open Science},
         year = 2020,
        month = dec,
       volume = {7},
       number = {12},
          eid = {200556},
        pages = {200556},
          doi = {10.1098/rsos.200556},
archivePrefix = {arXiv},
       eprint = {2004.00934},
 primaryClass = {astro-ph.GA},
       adsurl = {https://ui.adsabs.harvard.edu/abs/2020RSOS....700556H}
}

@ARTICLE{Zavala2021,
       author = {{Zavala}, J.~A. and {Casey}, C.~M. and {Manning}, S.~M. and {Aravena}, M. and {Bethermin}, M. and {Caputi}, K.~I. and {Clements}, D.~L. and {Cunha}, E. da and {Drew}, P. and {Finkelstein}, S.~L. and et al.},
        title = "{The Evolution of the IR Luminosity Function and Dust-obscured Star Formation over the Past 13 Billion Years}",
      journal = {\apj},
         year = 2021,
        month = mar,
       volume = {909},
       number = {2},
          eid = {165},
        pages = {165},
          doi = {10.3847/1538-4357/abdb27},
archivePrefix = {arXiv},
       eprint = {2101.04734},
 primaryClass = {astro-ph.GA},
       adsurl = {https://ui.adsabs.harvard.edu/abs/2021ApJ...909..165Z}
}

@ARTICLE{Walter2020,
       author = {{Walter}, Fabian and {Carilli}, Chris and {Neeleman}, Marcel and {Decarli}, Roberto and {Popping}, Gerg{\"o} and {Somerville}, Rachel S. and {Aravena}, Manuel and {Bertoldi}, Frank and {Boogaard}, Leindert and {Cox}, Pierre and et al.},
        title = "{The Evolution of the Baryons Associated with Galaxies Averaged over Cosmic Time and Space}",
      journal = {\apj},
         year = 2020,
        month = oct,
       volume = {902},
       number = {2},
          eid = {111},
        pages = {111},
          doi = {10.3847/1538-4357/abb82e},
archivePrefix = {arXiv},
       eprint = {2009.11126},
 primaryClass = {astro-ph.GA},
       adsurl = {https://ui.adsabs.harvard.edu/abs/2020ApJ...902..111W}
}

@ARTICLE{Madau2014,
       author = {{Madau}, Piero and {Dickinson}, Mark},
        title = "{Cosmic Star-Formation History}",
      journal = {\araa},
         year = 2014,
        month = aug,
       volume = {52},
        pages = {415-486},
          doi = {10.1146/annurev-astro-081811-125615},
archivePrefix = {arXiv},
       eprint = {1403.0007},
 primaryClass = {astro-ph.CO},
       adsurl = {https://ui.adsabs.harvard.edu/abs/2014ARA&A..52..415M}
}

@ARTICLE{Shivaei2015,
       author = {{Shivaei}, Irene and {Reddy}, Naveen A. and {Shapley}, Alice E. and {Kriek}, Mariska and {Siana}, Brian and {Mobasher}, Bahram and {Coil}, Alison L. and {Freeman}, William R. and {Sanders}, Ryan and {Price}, Sedona H. and et al.},
        title = "{The MOSDEF Survey: Dissecting the Star Formation Rate versus Stellar Mass Relation Using H{\ensuremath{\alpha}} and H{\ensuremath{\beta}} Emission Lines at z {\ensuremath{\sim}} 2}",
      journal = {\apj},
         year = 2015,
        month = dec,
       volume = {815},
       number = {2},
          eid = {98},
        pages = {98},
          doi = {10.1088/0004-637X/815/2/98},
archivePrefix = {arXiv},
       eprint = {1507.03017},
 primaryClass = {astro-ph.GA},
       adsurl = {https://ui.adsabs.harvard.edu/abs/2015ApJ...815...98S}
}

@ARTICLE{Boogaard2020,
       author = {{Boogaard}, Leindert A. and {van der Werf}, Paul and {Weiss}, Axel and {Popping}, Gerg{\"o} and {Decarli}, Roberto and {Walter}, Fabian and {Aravena}, Manuel and {Bouwens}, Rychard and {Riechers}, Dominik and {Gonz{\'a}lez-L{\'o}pez}, Jorge and et al.},
        title = "{The ALMA Spectroscopic Survey in the Hubble Ultra Deep Field: CO Excitation and Atomic Carbon in Star-forming Galaxies at z = 1-3}",
      journal = {\apj},
         year = 2020,
        month = oct,
       volume = {902},
       number = {2},
          eid = {109},
        pages = {109},
          doi = {10.3847/1538-4357/abb82f},
archivePrefix = {arXiv},
       eprint = {2009.04348},
 primaryClass = {astro-ph.GA},
       adsurl = {https://ui.adsabs.harvard.edu/abs/2020ApJ...902..109B}
}

@ARTICLE{Accurso2017,
       author = {{Accurso}, G. and {Saintonge}, A. and {Catinella}, B. and {Cortese}, L. and {Dav{\'e}}, R. and {Dunsheath}, S.~H. and {Genzel}, R. and {Gracia-Carpio}, J. and {Heckman}, T.~M. and {Jimmy} and et al.},
        title = "{Deriving a multivariate {\ensuremath{\alpha}}$_{CO}$ conversion function using the [C II]/CO (1-0) ratio and its application to molecular gas scaling relations}",
      journal = {\mnras},
         year = 2017,
        month = oct,
       volume = {470},
       number = {4},
        pages = {4750-4766},
          doi = {10.1093/mnras/stx1556},
archivePrefix = {arXiv},
       eprint = {1702.03888},
 primaryClass = {astro-ph.GA},
       adsurl = {https://ui.adsabs.harvard.edu/abs/2017MNRAS.470.4750A}
}

@ARTICLE{DaCunha2013,
       author = {{da Cunha}, Elisabete and {Groves}, Brent and {Walter}, Fabian and {Decarli}, Roberto and {Weiss}, Axel and {Bertoldi}, Frank and {Carilli}, Chris and {Daddi}, Emanuele and {Elbaz}, David and {Ivison}, Rob and et al.},
        title = "{On the Effect of the Cosmic Microwave Background in High-redshift (Sub-)millimeter Observations}",
      journal = {\apj},
         year = 2013,
        month = mar,
       volume = {766},
       number = {1},
          eid = {13},
        pages = {13},
          doi = {10.1088/0004-637X/766/1/13},
archivePrefix = {arXiv},
       eprint = {1302.0844},
 primaryClass = {astro-ph.CO},
       adsurl = {https://ui.adsabs.harvard.edu/abs/2013ApJ...766...13D}
}

@ARTICLE{Sanders2023,
       author = {{Sanders}, Ryan L. and {Shapley}, Alice E. and {Jones}, Tucker and {Shivaei}, Irene and {Popping}, Gerg{\"o} and {Reddy}, Naveen A. and {Dav{\'e}}, Romeel and {Price}, Sedona H. and {Mobasher}, Bahram and {Kriek}, Mariska and et al.},
        title = "{CO Emission, Molecular Gas, and Metallicity in Main-sequence Star-forming Galaxies at z {\ensuremath{\sim}} 2.3}",
      journal = {\apj},
         year = 2023,
        month = jan,
       volume = {942},
       number = {1},
          eid = {24},
        pages = {24},
          doi = {10.3847/1538-4357/aca46f},
archivePrefix = {arXiv},
       eprint = {2204.06937},
 primaryClass = {astro-ph.GA},
       adsurl = {https://ui.adsabs.harvard.edu/abs/2023ApJ...942...24S}
}

@ARTICLE{Kriek2015,
       author = {{Kriek}, Mariska and {Shapley}, Alice E. and {Reddy}, Naveen A. and {Siana}, Brian and {Coil}, Alison L. and {Mobasher}, Bahram and {Freeman}, William R. and {de Groot}, Laura and {Price}, Sedona H. and {Sanders}, Ryan and et al.},
        title = "{The MOSFIRE Deep Evolution Field (MOSDEF) Survey: Rest-frame Optical Spectroscopy for \raisebox{-0.5ex}\textasciitilde1500 H-selected Galaxies at 1.37 < z < 3.8}",
      journal = {\apjs},
         year = 2015,
        month = jun,
       volume = {218},
       number = {2},
          eid = {15},
        pages = {15},
          doi = {10.1088/0067-0049/218/2/15},
archivePrefix = {arXiv},
       eprint = {1412.1835},
 primaryClass = {astro-ph.GA},
       adsurl = {https://ui.adsabs.harvard.edu/abs/2015ApJS..218...15K}
}

@ARTICLE{Shivaei22,
       author = {{Shivaei}, Irene and {Popping}, Gerg{\"o} and {Rieke}, George and {Reddy}, Naveen and {Pope}, Alexandra and {Kennicutt}, Robert and {Mobasher}, Bahram and {Coil}, Alison and {Fudamoto}, Yoshinobu and {Kriek}, Mariska and {Lyu}, Jianwei and {Oesch}, Pascal and {Sanders}, Ryan and {Shapley}, Alice and {Siana}, Brian},
        title = "{Infrared Spectral Energy Distributions and Dust Masses of Sub-solar Metallicity Galaxies at z   2.3}",
      journal = {\apj},
         year = 2022,
        month = mar,
       volume = {928},
       number = {1},
          eid = {68},
        pages = {68},
          doi = {10.3847/1538-4357/ac54a9},
archivePrefix = {arXiv},
       eprint = {2201.04270},
 primaryClass = {astro-ph.GA},
       adsurl = {https://ui.adsabs.harvard.edu/abs/2022ApJ...928...68S}
}

@ARTICLE{Madden13,
       author = {{Madden}, S.~C. and {R{\'e}my-Ruyer}, A. and {Galametz}, M. and {Cormier}, D. and {Lebouteiller}, V. and {Galliano}, F. and {Hony}, S. and {Bendo}, G.~J. and {Smith}, M.~W.~L. and {Pohlen}, M. and {Roussel}, H. and {Sauvage}, M. and {Wu}, R. and {Sturm}, E. and {Poglitsch}, A. and {Contursi}, A. and {Doublier}, V. and {Baes}, M. and {Barlow}, M.~J. and {Boselli}, A. and {Boquien}, M. and {Carlson}, L.~R. and {Ciesla}, L. and {Cooray}, A. and {Cortese}, L. and {de Looze}, I. and {Irwin}, J.~A. and {Isaak}, K. and {Kamenetzky}, J. and {Karczewski}, O. {\L}. and {Lu}, N. and {MacHattie}, J.~A. and {O'Halloran}, B. and {Parkin}, T.~J. and {Rangwala}, N. and {Schirm}, M.~R.~P. and {Schulz}, B. and {Spinoglio}, L. and {Vaccari}, M. and {Wilson}, C.~D. and {Wozniak}, H.},
        title = "{An Overview of the Dwarf Galaxy Survey}",
      journal = {\pasp},
         year = 2013,
        month = jun,
       volume = {125},
       number = {928},
        pages = {600},
          doi = {10.1086/671138},
archivePrefix = {arXiv},
       eprint = {1305.2628},
 primaryClass = {astro-ph.GA},
       adsurl = {https://ui.adsabs.harvard.edu/abs/2013PASP..125..600M}
}

@ARTICLE{Davies17,
       author = {{Davies}, J.~I. and {Baes}, M. and {Bianchi}, S. and {Jones}, A. and {Madden}, S. and {Xilouris}, M. and {Bocchio}, M. and {Casasola}, V. and {Cassara}, L. and {Clark}, C. and {De Looze}, I. and {Evans}, R. and {Fritz}, J. and {Galametz}, M. and {Galliano}, F. and {Lianou}, S. and {Mosenkov}, A.~V. and {Smith}, M. and {Verstocken}, S. and {Viaene}, S. and {Vika}, M. and {Wagle}, G. and {Ysard}, N.},
        title = "{DustPedia: A Definitive Study of Cosmic Dust in the Local Universe}",
      journal = {\pasp},
         year = 2017,
        month = apr,
       volume = {129},
       number = {974},
        pages = {044102},
          doi = {10.1088/1538-3873/129/974/044102},
archivePrefix = {arXiv},
       eprint = {1609.06138},
 primaryClass = {astro-ph.GA},
       adsurl = {https://ui.adsabs.harvard.edu/abs/2017PASP..129d4102D}
}

\appendix
\section{ACE DGR$_{\rm mol}$ catalogue}
In Table~\ref{tab:physical_sample_by_metallicity}, we show properties used in the observable and physical dust-to-molecular gas analysis for the sample of 25 ACE galaxies. 
\label{tablerefproperties}
\begin{table*}
\centering
\caption{Properties used in the observable and physical dust-to-gas analysis for the ACE sample, sorted by metallicity. Horizontal rules mark the metallicity bins $12+\log(\mathrm{O/H}) < 8.38$ ($N=12$), $8.38 \leq 12+\log(\mathrm{O/H}) < 8.48$ ($N=6$), and $12+\log(\mathrm{O/H}) \geq 8.48$ ($N=7$). Masses and luminosities are shown as $\log_{10}$ values; entries prefixed with $<$ are 3$\sigma$ upper limits.}
\label{tab:physical_sample_by_metallicity}
\renewcommand{\arraystretch}{1.2}
\begin{tabular}{llllllll}
\toprule
V4ID & V2ID & $z$ & $\log L_{\nu,873}$ & $\log L'_{\mathrm{CO(3-2)}}$ & $\log M_{\mathrm{dust}}$ & $\log M_{\rm mol}$ & $12+\log(\mathrm{O/H})$ \\
 & & & $[\mathrm{erg\ s^{-1}\ Hz^{-1}}]$ & $[\mathrm{K\ km\ s^{-1}\ pc^2}]$ & $[M_\odot]$ & $[M_\odot]$ & \\
\midrule
3626 & 3623 & 2.325 & $<\,31.15$ & $<\,8.98$ & $<\,7.75$ & $<\,10.56$ & $8.19^{+0.03}_{-0.03}$ \\
6283 & 6486 & 2.224 & $<\,31.10$ & $<\,8.99$ & $<\,7.74$ & $<\,10.57$ & $8.22^{+0.02}_{-0.03}$ \\
19439 & 20331 & 2.466 & $<\,31.24$ & $<\,9.00$ & $<\,7.79$ & $<\,10.59$ & $8.22^{+0.02}_{-0.02}$ \\
8515 & 8824 & 2.454 & $<\,31.16$ & $<\,9.11$ & $<\,7.72$ & $<\,10.64$ & $8.27^{+0.02}_{-0.03}$ \\
3773 & 3784 & 2.425 & $<\,31.21$ & $<\,9.06$ & $<\,7.78$ & $<\,10.57$ & $8.27^{+0.03}_{-0.03}$ \\
25229 & 26417 & 2.181 & $<\,31.08$ & $9.02^{+0.11}_{-0.15}$ & $<\,7.73$ & $10.45^{+0.14}_{-0.14}$ & $8.29^{+0.03}_{-0.03}$ \\
3666 & 3692 & 2.086 & $31.45^{+0.11}_{-0.14}$ & $9.02^{+0.09}_{-0.11}$ & $8.15^{+0.27}_{-0.21}$ & $10.52^{+0.10}_{-0.10}$ & $8.29^{+0.02}_{-0.02}$ \\
19985 & 20891 & 2.188 & $31.51^{+0.05}_{-0.06}$ & $9.19^{+0.08}_{-0.11}$ & $8.17^{+0.26}_{-0.18}$ & $10.63^{+0.10}_{-0.10}$ & $8.31^{+0.01}_{-0.01}$ \\
24020 & 25153 & 2.092 & $<\,31.02$ & $<\,8.89$ & $<\,7.70$ & $<\,10.29$ & $8.32^{+0.03}_{-0.03}$ \\
6750 & 6976 & 2.127 & $31.04^{+0.11}_{-0.16}$ & $<\,8.93$ & $7.72^{+0.27}_{-0.23}$ & $<\,10.31$ & $8.35^{+0.02}_{-0.02}$ \\
9971 & 10394 & 2.411 & $31.43^{+0.10}_{-0.14}$ & $9.36^{+0.10}_{-0.13}$ & $8.02^{+0.28}_{-0.23}$ & $10.70^{+0.11}_{-0.11}$ & $8.36^{+0.02}_{-0.02}$ \\
9393 & 9773 & 2.413 & $31.46^{+0.10}_{-0.12}$ & $9.07^{+0.10}_{-0.12}$ & $8.04^{+0.29}_{-0.23}$ & $10.41^{+0.11}_{-0.12}$ & $8.37^{+0.02}_{-0.02}$ \\
\midrule
22193 & 23203 & 2.465 & $<\,31.19$ & $<\,9.03$ & $<\,7.74$ & $<\,10.25$ & $8.44^{+0.04}_{-0.04}$ \\
24763 & 25825 & 2.464 & $31.71^{+0.06}_{-0.07}$ & $9.32^{+0.08}_{-0.09}$ & $8.28^{+0.28}_{-0.21}$ & $10.55^{+0.12}_{-0.12}$ & $8.45^{+0.05}_{-0.05}$ \\
5901 & 6069 & 2.396 & $31.32^{+0.09}_{-0.12}$ & $<\,9.04$ & $7.91^{+0.28}_{-0.22}$ & $<\,10.25$ & $8.45^{+0.02}_{-0.03}$ \\
16594 & 17482 & 2.286 & $31.24^{+0.10}_{-0.13}$ & $9.24^{+0.08}_{-0.10}$ & $7.87^{+0.27}_{-0.23}$ & $10.43^{+0.10}_{-0.10}$ & $8.45^{+0.03}_{-0.03}$ \\
5814 & 5988 & 2.127 & $31.63^{+0.07}_{-0.08}$ & $9.46^{+0.07}_{-0.08}$ & $8.31^{+0.26}_{-0.19}$ & $10.67^{+0.08}_{-0.08}$ & $8.45^{+0.02}_{-0.02}$ \\
5094 & 5242 & 2.171 & $31.58^{+0.06}_{-0.07}$ & $9.68^{+0.08}_{-0.11}$ & $8.25^{+0.26}_{-0.19}$ & $10.84^{+0.12}_{-0.13}$ & $8.47^{+0.05}_{-0.06}$ \\
\midrule
21955 & 22954 & 2.468 & $31.38^{+0.09}_{-0.12}$ & $9.33^{+0.09}_{-0.12}$ & $7.95^{+0.29}_{-0.23}$ & $10.46^{+0.11}_{-0.12}$ & $8.50^{+0.03}_{-0.03}$ \\
8280 & 8593 & 2.494 & $31.90^{+0.04}_{-0.04}$ & $9.57^{+0.05}_{-0.05}$ & $8.46^{+0.27}_{-0.19}$ & $10.65^{+0.09}_{-0.10}$ & $8.51^{+0.05}_{-0.05}$ \\
4497 & 4611 & 2.441 & $31.31^{+0.10}_{-0.12}$ & $9.33^{+0.08}_{-0.10}$ & $7.89^{+0.28}_{-0.23}$ & $10.42^{+0.11}_{-0.11}$ & $8.51^{+0.04}_{-0.04}$ \\
19013 & 19876 & 2.457 & $31.59^{+0.06}_{-0.08}$ & $9.26^{+0.10}_{-0.12}$ & $8.15^{+0.28}_{-0.21}$ & $10.35^{+0.12}_{-0.12}$ & $8.52^{+0.03}_{-0.03}$ \\
13701 & 13985 & 2.166 & $31.76^{+0.05}_{-0.05}$ & $9.47^{+0.08}_{-0.09}$ & $8.42^{+0.27}_{-0.19}$ & $10.52^{+0.09}_{-0.10}$ & $8.55^{+0.02}_{-0.03}$ \\
3324 & 3284 & 2.307 & $31.62^{+0.06}_{-0.06}$ & $9.39^{+0.08}_{-0.10}$ & $8.23^{+0.27}_{-0.19}$ & $10.40^{+0.11}_{-0.12}$ & $8.57^{+0.05}_{-0.05}$ \\
13296 & 13899 & 2.167 & $31.42^{+0.09}_{-0.11}$ & $9.35^{+0.08}_{-0.10}$ & $8.09^{+0.27}_{-0.21}$ & $10.26^{+0.12}_{-0.12}$ & $8.61^{+0.05}_{-0.05}$ \\
\bottomrule
\end{tabular}
\renewcommand{\arraystretch}{1.0}
\end{table*}

\section{[C\,{\sc ii}] litereature comparison}
\label{appendixCII}
\begin{figure*}[h!]
    \centering
    \includegraphics[width=0.95\textwidth]{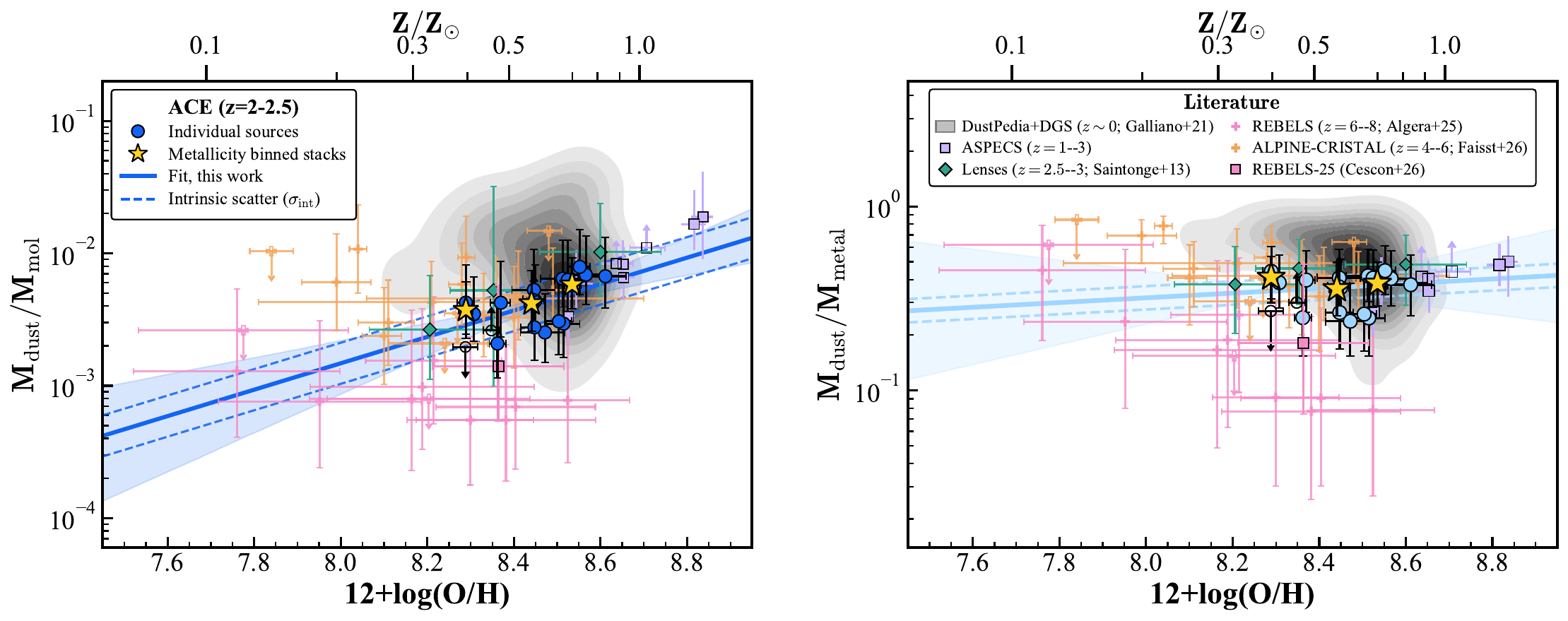}
  \caption{\textbf{Left:}Dust-to-gas ratio, expressed as \(M_{\rm dust}/M_{\rm mol}\), as a function of metallicity for the ACE sample at \(z=2\)--2.5, compared with local and high-redshift literature measurements. \textbf{Right:} Dust-to-metal ratio, expressed as \(M_{\rm dust}/M_{\rm metal}\), as a function of metallicity. Markers for ACE and CO literature comparison samples are the same as in Figure~\ref{fig:MdustMmol} and Figure~\ref{fig:MdustMmetal}. Literature comparison samples include DustPedia+DGS at \(z\sim0\), ASPECS at \(z=1\)--3, and higher-redshift [C\,{\sc ii}] samples from ALPINE at \(z\sim4-6\) and REBELS at \(z\sim6-8\), extending the comparison to lower metallicities.}
    \label{fig:MdustMgasCII}
\end{figure*}
To examine whether the lower $\rm DGR_{mol}$ values seen in the $z>4$ samples are related to metallicity, we show the ALPINE-CRISTAL and REBELS-IFU measurements in the $\rm DGR_{mol}$--metallicity plane in the left panel of Figure~\ref{fig:MdustMgasCII}. The metallicities are determined using the same strong-line calibration as for ACE \citep{Sanders2026}. For the REBELS-IFU galaxies, however, we use only the emission-line ratios considered by \cite{Rowland2026}, as these diagnostics are found to be more appropriate at $z>6$ than the combination of multiple line ratios used the lower-redshift samples.   As described in Section~\ref{redshiftevo} we adopt the molecular gas masses derived using the the constant $\alpha_{\rm [CII]}$ conversion factor from \citet{Zanella2018}. For ALPINE-CRISTAL, we recompute the dust masses from the single-band ALMA measurements using the assumptions adopted for ACE, whereas for REBELS--IFU we use the dust masses reported by \citet{Algera2026}. The ALPINE-CRISTAL galaxies broadly overlap with the ACE measurements, while the REBELS--IFU sources generally occupy lower $\rm DGR_{mol}$ values. Neither sample, however, shows a clear trend with metallicity. We also show, REBELS-25 at ($z=7.3$), for which a CO-based molecular gas mass is available \citep{Cescon2026}. REBELS-25, shown with a different symbol from the [C{\sc ii}]-based REBELS sample, has a metallicity of $12+\log({\rm O/H})=8.36\pm0.15$ and lies in the same general region of the dust-to-molecular-gas plane as the other REBELS galaxies, and at fixed metallicity just below the ACE measurements. 

The uncertainties for the [C{\sc ii}]-based samples are substantially larger than those for ACE and the other CO-based literature samples, and their interpretation is further limited by the uncertainty in $\alpha_{\rm [CII]}$ (see e.g. \citet{Sommovigo2022, Algera2026}). Simulations suggest that $\alpha_{\rm [CII]}$ may vary with metallicity \citep{Vallini2025}, analogous to the metallicity dependence commonly adopted for $\alpha_{\rm CO}$, but the currently available predictions are calibrated for conversion to total gas mass and are based on the intrinsic [C{\sc ii}] luminosities of the simulations. We therefore do not adopt a metallicity-dependent $\alpha_{\rm [CII]}$ here. When included in the left panel of Figure~\ref{fig:MdustMgasCII}, the [C{\sc ii}]-based data suggest possible mild evolution toward lower $\rm DGR_{mol}$ at $z\gtrsim4$, driven primarily by the REBELS sample. Thus, although the current observations do not reveal a strong metallicity dependence within the high-redshift [C{\sc ii}] samples, they may indicate a lower normalization of $\rm DGR_{mol}$ at the highest redshifts.

A similar pictures emerges for the dust-to-metal ratio shown in the right panel of Figure~\ref{fig:MdustMgasCII}, at fixed metallicity, the REBELS galaxies generally lie below the ACE sample indicating lower fraction of metals locked into dust. In contrast to the ($\rm DGR_{mol}$), both REBELS and ALPINE appear to show a decrease in dust-to-metal ratio with increasing metallicity that is not seen in ACE and the other cosmic noon samples. 

\section{From molecular gas to total gas}
\label{totalgasappendix}
\begin{figure}[h!]
    \centering
    \includegraphics[width=\linewidth]{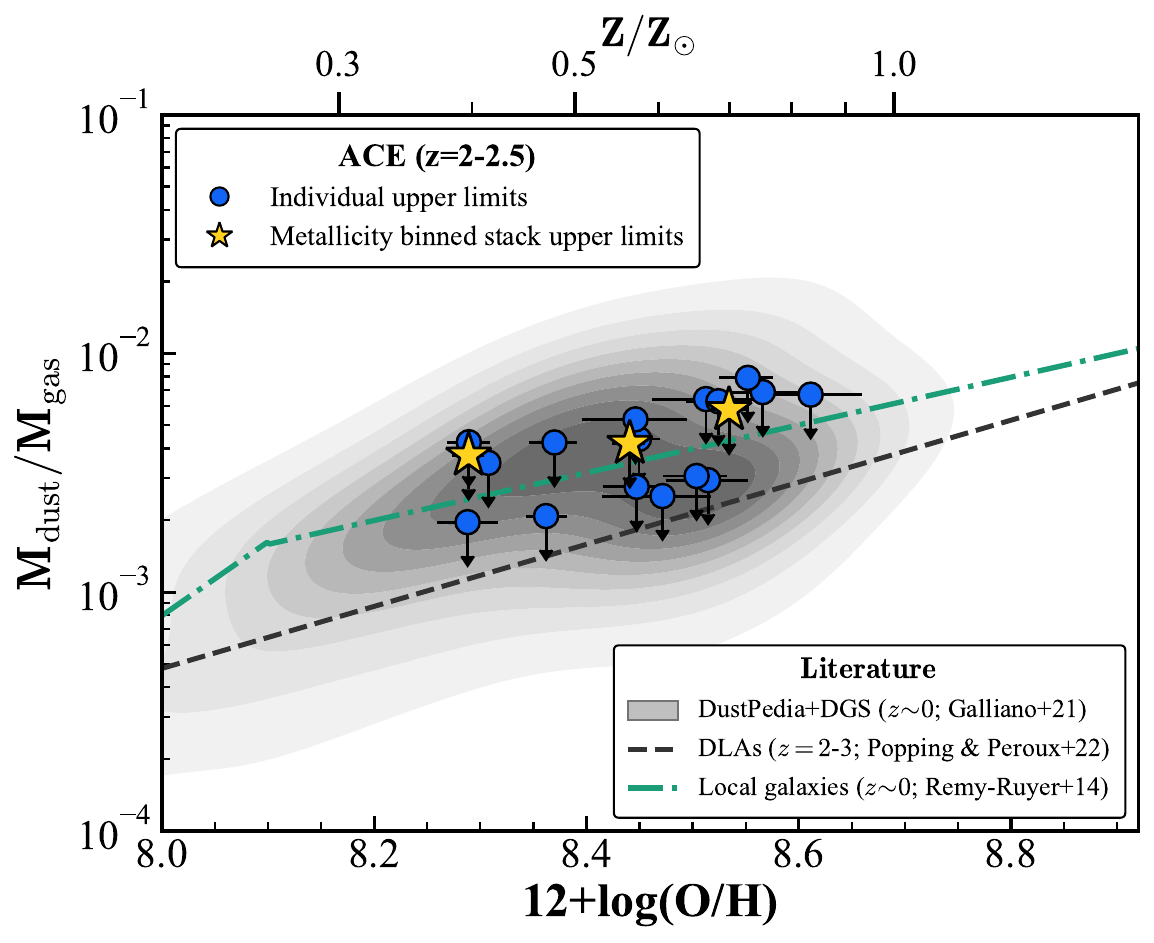}
    \caption{Dust-to-gas ratio, expressed as $M_{\rm dust}/M_{\rm gas}$, as a function of  metallicity for the ACE sample at $z=2-2.5$. The ACE sample is shown with the same symbols as in Figure~\ref{fig:MdustMmol}. For comparison, we also show the local DustPedia+DGS observations at $z\sim0$ from \citet{Galliano2021} (grey contours), the broken power-law relation from \citet{RemyRuyer2014} (green dash-dotted curve), and DLA measurements at $z=2-3$ from \citet{PoppingPeroux22} (black dashed line). While the ACE points still reflect $\rm DGR_{mol}$, the local literature samples include the atomic gas component in their gas masses. Despite these different gas definitions, the ACE relation shows little offset from the local observations.}
    \label{fig:MdustMgas}
\end{figure}
Most literature studies consider the dust-to-total-gas ratio, where the total gas mass is defined as $M_{\rm gas}=M_{\rm mol}+M_{\rm atomic}$. In Fig.~\ref{fig:MdustMgas}, we therefore place the ACE sample in the $\rm DGR_{tot}$--metallicity plane to facilitate comparison with local-Universe samples, for which both molecular and atomic gas measurements are available, as well as with absorber-based constraints at similar redshift. For the ACE galaxies, only the molecular gas component is directly constrained, so the ACE points shown in this plane are upper limits on the total $\rm DGR_{tot}$ ratio. By contrast, the local comparison samples are plotted using their literature values for the total gas mass, such that their ratios correspond to $\rm DGR_{tot}$.

We compare the ACE galaxies to the local DustPedia+DGS sample \citep{Galliano2021}, which is shown here including the atomic gas component, to the broken power-law relation of \citet{RemyRuyer2014}, and to the DLA relation of \citet{PoppingPeroux22}. The \cite{RemyRuyer2014} sample comprises 126 nearby galaxies drawn from the DGS \citep{Madden2013}, Key Insights on Nearby Galaxies: a Far-Infrared Survey with Herschel \citep[KINGFISH;][]{Kennicutt2011}, and an additional literature subsample from \citet{Galametz2011}. Using a metallicity-dependent $\alpha_{\rm CO}$ \citep{Schruba2012}, they find a turnover at $12+\log({\rm O/H})\simeq 8.1$; below this metallicity, where CO detections become rare, they estimate molecular gas masses for galaxies without CO detections by adopting $M_{\rm mol}=0.68\,M_{\rm HI}$, so that their total gas masses include both the measured atomic gas and an inferred molecular component. Their dust masses were derived using the SED-modelling framework of \citet{Galliano2011}. Our ACE sample does not extend to sufficiently low metallicities to test or quantify such a turnover directly. We also include the fit lines from damped Lyman-$\alpha$ absorbers at cosmic noon, which lie systematically below the local relation at fixed metallicity. This offset is likely not driven by the metallicity or atomic gas contribution alone, but also by the very different selection function of absorber samples. Unlike ACE, which probes the dust-rich molecular ISM of actively star-forming galaxies, absorbers measure neutral gas along sightlines that may intersect outer disc gas or circumgalactic material and are therefore more sensitive to diffuse, H\,{\sc i}-dominated reservoirs. Since DLA sightlines are not guaranteed to trace the central ISM, and may also be associated with lower-mass host galaxies on average \citep{Khare2007}, their lower normalization could reflect either a host-mass effect, an environmental effect, or a combination of both. 

Despite the different gas definitions, the ACE upper limits lie close to the local $\rm DGR_{tot}$--metallicity relation. Because the atomic gas content of the ACE galaxies is  unknown, this comparison remains provisional: if these galaxies host substantial H\,{\sc i} reservoirs, their positions would shift downward in the $\rm DGR_{tot}$ plane, potentially moving them below the local relation and into closer agreement with absorber-based constraints at comparable redshift. Models and simulations predict substantial but uncertain atomic gas contributions at these redshifts, ranging from $\sim25\%$ to $\gtrsim50\%$ of the cold gas reservoir depending on galaxy properties such as stellar mass and metallicity, as well as on the adopted H\,{\sc i} and H$_2$ partitioning prescriptions \citep[e.g.][]{Lagos2011,Popping2014,Popping2015model,Morselli2021}. For galaxies with $M_\star\sim10^{10},M_\odot$, comparable to the median stellar mass of the ACE sample, semi-analytic and semi-empirical models predict an H\,{\sc i} contribution of order $\sim25\%$ at $z\sim2$ \citep{Lagos2011,Popping2014,Popping2015model}. In the local Universe, H\,{\sc i} typically extends well beyond the stellar and molecular disks of galaxies \citep{Saintonge2022}, and a similar picture may be expected at higher redshifts, with extended H\,{\sc i} reservoirs surrounding the central molecular, star-forming regions \citep{Walter2020}. Thus, the total H\,{\sc i} content of a galaxy may be substantially larger than the atomic gas associated with the region traced by its CO and dust emission. Future H\,{\sc i} surveys with the Square Kilometre Array (SKA) will provide increasingly direct constraints on the atomic gas content of galaxies out to $z\sim1$ \citep{StaveleySmith2015}. At cosmic noon, where individual H\,{\sc i} detections will remain challenging, deep observations and stacking with the SKA, will provide complementary constraints on the atomic gas reservoir at $z\sim1-3$ \citep{MessiasSKA2024}.

\end{document}